\documentclass[fp,twocolumn]{jpsj3}
\usepackage{bm}
\usepackage{amsmath}
\usepackage{txfonts}
\usepackage{graphicx}
\usepackage{color}

\title{Developments in the Tensor Network --- from Statistical Mechanics to Quantum Entanglement}

\author{
Kouichi Okunishi$^1$\thanks{okunishi@phys.sc.niigata-u.ac.jp}, 
Tomotoshi Nishino$^2$\thanks{nishino@kobe-u.ac.jp} 
 and  Hiroshi Ueda$^{3,4,5}$\thanks{h\_ueda@qiqb.osaka-u.ac.jp} 
}
\inst{
$^1$Department of Physics, Faculty of Science, Niigata University, Niigata 950 -2181, Japan \\
$^2$Department of Physics, Graduate School of Science, Kobe University, Kobe 657-8501, Japan \\
$^3$Center for Quantum Information and Quantum Biology, Osaka University,  Toyonaka 560-0043, Japan \\
$^4$JST, PRESTO, Kawaguchi 332-0012, Japan \\
$^5$Computational Materials Science Research Team, RIKEN Center for Computational Science (R-CCS), Kobe 650-0047, Japan
}

\abst{
Tensor networks (TNs) have become one of the most essential building blocks for various fields of theoretical physics such as condensed matter theory, statistical mechanics, quantum information, and quantum gravity.
This review provides a unified description of a series of developments in the TN from the statistical mechanics side. 
In particular, we begin with the variational principle for the transfer matrix of the 2D Ising model, which naturally leads us to the matrix product state (MPS) and the corner transfer matrix (CTM).
We then explain how the CTM can be evolved to such MPS-based approaches as density matrix renormalization group (DMRG) and infinite time-evolved block decimation.
We also elucidate that the finite-size DMRG played an intrinsic role for incorporating various quantum information concepts in subsequent developments in the TN.
After surveying higher-dimensional generalizations like tensor product states or projected entangled pair states, we describe tensor renormalization groups (TRGs), which are a fusion of TNs and Kadanoff-Wilson type real-space renormalization groups, focusing on their fixed point structures.
We then discuss how the difficulty in TRGs for critical systems can be overcome in the tensor network renormalization and the multi-scale entanglement renormalization ansatz.  
}

\begin{document}
\maketitle

\section{Overview}

Tensor networks (TNs) have been providing deep insights for understanding essential physics embedded in quantum and classical many-body systems.
Also, several developments in TN simulation techniques incorporating the concept of quantum entanglement enable us to quantitatively analyze various interesting phenomena inherent in the many-body systems.
A main reason for such successes of the TNs is that it can provide a clear answer to a fundamental question in physics;
 How can we extract effective degrees of freedom representing essential physics embedded in a huge number of degrees of freedom/huge dimension of the Hilbert space of the many-body systems?
This question is deeply related to the concept of renormalization group (RG).
Thus,  theoretical backgrounds of the TN have been an essential issue from both viewpoints of theoretical physics and practical computational physics.

We focus on the theoretical background behind the TN formulation of quantum and classical many-body problems, rather than the technical aspects. 
Of course, the TN already has various styles and many applications\cite{Peschel1999,SchollwoeckRMP,Schollwoeck,Verstraete2008,Orus2014review, Montangero2018,Zeng_book, Biamonte2019lectures, Ran2020}, all of which we cannot cover in this review.
Thus, we particularly cut into the issue from such a statistical mechanical problem as variational approximation for two-dimensional (2D) Ising model, and then evolve the argument to quantum lattice systems.
This is because the path integral representation directly relates the $d$D quantum many-body problems with the ($d+1$)D classical lattice statistics. 
In other words, the transfer-matrix formalism of ($d+1$)D classical lattice models is mathematically equivalent to the imaginary-time formulation of quantum many-body systems, where the quantum Hamiltonian can be interpreted as an anisotropic limit of the transfer matrix.
For the classical lattice statistics, however, the space and (imaginary) time structure of the lattice can be treated on the equal footing, which provides a clear view for the matrix product state (MPS) decomposition of the maximal eigenvector of the transfer matrix (the ground-state wavefunction of the Hamiltonian for the quantum case). 
 After explaining the fundamentals of the TN structure for  2D classical or 1D quantum systems, we proceed to other related developments in TNs  such as higher dimensions, real/imaginary time evolution, tensor renormalization groups, etc. 
We think that this route would be the best approach to the unified understanding of the TN, with emphasizing the notion that the concept of many-body entanglement is equally relevant to the transfer matrix/Hamiltonian formulations for classical/quantum many-body systems.

In the TN formulation of classical spin systems, the maximal eigenvector of the transfer matrix (the ground-state wavefunction of the Hamiltonian for the quantum case) is represented as a contraction of local tensors with respect to auxiliary spin indices.
Then, there are two essential requirements to obtain the optimal TN state. 
 One is the variational principle for the TN state and the other is how to construct efficient RG-like transformations to extract effective degrees of freedom. 
A key ingredient satisfying these requirements is singular value decomposition (SVD) for a variational TN state through its matrix/tensor representation.
For the 2D classical model, the low-rank approximation based on the SVD was implicitly used in the form of a corner transfer matrix (CTM) in the bulk limit through the matrix eigenvalue problem\cite{Baxter1968, Baxter1978}. 
A combination of the SVD and a real-space RG for 1D quantum systems was explicitly introduced by S. ~R. ~White in the density matrix renormalization group (DMRG)\cite{White1992,White1993}.
After the DMRG, several TN algorithms assisted by quantum information ideas have been rapidly expanded to various aspects of many-body problems.
In particular, the application of SVD to a quantum many-body state is essentially equivalent to Schmidt decomposition\cite{Schmidt1907,Ekert1995} in the quantum information context, and entanglement entropy (EE) and entanglement spectrum respectively defined as the von Neumann entropy and the logarithm of the singular value spectrum provide very useful information for characterizing its entanglement structure.
Interestingly, such entanglement analysis inspired by quantum information has given a rise to significant feedback to the formulation of real-space RGs.
On the basis of the area law of EE with the log-correction term,   entanglement renormalizations, i.e.  multi-scale entanglement renormalization ansatz (MERA)\cite{MERA2007} and tensor network renormalization (TNR)\cite{TNR2015} were designed.
These entanglement RG approaches enable us to extract numerically exact critical phenomena in the framework of the real-space RG for the first time since Kadanoff's proposal in 1966.\cite{Kadanoff1966,Efrati2014}

In the following sections, we explain the TN from statistical mechanical viewpoint. 
In \S 2, we briefly summarize history of the TN and associated researches from the modern viewpoint.
If a reader is directly interested in the formulation of the TN,  he/she may skip this section, but a history of the TN often provides interesting and instructive information.
In \S 3, we introduce the square-lattice Ising model as a typical example of the TN. 
We discuss the variational evaluation of the partition function and the free energy on the basis of Baxter's CTM, where the MPS is introduced in a very natural way without passing through the Schmidt decomposition.
In \S 4, we explain the corner transfer matrix renormalization group (CTMRG), which is a prototype of modern TN approaches. 
We then proceed to discussions of a variety of MPS-type algorithms for 1D quantum systems in \S 5. 
In this section, we also mention the role of the DMRG in the context of the modern TNs.
In \S 6,  we consider higher-dimensional generalizations of the MPS:
tensor product states (TPS) or projected entangled pair states (PEPS).
In \S 7, we discuss the relation between TN approaches and real-space RGs, where we particularly focus on the fixed point structures of tensor-renormalization-group (TRG) type algorithms.
We then explain how the TNR and MERA overcame the difficulty in the TRGs for critical systems.
In \S 8, we briefly mention recent trends and possible developments in the TN.
In Appendix, we provide a list of useful TN packages.

\section{Tensor network history}

Around the beginning of the 21st century, the concept of MPS, which has been a fundamental part of the TN, was integrated from several pioneering studies independently developed in various research fields in theoretical physics. 
Let us begin with the chronological sequence in the earlier developments which led to the modern MPS formalism.

\subsection{Early Development of Matrix Product State}

To our knowledge, the earliest example of MPS dates back to the Kramers-Wannier approximation for the 2D Ising model in 1941.\cite{KW}
A key idea of the approximation was that a variational state for the row-to-row transfer matrix was represented as the thermal equilibrium state of the 1D Ising model under an effective magnetic field, which can be written as a product of $2\times2$ matrices of effective Boltzmann weight.
Thus, the variational state of the Kramers-Wannier type can be viewed as a prototype of the MPS\cite{KW_MPS}. 
A systematical generalization of the 1D variational state was proposed by Baxter in 1968 for an analysis of the dimer model on a square lattice,~\cite{Baxter1968} which is basically equivalent to  the infinite MPS nowadays. 
This variational state was constructed as a contraction of 3-leg tensors aligned in the row(or equivalently column) direction, where the local 3-leg tensor with the auxiliary degrees of freedom is a generalization of the effective Boltzmann weight in the Kramers-Wannier case.
Assuming the uniform MPS in the bulk limit, moreover, he derived a closed form of self-consistent equations for the corner transfer matrix (CTM).~\cite{Baxter1968, Baxter1978}
Note that the product of four CTMs is basically equivalent to the reduced density matrix under the half-infinite bipartitioning of a 1D quantum system. 
A recursive method optimizing the variational state through matrix diagonalization of CTMs was explained in  \S 13 of his textbook, \cite{BaxterBook} where some of the fundamental ideas of TNs  were presented about two decades earlier than any other else.

A quantum-system counterpart of the MPS  can be attributed to the valence-bond solid (VBS) state, which is the exact ground-state wavefunction of Affleck-Kennedy-Lieb-Tasaki (AKLT) chain proposed in 1987~\cite{AKLT1,AKLT2} for understanding physics of the Haldane conjecture.
In the VBS state,  physical $S=1$ spins are correlated or entangled through the auxiliary $S=1/2$ spins.
Then, an essential point is that the connectivity of the auxiliary $S=1/2$ spins can be represented in the matrix product form, which enables us to straightforwardly construct the MPS representation of the VBS state.
Using the uniform and finite-dimensional MPS,  Fannes {\it et al}  proved that the correlation length of the VBS states is always finite.~\cite{Fannes1989,Fannes1992,Fannes1996} 
The term `matrix product ground state' firstly appeared in Refs. \citen{Klumper1991, Klumper1992,Klumper1993}, where the VBS state was also explicitly written down in the MPS form.
Recently, the VBS state is well known as a typical example of the symmetry-protected topological (SPT) order/entanglement in quantum many-body systems\cite{Gu_Filtering2009,Pollmann2010}.

Here, it is worth mentioning that  the nonlocal string order characterizing the VBS state was originally proposed by  Rommelse and den Nijs for the disordered flat phase of a restricted solid on solid model~\cite{denNijs1,denNijs2}.
Moreover, the variational state in  Baxter's form was used for a precise estimation of the ground-state energy of $S = 1$ Heisenberg chain~\cite{Nightingale} in the context of the Haldane conjecture \cite{Haldane}, though the recursive formulation for the CTM was not directly applied to 1D quantum systems at that time.
Nevertheless, it is worth noting that the MPS structure of the eigenvector is assumed in the exact solution of the eight-vertex model\cite{Baxter1972}, and in the vertex operator approach to the XXZ chain\cite{Davies1993}.
The direct and explicit connection between the MPS and the algebraic Bethe ansatz\cite{Faddeev1995,Korepin1993} for 1D quantum integrable systems was revisited in recent works.\cite{Alcaraz2006,Katsura2010,Murg2012}
These suggest that the statistical mechanics viewpoints often played an intrinsic role in revealing quantum many-body physics.

\subsection{After Density Matrix Renormalization Group}

The modern stream of the TN began in 1992 with the invention of the DMRG by S.R. White.~\cite{White1992,White1993, preDMRG} 
After the success of DMRG for the $S = 1/2$ and $S = 1$ Heisenberg chains, it has been extensively applied to 1D quantum many-body problems and been established as a de-facto-standard numerical RG method for condensed matter physics.~\cite{Peschel1999,SchollwoeckRMP}
Strictly speaking, however, the DMRG is not a conventional real-space RG, although its name contains ``renormalization group''.
This is because no rescaling of the length scale is involved in its formulation where the bipartitioned system is iteratively updated with the combination of the low-rank approximation of SVD for the ground-state wavefunction and insertion of local two sites at the center of the system Hamiltonian.
 The detailed analysis of the iteration process in the DMRG was analyzed by  \"Ostlund and Rommer, and it was clarified that the DMRG can be viewed as  a variational method based on an MPS-type wavefunction.~\cite{Ostlund1995,Rommer1997} 
Then, some acceleration algorithms of DMRG were proposed in light of the MPS representation\cite{PWFRG,White1996,Hieida1997}. 
Here, it should be noted that the successive use of SVD in the DMRG corresponds to the Schmidt decomposition of the ground-state wavefunction in quantum information terminology, which drew much interest of quantum-information researchers to the MPS/TN around 2000.

The statistical mechanical counterpart of DMRG was formulated by Nishino for the transfer matrix of the 2D Ising model in 1995, regardless of the MPS formulation by Baxter.~\cite{NishinoDMRG1995, NishinoHOH1999} 
This approach was straightforwardly generalized to finite temperature problems of quantum spin chains\cite{Bursill1996,WangXiang1997,Shibata1997, Shibata2003,Okunishi1999}, on the basis of the quantum transfer matrix\cite{Suzuki1987} constructed through the Suzuki-Trotter decomposition~\cite{Trotter1959,Suzuki1976}.
Nevertheless, the asymmetricity of the quantum transfer matrix and the periodic boundary condition in the imaginary time direction is technically cumbersome. 
After 2000, thus, a research trend of the TN based on the Suzuki-Trotter decomposition turned to direct simulations of real/imaginary time evolution problems.

Meanwhile, the formulation of DMRG for the transfer matrix naturally stimulated us to clarify its relation to  Baxter's formulation of the CTM and MPS in 1996.~\cite{Okunishi1996} 
Unifying the DMRG and CTM, Nishino and Okunishi developed the CTMRG,~\cite{CTMRG1,CTMRG2} which is efficient for two-dimensional lattice models.
In addition, the spectrum of the CTM was clarified to be essential for understanding the entanglement spectrum for the setup of half-infinite bipartitioning in 1D quantum systems.\cite{Peschel_Kaulke_Legeza1999, OHA, Lefevre2008,Cho2017}
The algorithm of the CTMRG has been combined with flexible use of the SVD by  Orus and Vidal, inspired by subsequent developments in TN algorithms.~\cite{Orus2009} 
Recently, numerical convergence of the CTMRG in the thermodynamic limit was improved by Fishman et al.~\cite{Fishman2018}

In the field of the nonequilibrium statistical mechanics, the exact MPS description  of stochastic processes on 1D lattices was  formulated independently at almost the same timing as the appearance of DMRG.
 Derrida introduced the ``matrix product ansatz'' for steady states of asymmetric exclusion processes in 1993,~\cite{Derrida1993} and a variety of extensions have been proposed.\cite{Derrida1998,ASEP2007}
In the context of TN algorithms, the DMRG was firstly extended to the asymmetric exclusion process \cite{Hieida1998,Nagy2002} and reaction-diffusion processes\cite{Carlon1999}. 
A particular point for the stochastic process is that the transition matrix is usually asymmetric and the norm of its eigenvector is defined by the $1$-norm, in contrast by the $2$-norm (Euclidean norm) usual in quantum mechanical problems.\cite{pnorm}
This suggests that how to construct the reduced density matrix has been a nontrivial problem for the stochastic process\cite{Enss2001}. 
Recently, this problem was revisited in light of various developments in TN algorithms,\cite{Johnson2010,Harada2019} where possible ways of tensor constructions are carefully examined depending on the physical property of steady states.


As mentioned above, the MPS formalism in the DMRG and the CTM is basically equivalent, as far as the uniform bulk limit is concerned.
However, we should remark that the finite-system-size DMRG has played a more significant role than the CTM approach for the development of TNs in the 21st century.
This is because the finite-system-size algorithm of DMRG possesses two particular features, which are not involved in the Baxter-type variational formulation based on the thermodynamic limit. 
The first one is that the finite-system size algorithm established the position-dependent update scheme of local tensors, which led us to more flexible tensor-construction methods based on SVD.
The other is that the finite-system-size algorithm can treat 1D quantum systems  with long-range interactions up to moderate chain length within realistic computational cost.
Through mapping to effective 1D quantum systems with long-range interactions,  the application range of the DMRG was expanded to a wide variety of quantum systems such as finite-size 2D quantum system\cite{Liang1994}, bosonic systems\cite{Jeckelmann1998},dynamical quantities\cite{Jeckelmann2002}, random systems\cite{Hida1996}, momentum space\cite{Xiang1996}, quantum Hall systems\cite{Shibata2001}, quantum chemistry, etc.\cite{White1999} 
We think that the development of the TN algorithms  was certainly inspired by these features of the finite-system-size DMRG.

\subsection{MPS and quantum information}

In the 21st century, quantum many-body physics met the concept of quantum entanglement originating from quantum information.
In particular, the EE provides a useful marker to extract nonlocal quantum correlations between a subsystem and its complement in the total wavefunction. 
For instance, extensive analyses of 1D quantum many-body systems based on the EE provided renewal understanding of quantum phase transitions\cite{Vidal_Latorre2003, Calabrese2004} and SPT orders \cite{Gu_Filtering2009,Pollmann2010}, complementarily to conventional physical quantities such as order parameters, correlation functions, etc.
 Moreover, the EE is also easy to handle through the SVD of wavefunctions in the framework of the MPS.
 Thus, such intensive entanglement analyses stimulated researchers to several MPS algorithms such as time-evolved block decimation (TEBD)\cite{TEBD}, infinite TEBD\cite{iTEBD},  variational uniform matrix product state algorithm (VUMPS)\cite{Zauner-Stauber2018} as well as  time-dependent DMRG\cite{Daley2004,WhiteFeiguin2004}.
Here, it is worth mentioning that, in some early works, the EE was eventually used for setting up effective sweeping pathways in finite-size DMRG computations, without calling ``entanglement entropy" .\cite{Xiang2001, Legeza2003,Legeza2004}

When looking back to these MPS-based algorithms from the modern perspective, we have two intrinsic theoretical backgrounds;
The first one is the statistical/quantum mechanical variational principle for extracting the nature of bulk systems, where the self-consistent matrix/tensor equations satisfied in the thermodynamic limit are primarily  deduced, as in the case of Baxter's CTM.
The other is of course the quantum information viewpoint, where the entanglement among quantum-mechanical particles/states is a primal problem to be analyzed. 
The most significant benchmark model for understanding the quantum many-body entanglement has been the AKLT chain, where singlet pairs of the auxiliary spins in the VBS state can be interpreted as a nontrivial accumulation of Bell pairs.
Accordingly, the MPS from quantum information was constructed as a generalization of the VBS-type ground state, and the RG transformations in the MPS algorithms can be interpreted as sequential operations to control entanglements among auxiliary spin degrees of freedom.

Of course, the above two standpoints should be consistent with each other if the tensor optimization is properly done. 
For example, it is well known that the DMRG generates the exact MPS for the AKLT model.
Also, the row-to-row transfer-matrix formulation in the statistical mechanics is basically equivalent to  the matrix product operator (MPO) in the quantum information context.\cite{McCulloch2007}
In our view, the TN algorithm inspired by quantum information tends to use more flexible operation of local tensors, while those based on statistical mechanics are more careful about the stability of its global fixed point.
In this review, we begin with the fixed point variational equations for the MPS and then discuss their relation to the modern MPS-type algorithms, with putting a special emphasis on the role of finite-system-size DMRG to bridge the gaps between the above two backgrounds.

In addition, the quantum information viewpoint has played a crucial role in the generalization of TN algorithms beyond the MPS. 
In particular, the area law of EE\cite{Eisert_RMP2010} provides a guiding principle to design the connectivity of tensors in general TN algorithms.
It is well established that the ground states of 1D gapful systems can be well approximated by the tree-type TN states including MPS.
This is generally the case of the TN algorithms for higher dimensional systems, as will be explained in the next subsection.
For  critical systems, meanwhile, the log-correction to the area law of EE dwarfs the capacity of the tree-type TN states.
 In order to settle the log-correction problem,  the idea of disentangler that intrinsically changes the connectivity of tensors was introduced in the MERA\cite{MERA2007,MERA2008}.  
The development of the MERA network led us to the curious connection of the TN to the holography of quantum gravity\cite{RT_PRL2006,RT_JHEP2006}.

%

\subsection{higher dimensions}

Inspired by the success of DMRG for 1D systems, several numerical RG techniques have been examined for 2D quantum systems or equivalently 3D classical models so far.
The first trial was a naive extension of CTMRG to the 3D Ising model \cite{CTTRG}, which was also mentioned in Baxter's book \cite{BaxterBook}.
However, the result of this approach was not so good, mainly because the decay of spectra of reduced density matrices for 1D or 2D cuts in the 3D lattice was very slow, compared with the CTMRG for the 2D Ising model. 
At that time, whether the optimization scheme of tensors or the TN structure could be the reason for such not so good accuracy was not clear.
Thus,  Okunishi and Nishino directly examined the Kramers-Wannier variational approximation for the layer-to-layer transfer matrix of the 3D Ising model~\cite{KW3D}, focusing on the origin of the CTM variation.
The estimated transition temperature was much better than the expected, despite of only two variational parameters contained.
This result suggests that the variational state based on the statistical lattice model also works well for higher dimensional systems.
In analogy with Kramers-Wannier approximation, they formulated a direct variational algorithm for a trial state constructed as a product of local plaquette tensors, incorporating the CTMRG for the double-layered environment tensors. \cite{TPVA1,TPVA2}
Further,  the tensor product state (TPS) consisting of vertex tensors carrying auxiliary degrees of freedom with $D = 2\, , 3$\cite{TPS1,TPS2} was introduced, which systematically improved the accuracy of estimated transition temperatures.
Note that the TPS algorithm particularly for 5-leg vertex tensors is basically equivalent to the variational update scheme \cite{Corboz2016} for ``infinite projected entangled pair state'' (iPEPS) for quantum cases.\cite{iPEPS}


For quantum systems, meanwhile, the higher dimensional version of the VBS state was already included in the AKLT paper.\cite{AKLT2}
For instance, the 2D VBS state can be represented as a contraction of local tensors with respect to auxiliary spins.
However, how to efficiently contract such a 2D array of tensors was a nontrivial problem at that time, in contrast to the 1D chain.  
To our best knowledge, the TN approach to the 2D quantum system was initiated for the anisotropic version of the $S=3/2$ Honeycomb lattice AKLT model\cite{Niggemann1997}, where a variant of DMRG was used for evaluation of the double-layered 2D classical model associated with the norm of the 2D VBS state.~\cite{Hieida1999}
A generalization of the TPS-based algorithm for quantum spin systems,~\cite{qTPVA} and an optimization algorithm through the vertical reduced density matrix in the imaginary time direction~\cite{VDMA,Maeshima2004} were examined.
In the context of quantum information,  Verstraete and Cirac also proposed the variational state of 5-leg tensors as a generalization of the 2D VBS state for weakly entangled 2D finite-size quantum systems, which is now well established as PEPS.~\cite{PEPS}

In PEPS algorithms~\cite{PEPS,Murg2007,iPEPS},  the optimal tensor is computed through the environment tensor similar to the TPS algorithm for the classical system, so as to minimize the distance between an approximated PEPS and a targeted state generated by the imaginary time evolution.
This update scheme combined with the imaginary-time evolution to calculate the ground-state wavefunction of 2D quantum systems is usually called ``full update".
As mentioned before, on the other hand,  the update scheme of the local tensor based on the direct variation for the bulk ground-state energy is called ``variational update" in the PEPS literature\cite{Corboz2016}.
Moreover, a cheaper but less accurate version of the tensor optimization scheme, which is called ``simple update"\cite{Xiang_simple2008,Orus2009}, is also used to prepare a good initial tensor for the full- or variational-update schemes.
Recently, several optimization algorithms of TPS/PEPS have been widely used as a standard numerical tool for analyzing 2D quantum models and 3D statistical systems.%

In addition, we should notice that the PEPS has been extensively used for analyzes of (symmetry protected) topological states of 2D quantum systems\cite{Gracia2008, Gu_String2009,Schuch2010,Schuch2011,Zeng_book,Cirac2020,Wei2016,Wei2020}, as can be expected from its origin.
Also the PEPS/TPS representation of 2D VBS states and their extensions was utilized for describing the  measurement-based quantum computation\cite{MBQC2001,GrossEisert2007,VerstraeteCirac2004,FujiiMorimae2012}.
In this sense, the TN formulation for 2D quantum systems made a certain contribution to the development of quantum many-body physics beyond the framework of the variational calculation of the ground state.

\subsection{Real-Space Renormalization revisited}

The real-space RG~\cite{Kadanoff1966,Efrati2014,Burkhardt1982} has been an important concept in physics of many-body problems for a long time.
However, such a conventional real-space RG as block-spin transformation often fails in estimating correct scaling dimensions of second-order quantum/thermal phase transitions.
In accordance with the success of SVD in the DMRG/CTMRG,  Levin and Nave formulated a simple real-space RG scheme based on SVD ---tensor renormalization group (TRG) method~\cite{TRG2007}---, which explicitly accompanies rescaling of the lattice space in contrast to the DMRG/CTMRG approaches.
Moreover, the TRG-based scheme combined with the higher-order SVD, which is often abbreviated as HOTRG,~\cite{HOTRG2012} was also presented.  
Since the higher dimensional extension of the HOTRG is straightforward, it is often used for lattice gauge models\cite{Yoshimura2018,Akiyama2019}.
The accuracy of these TRG-based methods is significantly improved compared with the conventional block-spin-transformation approach.
For example, the transition temperature of the 3D Ising model estimated by the HOTRG is comparable with recent Monte Carlo simulations.\cite{Wang2014}
However,  the above TRG methods always generate  tree-type TN states and their fixed points are characterized by  corner double line (CDL) tensors\cite{TRG2007,Gu_TRG2008}, which bring a certain length scale determined by a number of retained bases even at a critical point.
Here, we note that the CDL tensor for the 2D classical system has basically the same structure as the corresponding CTMs.\cite{Ueda2014}
Thus, the TRG approaches are not capable of representing the log-correction to the area law of EE associated with critical phenomena, although the use of SVD certainly contributed to improving the reliability of the real-space RG.


For retracting the log-correction to the area-law of EE,  the MERA  was formulated by G. Vidal for 1D quantum systems, where a unitary operator bridging tensors in the layered tree network structure ---the concept of disentangler--- was first introduced.\cite{MERA2007,MERA2008,Evenbly2009} 
For 2D classical systems, then, a TRG-based algorithm equipped with the disentangler, which is named tensor-network renormalization (TNR), was integrated by Evenbly and Vidal \cite{TNR2015}.
The development of the TNR was achieved on the basis of several insights into the entanglement controlling associated with the MERA.
From the RG viewpoint, however, we would like to mention that the TNR involves a more natural framework of the RG transformation,  whereas the MERA algorithm can be viewed as a finite-size-system version of the TNR\cite{TNRtoMERA}.
This is because tensors in the TNR can be optimized in a quasi-local way, while the variational optimization of tensors in the MERA basically consults the global energy minimization.
 
A key role of the disentangler in the TNR is scale-dependent filtering of short-range entanglements,  which enables us to suppress the CDL decoupling in TNR iterations at the critical point.
In other words, the TNR generates a scale-invariant TN capable of representing the log-correction to the area law of EE.
The critical fluctuation can be properly taken into account and the correct critical indices of the 2D Ising model were extracted from the fixed point of the TNR.
We note that, recently,  several entanglement filtering techniques correctly describing critical phenomena were also proposed.~\cite{LoopTNR2017,Harada2018,GILT2018} 

As mentioned above,  the algorithm of MERA for quantum systems is composed of step-by-step updating of tensors in its network so as to minimize the total ground-state energy, in contrast to the TNR which is a one-way algorithm to flow to the bulk fixed point.
Instead,  the connectivity of tensors among different scaled layers is more visible in the MERA network, where one can explicitly confirm that the entanglement of a certain subsystem can be supported by the scalable loop structure of tensor legs.
This particular feature of the MERA leads us to the correspondence to the minimal surface in the Ryu-Takayanagi formula of EE,\cite{RT_PRL2006,Swingle2012} which attracts much interest from quantum information and quantum gravity sides. 
For quantum field theories, an interesting correspondence between continuous MERA and AdS geometry was actually suggested.\cite{cMERA2013,Nozaki2012}
 In addition, the flexible arrangement of disentanglers in the MERA allows us to construct a lattice implementation of 2D conformal field theories (CFTs)\cite{Pfeifer2009,Evenbly2016}.
In this sense, we think that TNR and MERA  can be a milestone in the context of physics of the real-space RG.
Since the computational cost of TNR or MERA is relatively high compared with the MPS-type formulations, on the other hand,  there are fewer applications of them to practical condensed matter problems.

\section{MPS and CTM: variational principle for the transfer matrix}
\label{Sec_2}

\begin{figure}[bt]
\begin{center} \includegraphics[width = 8.0 cm]{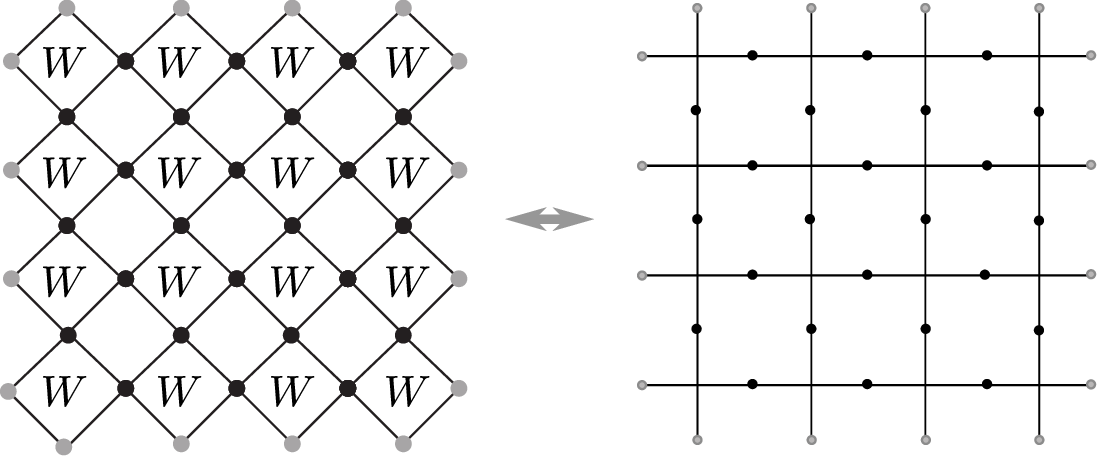} \end{center}
\caption{A square lattice Ising model.
(left) The Boltzmann weight is defined on the plaquette of the $45^\circ$-rotated square lattice.
(right) The vertex-model representation of the Boltzmann weight defined on a dual lattice.
The relation between two representations is given by Eq. (\ref{vertex_weight}). 
We basically consider the vertex model, which is usually easy to see the correspondence to quantum spin systems.}
\label{Fig1}
\end{figure}

 Let us consider a square-lattice Ising model as a typical example,  which provides the most fundamental insight for understanding  physics of the TN.
The Boltzmann weight of the Ising model is defined for nearest-neighboring spins.
For later convenience in discussing the connection to quantum systems, however, we adopt a vertex-model representation of the Boltzmann weight, which is a 4-leg local tensor carrying edge spin variables instead of spins on the lattice sites. 
There are several approaches to map the Ising model to the vertex model. 
Here, we present a simple approach based on the diagonal lattice in the left panel of Fig. \ref{Fig1}, where black dots represent the Ising spins and gray dots indicates boundary spins.

We then regard the plaquette as a unit of the Boltzmann weight containing four Ising spins at the corners. 
Explicitly, the energy contained in the plaquette is written as 
\begin{equation}
\varepsilon = - J (t s +  s t' + t' s' + s' t) \, ,
\end{equation}
where $J (>0)$ is a coupling constant and $s$, $s'$, $t$, and  $t'$ are the Ising spins taking $+$ or $-$.
Then the local Boltzmann weight on the plaquette is expressed as
\begin{align}
W( t s \,  t' s')^{~} 
=\hspace{3mm}{\setlength\unitlength{2mm}
\begin{picture}(4,5)(0,-0.3)
\put(0.0,0.0){\line(1,1){2.0}}
\put(2.0,2.0){\line(1,-1){2.0}}
\put(2.0,-2.0){\line(1,1){2.0}}
\put(0.0,0.0){\line(1,-1){2.0}}
\put(0.0,0.0){\circle*{0.3}}
\put(2.0,2.0){\circle*{0.3}}
\put(2.0,-2.0){\circle*{0.3}}
\put(4.0,0.0){\circle*{0.3}}
\put(2,-3.2){\makebox(0,0)[b]{\scriptsize \mbox{$s$}}}
\put(4.4,0){\makebox(0,0)[l]{\scriptsize \mbox{$t'$}}}
\put(2,2.5){\makebox(0,0)[b]{\scriptsize \mbox{$s'$}}}
\put(-0.4,0){\makebox(0,0)[r]{\scriptsize \mbox{$t$}}}
\put(2,0){\makebox(0,0)[c]{\mbox{$W$}}}
\end{picture}
}
\hspace{3mm}
=\hspace{3mm}{\setlength\unitlength{2mm}
\begin{picture}(4,5)(0,-0.3)
\put(2,-2){\line(0,1){4}}
\put(4.0,0.0){\line(-1,0){4}}
\put(2,-3.2){\makebox(0,0)[b]{\scriptsize \mbox{$s$}}}
\put(4.4,0){\makebox(0,0)[l]{\scriptsize \mbox{$t'$}}}
\put(2,2.5){\makebox(0,0)[b]{\scriptsize \mbox{$s'$}}}
\put(-0.4,0){\makebox(0,0)[r]{\scriptsize \mbox{$t$}}}
\end{picture}
}\hspace{3mm} . 
\label{vertex_weight}
\\ \nonumber
\end{align}
%
The last term corresponds to the vertex representation of the Boltzmann weight, which is viewed as a {\it 4-leg tensor} with edge spin degrees of freedom.
Explicitly, we have
\begin{eqnarray}
&& W(+\!+\! + +)^{~} = W(-\!-\!--)^{~} = \frac{1}{W(+\!-\!+-)^{~}} = e^{4\beta J} \, , \nonumber\\
&& W(+\!-\!--)^{~} = W(-\!+\!++) = W(+\!+\!--) = 1 \, ,
\end{eqnarray}
and their cyclic permutations with respect to the spin indices, where $\beta $ denotes an inverse temperature.
Note that the empty plaquettes in Fig. \ref{Fig1} do not contribute to the partition function of the entire system.
We can then map the diagonal lattice Ising model to the square lattice vertex model as depicted in the right panel of Fig. \ref{Fig1}.
An interesting aspect of the vertex model is that its partition function is represented as a contraction of all of the local 4-leg tensors included in the system, implying the Ising model itself can be viewed as an example of the TN.

For a practical calculation of the partition function in the bulk limit, we introduce the row-to-row transfer matrix.
Let us begin with the notation of tensors and their graphical representation used in this section. 
For instance, we write the local tensor of the Boltzmann weight at $n$ site as 
\begin{align}
W_n \equiv W(t_n s_n\, t_{n+1} s'_{n})
=\hspace{3mm}
{\setlength\unitlength{1.8mm}
\begin{picture}(4,4)(0,-0.3)
\put(2,-2){\line(0,1){4}}
\put(4.0,0.0){\line(-1,0){4}}
\put(2,-3.2){\makebox(0,0)[b]{\scriptsize \mbox{$s_n$} }}
\put(4.4,0){\makebox(0,0)[l]{\scriptsize \mbox{$t_{n+1}$} }}
\put(2,2.5){\makebox(0,0)[b]{\scriptsize \mbox{$s_n'$} }}
\put(-0.4,0){\makebox(0,0)[r]{\scriptsize \mbox{$t_n$} }}
\end{picture}
} \qquad , 
\label{eq_bwn}
\\ \nonumber
\end{align} 
where the leg indices of the tensor $W_n$ are omitted for simplicity. 
If we specify a tensor element with explicit leg-indices, we assign the brackets for the tensor, like $W(t_n s_n\, t_{n+1} s'_{n})$. 

Using Eq. (\ref{eq_bwn}), we write the tow-to-row transfer matrix as
\begin{align}
T& = \sum_{ \{t\} } \prod_{n=1}^N W_n  =
\hspace{3mm}
{\setlength\unitlength{1.8mm}
\begin{picture}(18,4)(0,-0.3)
\put(2,-1.9){\line(0,1){3.8}}
\put(5,-1.9){\line(0,1){3.8}}
\put(8,-1.9){\line(0,1){3.8}}
\put(10.0,0.0){\line(-1,0){10}}
\put(16,-1.9){\line(0,1){3.8}}
\put(18.0,0.0){\line(-1,0){3.8}}
\put(2,-3.2){\makebox(0,0)[b]{\scriptsize \mbox{$s_1$}}}
\put(18.4,0){\makebox(0,0)[l]{\scriptsize \mbox{$t_{N+1}$}}}
\put(2,2.5){\makebox(0,0)[b]{\scriptsize \mbox{$s_1'$}}}
\put(-0.4,0){\makebox(0,0)[r]{\scriptsize \mbox{$t_1$}}}
\put(5,-3.2){\makebox(0,0)[b]{\scriptsize \mbox{$s_2$}}}
\put(5,2.5){\makebox(0,0)[b]{\scriptsize \mbox{$s_2'$}}}
\put(8,-3.2){\makebox(0,0)[b]{\scriptsize \mbox{$s_3$}}}
\put(8,2.5){\makebox(0,0)[b]{\scriptsize \mbox{$s_3'$}}}
\put(11,0.0){\makebox(0,0)[l]{$\dots$}}
\put(16,-3.2){\makebox(0,0)[b]{\scriptsize \mbox{$s_N$}}}
\put(16,2.5){\makebox(0,0)[b]{\scriptsize \mbox{$s_N'$}}}
\end{picture}
\label{eq_def_rtrtm}
}\hspace{5mm} , 
\end{align}
which transfers the spins in the row direction (vertical direction in Fig. \ref{Fig1}).
Note that the connected lines between vertices in the diagram indicate summations with respect to the corresponding leg indices.
If $t_1 = t_{N+1}$, the boundary condition of $T$ is periodic.
For the ferromagnetic boundary case, $W_1$ and $W_N$ are respectively replaced with $G_1$ and $G_N$ defined by
\begin{align}
G_1 \equiv 
W_1\Big|_{t_1=+}=
{\setlength\unitlength{1.8mm}
\begin{picture}(2,5)(0,-0.3)
\put(1,-2){\line(0,1){4}}
\put(3,0.0){\line(-1,0){2}}
\put(1,-3.2){\makebox(0,0)[b]{\scriptsize \mbox{$s_1$} }}
\put(3.4,0){\makebox(0,0)[l]{\scriptsize \mbox{$t_{2}$} }}
\put(1,2.5){\makebox(0,0)[b]{\scriptsize \mbox{$s_1'$} }}
\end{picture}
}\hspace{5mm} , \quad 
G_N \equiv 
W_{N}\Big|_{t_{N+1}=+}=
\hspace{2mm}
{\setlength\unitlength{1.8mm}
\begin{picture}(2,5)(0,-0.3)
\put(3.1,-2){\line(0,1){4}}
\put(3.1,0.0){\line(-1,0){2}}
\put(3.1,-3.2){\makebox(0,0)[b]{\scriptsize \mbox{$s_N$} }}
\put(-0.4,0){\makebox(0,0)[l]{\scriptsize \mbox{$t_{N}$} }}
\put(3.1,2.5){\makebox(0,0)[b]{\scriptsize \mbox{$s_N'$} }}
\end{picture}
} \hspace{6mm} ,
\\ \nonumber
\end{align}
which are illustrated as 3-leg boundary tensors.

The bulk partition function of the Ising model can be basically evaluated as the maximum eigenvalue $\Lambda$ of $T$ with a sufficiently large $M$.
Let us write the eigenstate of $T$ corresponding to $\Lambda$ as $|\Psi\rangle$\cite{ketspace}.
We can then construct $|\Psi\rangle$ as follows,
\begin{align}
|\Psi \rangle = \lim_{M\to \infty} |\Psi^{(M)}\rangle \, ,
\label{eq_MPS_inf}
\end{align}
with 
\begin{align}
 |\Psi^{(M)}\rangle \equiv  T^{M/2-1} |\Psi^{(0)}\rangle \, ,
\label{eq_MPS_fin}
\end{align}
where $|\Psi^{(0)}\rangle$ is a certain ``initial state" that does not orthogonal to $|\Psi\rangle $.
Here, note that $M$ is an even integer corresponding to the linear dimension of the system in the row direction.\cite{mhalf}
In the low-temperature ordered phase,  we usually set up the ferromagnetic ``initial condition" for $|\Psi^{(0)} \rangle$ with the use of the 3-leg tensor  %
\begin{align}
F_n^{(0)} \equiv W_n\Big|_{s_n = +}=
\hspace{3mm}
{\setlength\unitlength{1.8mm}
\begin{picture}(4,4)(0,-0.3)
\put(2,-1){\line(0,1){1.9}}
\put(3.8,-1){\line(-1,0){3.8}}
\put(-1,-1){\makebox(0,0)[l]{\scriptsize \mbox{$t_{n}$}}}
\put(4.2,-1){\makebox(0,0)[l]{\scriptsize \mbox{$t_{n+1}$}}}
\put(2,1.5){\makebox(0,0)[b]{\scriptsize \mbox{$s_n'$}}}
\end{picture}
}  \qquad \, .
\end{align}
Using $F_n^{(0)}$ tensor, we can explicitly write $|\Psi^{(0)}\rangle$ as a product form
\begin{equation}
|\Psi^{(0)} \rangle   = \sum_{\{t\}} \prod_{n=1}^N F^{(0)}_n = 
\hspace{3mm}
{\setlength\unitlength{1.8mm}
\begin{picture}(18,4)(0,-0.3)
\put(2,-1){\line(0,1){1.9}}
\put(5,-1){\line(0,1){1.9}}
\put(8,-1){\line(0,1){1.9}}
\put(10.0,-1){\line(-1,0){10}}
\put(16,-1){\line(0,1){1.9}}
\put(18.0,-1){\line(-1,0){3.8}}
\put(18.4,-1){\makebox(0,0)[l]{\scriptsize \mbox{$t_{N+1}$}}}
\put(2,1.5){\makebox(0,0)[b]{\scriptsize \mbox{$s_1'$}}}
\put(-0.4,-1){\makebox(0,0)[r]{\scriptsize \mbox{$t_1$}}}
\put(5,1.5){\makebox(0,0)[b]{\scriptsize \mbox{$s_2'$}}}
\put(8,1.5){\makebox(0,0)[b]{\scriptsize \mbox{$s_3'$}}}
\put(11,-1){\makebox(0,0)[l]{$\dots$}}
\put(16,1.5){\makebox(0,0)[b]{\scriptsize \mbox{$s_N'$}}}
\end{picture}
}\hspace{5mm}  ,
\label{boundaryMPS}
\end{equation}
where  the lines connecting $F_n^{(0)}$ tensors also represent the summation of $\{t\}$ spins.
Note that this is a simplest example of the MPS.
If the periodic boundary is the case, $t_1 = t_{N+1}$ in Eq. (\ref{boundaryMPS}).
For the ferromagnetic boundary case,  $F_1$ and $F_N$ are respectively replaced with $C_1$ and $C_N$ defined by
\begin{align}
C_1^{(0)} \equiv 
F_1^{(0)}\Big|_{t_1=+}=
{\setlength\unitlength{1.8mm}
\begin{picture}(2,5)(0,-0.3)
\put(1,-1){\line(0,1){2}}
\put(3,-1){\line(-1,0){2}}
\put(3.4,-1){\makebox(0,0)[l]{\scriptsize \mbox{$t_{2}$} }}
\put(1,1.5){\makebox(0,0)[b]{\scriptsize \mbox{$s_1'$} }}
\end{picture}
}\hspace{5mm} , \quad 
C_N^{(0)} \equiv 
F_N^{(0)}\Big|_{t_{N+1}=+}=
\hspace{2mm}
{\setlength\unitlength{1.8mm}
\begin{picture}(2,5)(0,-0.3)
\put(3.1,-1){\line(0,1){2}}
\put(3.1,-1){\line(-1,0){2}}
\put(-0.4,-1){\makebox(0,0)[l]{\scriptsize \mbox{$t_{N}$} }}
\put(3.1,1.5){\makebox(0,0)[b]{\scriptsize \mbox{$s_N'$} }}
\end{picture}
} \hspace{5mm} ,
\label{eq_ctm_step1}
\end{align}
which are illustrated as 2-leg corner tensors.

\begin{figure}
\begin{center} \includegraphics[width = 8.0 cm]{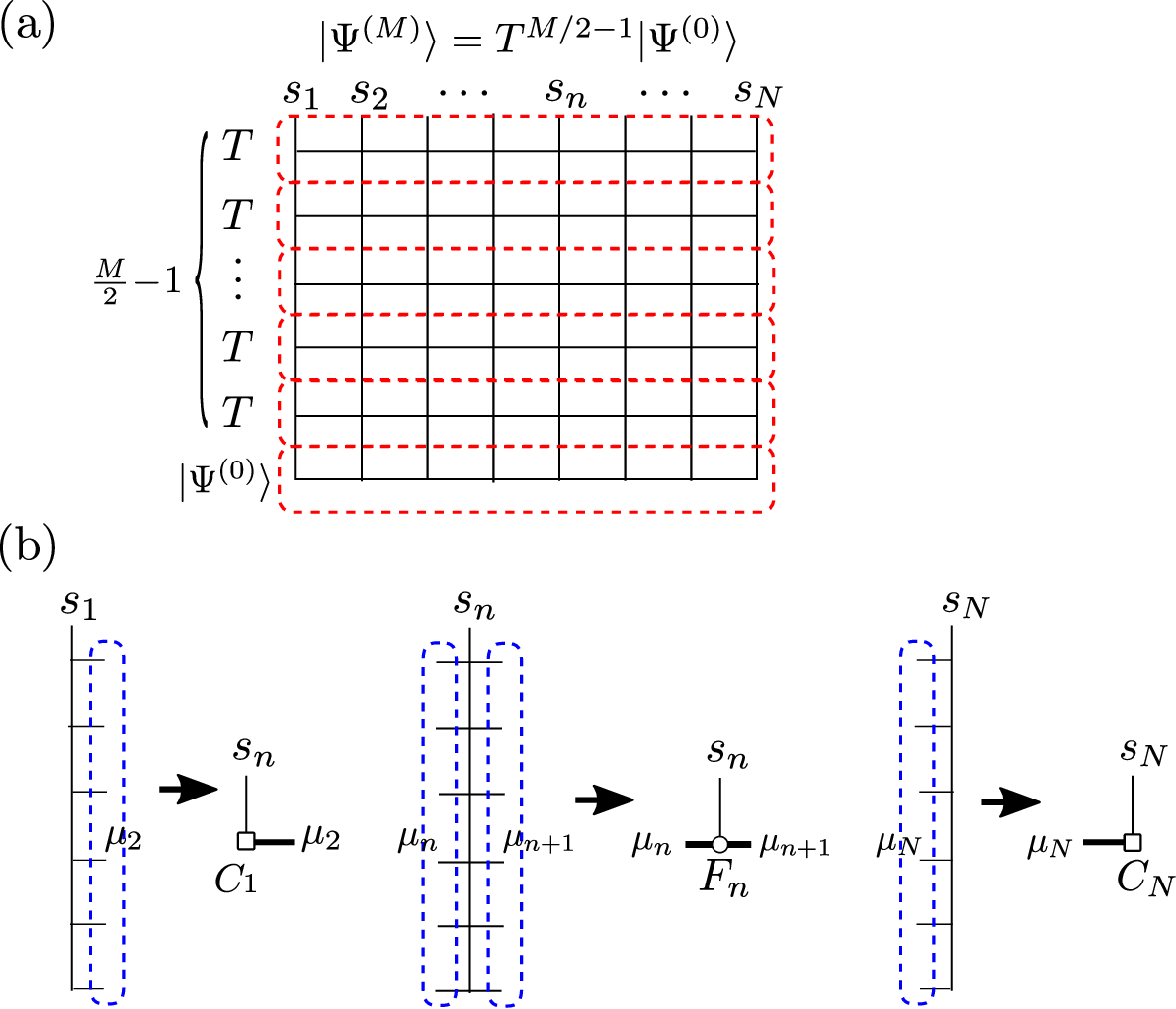} \end{center}
\caption{(Color online) MPS decomposition of the trial state $|\Psi^{(M)}\rangle = T^{M/2-1} |\Psi^{(0)}\rangle $.
(a) $|\Psi^{(M)}\rangle $ is represented as a contraction of vertex weights on the lower-half plane with an ``initial state'' $|\Psi^{(0)}\rangle$.
(b) Then, $|\Psi^{(M)}\rangle $ is decomposed into a product of $F$ tensors and corner tensors $C$, by regarding  spins surrounded by blue dotted lines in the column direction as collective spin variables $\{\mu\}$.}
\label{fig_mps}
\end{figure}

Let us set up a variational state for the transfer matrix $T$.
According to Eq. (\ref{eq_MPS_inf}), $|\Psi \rangle $ is given by transfer-matrix multiplications to $|\Psi^{(0)}\rangle$, which is equivalent to the Ising model defined on the half-infinite plane with a set of initial tensors $\{ F^{(0)} \}$ and $\{C^{(0)}\}$, as depicted in Fig. \ref{fig_mps}(a) .
In a practical situation, we can use $|\Psi^{(M)} \rangle $ with a finite but sufficiently large $M$ as an approximation of $|\Psi\rangle$.
As shown in Fig. \ref{fig_mps}(b), then, we can represent $|\Psi^{(M)}\rangle$ as a contraction of the column tensors $F_n^{}$, where $M$ basically indicates the number of layers stacked in the row direction.
Note that $F_n^{}$ can be viewed as a generalization of the initial tensor $F_n^{(0)}$.
The dimension of $F_n^{}$ is $2\times 2^{M/2}\times 2^{M/2}$, which originates from the $s$ spin at the upper surface layer and $\{t\}$ spins in the column direction in Fig. \ref{fig_mps}.
A key idea is that $F_n$ with truncated $\{t\}$-spin degrees of freedom can be used as a good approximation of $F_n^{}$ with $M\to \infty$.
Explicitly, we write 
\begin{align}
F_n \equiv F_n(s'_n|\mu_n\mu_{n+1})=
\hspace{3mm}
{\setlength\unitlength{1.8mm}
\begin{picture}(4,4)(0,-0.3)
\put(1.9,-0.7){\line(0,1){1.8}}
\linethickness{1.5pt}
\put(3.8,-1){\line(-1,0){1.6}}
\put(1.6,-1){\line(-1,0){1.7}}
\put(-1.5,-1){\makebox(0,0)[l]{\scriptsize \mbox{$\mu_{n}$}}}
\put(4.2,-1){\makebox(0,0)[l]{\scriptsize \mbox{$\mu_{n+1}$}}}
\put(2,1.5){\makebox(0,0)[b]{\scriptsize \mbox{$s'_n$}}}
\put(1.4,-1.5){\mbox{$\circ$}}
\end{picture}
} \hspace{10mm} ,
\label{eq_frg}
\end{align}
where $s'_n \in \pm 1$, and $\mu_n$ and $\mu_{n+1}$ denote indices of $m$-state block-spin variables with  $m \ll 2^{M/2}$. 
In the diagram of Eq. (\ref{eq_frg}), we have assigned a circle symbol at the joint of the vertical line and the thick horizontal line representing the $m$ state block-spin variables.
Using $F_n$, we can construct  a variational state as
\begin{align}
|\tilde{\Psi}\rangle  \equiv \sum_{\{\mu\}} \prod_{n=1}^N F^{}_n = 
\hspace{3mm}
{\setlength\unitlength{1.8mm}
\begin{picture}(18,4)(0,-0.3)
\put(2,-0.7){\line(0,1){1.6}}
\put(5,-0.7){\line(0,1){1.6}}
\put(8,-0.7){\line(0,1){1.6}}
\put(16,-0.7){\line(0,1){1.6}}
\linethickness{1.5pt}
\multiput(2.3,-1)(3,0){2}{\line(1,0){2.4}}
\put(1.7,-1){\line(-1,0){1.7}}
\put(10,-1){\line(-1,0){1.7}}
\put(18,-1){\line(-1,0){1.7}}
\put(15.7,-1){\line(-1,0){1.7}}
\multiput(1.5,-1.5)(3,0){3}{\mbox{$\circ$}}
\put(15.5,-1.5){\mbox{$\circ$}}
\put(18.4,-1){\makebox(0,0)[l]{\scriptsize \mbox{$\mu_{N+1}$}}}
\put(2,1.5){\makebox(0,0)[b]{\scriptsize \mbox{$s_1'$}}}
\put(-0.4,-1){\makebox(0,0)[r]{\scriptsize \mbox{$\mu_1$}}}
\put(5,1.5){\makebox(0,0)[b]{\scriptsize \mbox{$s_2'$}}}
\put(8,1.5){\makebox(0,0)[b]{\scriptsize \mbox{$s_3'$}}}
\put(11,-1){\makebox(0,0)[l]{$\dots$}}
\put(16,1.5){\makebox(0,0)[b]{\scriptsize \mbox{$s_N'$}}}
\end{picture}
}\hspace{7mm}  , \label{variationalMPS}
\end{align} 
which is clearly in the MPS form reflecting the transfer-matrix structure of the square lattice.
For the periodic boundary, $\mu_1=\mu_{N+1}$.
For the open boundary, $F_1$ and $F_N$ should be replaced with the boundary corner tensors
\begin{align}
C_1^{} \equiv  C_1(s'_1|\mu_2)= 
{\setlength\unitlength{1.8mm}
\begin{picture}(2,5)(0,-0.3)
\put(1,-0.5){\line(0,1){1.7}}
\linethickness{1.5pt}
\put(3.3,-1){\line(-1,0){1.8}}
\put(0.4,-1.5){\mbox{$\square$}}
\put(3.4,-1){\makebox(0,0)[l]{\scriptsize \mbox{$\mu_{2}$} }}
\put(1,1.5){\makebox(0,0)[b]{\scriptsize \mbox{$s_1'$} }}
\end{picture}
}\hspace{5mm} , \quad 
C_N^{} = C_N(s'_N|\mu_N)= 
\hspace{2mm}
{\setlength\unitlength{1.8mm}
\begin{picture}(2,5)(0,-0.3)
\put(3.1,-0.5){\line(0,1){1.7}}
\linethickness{1.5pt}
\put(2.7,-1){\line(-1,0){1.8}}
\put(2.5,-1.5){\mbox{$\square$}}
\put(-0.7,-1){\makebox(0,0)[l]{\scriptsize \mbox{$\mu_{N}$} }}
\put(3.1,1.5){\makebox(0,0)[b]{\scriptsize \mbox{$s_N'$} }}
\end{picture}
} \hspace{5mm} ,
\label{eq_ctm_step2}
\end{align}
for which we have assigned a square symbol.

We consider the variation of $T$ with respect to $|\tilde{\Psi}\rangle$.
In the following, we basically consider the transfer matrix with the fixed boundary, implying that the boundary tensors $G_1$ and $G_N$ are attached at the left and right edges of $T$.
 We write the expectation value of $T$ for the variational state  $|\tilde{\Psi}\rangle$ as
\begin{align}
\Lambda =  \lambda / \lambda' \, ,
\end{align}
with
\begin{align}
\lambda = \langle \tilde{\Psi}| T |\tilde{\Psi}\rangle \,, \quad \lambda' = \langle\tilde{\Psi} |\tilde{\Psi}\rangle \,. 
\label{eq_lambdas}
\end{align}

Then, we have to take variation of the above expression of $\Lambda$ with respect to any element of the tensor $F_n$, which is basically equivalent to the variational principle of the MPS for the 1D quantum system.
However, it is still a tough problem to precisely evaluate  $\lambda$ and $\lambda'$. 
An important idea by Baxter, which leads to the CTM, is that the variational principle can be used for evaluations of $\lambda$ and $\lambda'$ again.\cite{Baxter1968,Baxter1978}

\begin{figure}[tb]
\begin{center} \includegraphics[width = 8.0 cm]{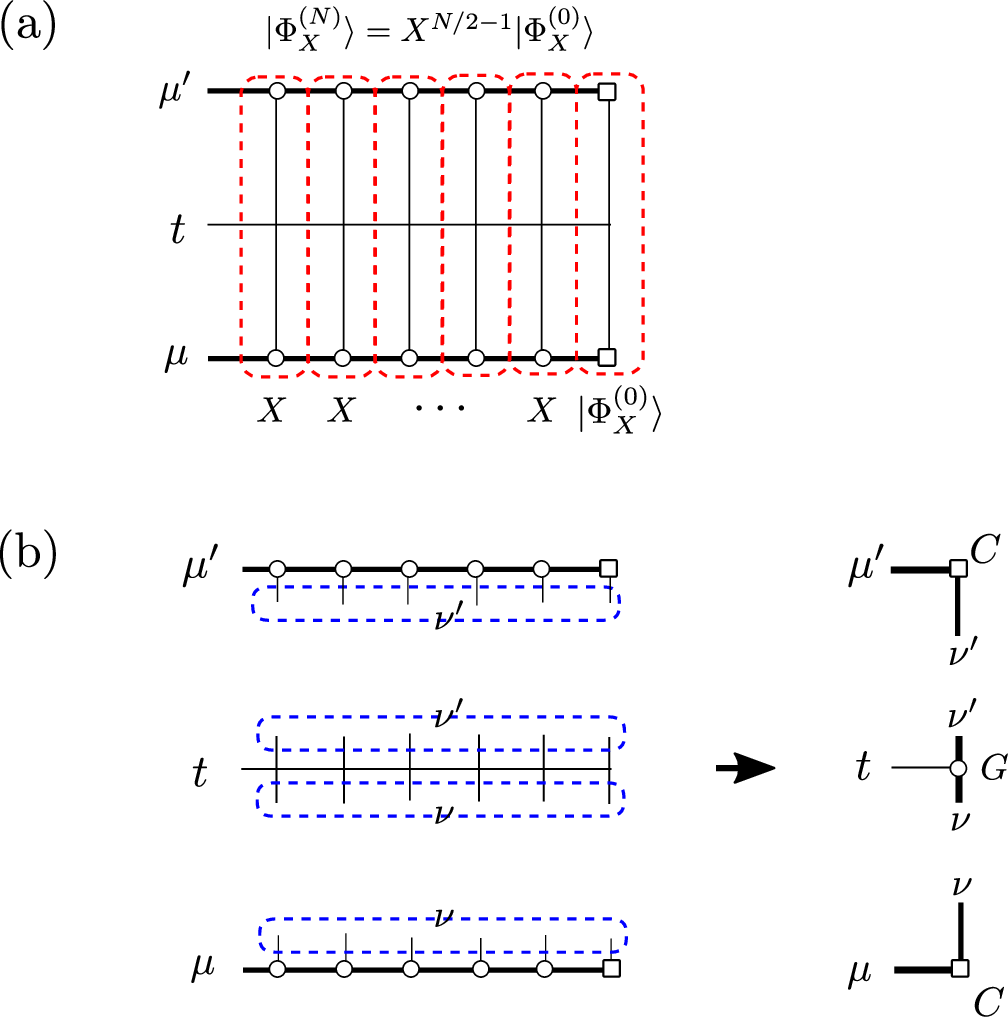} \end{center}
\caption{(Color online) CTM decomposition of the trial state $|\Phi_X^{(N)}\rangle = X^{N/2-1} |\Phi_X^{(0)}\rangle $.
(a) $|\Phi_X^{(N)}\rangle $ is represented as a configuration sum of vertex weights on the right-half plane with an ``boundary" state $|\Phi_X^{(0)}\rangle $.
(b)  $|\Phi_X^{(N)}\rangle $ is decomposed into a product of $G$ tensor and the corner tensors $C$, by regarding  spins surrounded by blue dotted lines in the row direction  as collective spin variables $\nu$ or $\nu'$.}
\label{fig_mps_ctm}
\end{figure}

In order to resolve the transfer-matrix structures embedded in $\lambda$ and $\lambda'$, we further  define a set of tensors $X_n$ as   
\begin{align}
 {X_n}&(t_{n+1} \mu_{n+1}\mu'_{n+1} |t_{n}\mu_{n}\mu'_{n}) \equiv \nonumber \\
 & \sum_{s_n,s_n'} F_n(s'_n|\mu'_n \mu'_{n+1}) W(t_n s_n t_{n+1}'s_n') F_n (s_n|\mu_n \mu_{n+1}) \, ,
\end{align}
for $1<n<N$ and
\begin{align}
{X_1}(t_2 \mu_{2}\mu'_{2})& \equiv \sum_{s_1,s_1'} C_1(s'_1|\mu'_{2}) G_1( t_{2}|s_1 s_1') C_1 (s_1|\mu_{2}) \, ,\\ 
{X_N}(t_N \mu_{N}\mu'_{N}) &\equiv \sum_{s_N,s_N'} C_N(s'_N|\mu'_N) G_N(t_N|s_N s_N') C_N (s_N|\mu_N ) \, ,
\end{align}
for $n=1$ and $N$.
The diagrammatic representation of $X_n$ is 
\begin{align}
X_1 =  \hspace{4mm}
{\setlength\unitlength{1.6mm}
\begin{picture}(4,3)(0,-0.3)
\put(0,-3.5){\line(0,1){7}}
\put(2.0,0){\line(-1,0){2}}
\put(-0.7,-4.6){\mbox{$\square$}}
\put(-0.7,3.4){\mbox{$\square$}}
\linethickness{1.5pt}
\put(2.3,-4){\line(-1,0){1.8}}
\put(2.3,4){\line(-1,0){1.8}}
\put(-1.2,4.5){\makebox(0,0)[b]{\scriptsize \mbox{$C_1$}}}
\put(-1.2,-5.6){\makebox(0,0)[b]{\scriptsize \mbox{$C_1$}}}
\put(-1.3,-0.5){\makebox(0,0)[b]{\scriptsize \mbox{$G_1$}}}
\end{picture}
}, \quad 
X_n =  {\setlength\unitlength{1.6mm}
\begin{picture}(4,6)(0,-0.3)
\put(2,-3.6){\line(0,1){7.2}}
\put(4.0,0){\line(-1,0){4}}
\put(1.5,-4.6){\mbox{$\circ$}}
\put(1.5,3.4){\mbox{$\circ$}}
\linethickness{1.5pt}
\put(4,-4){\line(-1,0){1.6}}
\put(4,4){\line(-1,0){1.6}}
\put(1.7,-4){\line(-1,0){1.6}}
\put(1.7,4){\line(-1,0){1.6}}
\put(2,4.6){\makebox(0,0)[b]{\scriptsize \mbox{$F_n$}}}
\put(2,-5.9){\makebox(0,0)[b]{\scriptsize \mbox{$F_n$}}}
\put(3.4,-1.6){\makebox(0,0)[b]{\scriptsize \mbox{$W_n$}}}
\end{picture}
}\,\, , \quad
X_N={\setlength\unitlength{1.6mm}
\begin{picture}(4,3)(0,-0.3)
\put(2,-3.5){\line(0,1){7}}
\put(2,0){\line(-1,0){2}}
\linethickness{1.5pt}
\put(1.5,-4){\line(-1,0){1.6}}
\put(1.5,4){\line(-1,0){1.6}}
\put(1.3,-4.6){\mbox{$\square$}}
\put(1.3,3.4){\mbox{$\square$}}
\put(3.6,4){\makebox(0,0)[b]{\scriptsize \mbox{$C_N$}}}
\put(3.6,-5.6){\makebox(0,0)[b]{\scriptsize \mbox{$C_N$}}}
\put(3.6,-0.5){\makebox(0,0)[b]{\scriptsize \mbox{$G_N$}}}
\end{picture}
}\,\; ,\label{eq_defX}
 \\ \nonumber
\end{align}
which visualizes that $X_n$ effectively plays the role of a transfer matrix in the column direction, and $X_{1, N}$ corresponds to the boundary tensor.
 Using the above $X_n$, we obtain a compact expression of $\lambda$,  
\begin{align}
& \lambda = \sum_{\{t, \mu,\mu'\}} \prod_{n=1}^{N} X_n  = \hspace{1mm}
{\setlength\unitlength{1.6mm}
\begin{picture}(20,6)(0,-0.3)
\put(2,-3.5){\line(0,1){7}}
\put(5,-3.6){\line(0,1){7.2}}
\put(8,-3.6){\line(0,1){7.2}}
\put(17,-3.6){\line(0,1){7.2}}
\put(20,-3.5){\line(0,1){7}}
\put(10.0,0){\line(-1,0){8}}
\put(20.0,0){\line(-1,0){5.8}}
\put(1.3,-4.6){\mbox{$\square$}}
\put(1.3,3.4){\mbox{$\square$}}
\multiput(4.5,-4.6)(2.9,0){2}{\mbox{$\circ$}}
\multiput(4.5,3.4)(2.9,0){2}{\mbox{$\circ$}}
\put(16.5,-4.6){\mbox{$\circ$}}
\put(16.5,3.4){\mbox{$\circ$}}
\put(19.3,-4.6){\mbox{$\square$}}
\put(19.3,3.4){\mbox{$\square$}}
\linethickness{1.5pt}
\put(10,-4){\line(-1,0){1.7}}
\put(7.6,-4){\line(-1,0){2.2}}
\put(4.7,-4){\line(-1,0){2.2}}
\put(19.5,-4){\line(-1,0){2.1}}
\put(16.7,-4){\line(-1,0){2.1}}
\put(10,4){\line(-1,0){1.7}}
\put(7.6,4){\line(-1,0){2.2}}
\put(4.7,4){\line(-1,0){2.2}}
\put(19.5,4){\line(-1,0){2.1}}
\put(16.7,4){\line(-1,0){2.1}}
\put(2,-6){\makebox(0,0)[b]{\scriptsize \mbox{$C_1$}}}
\put(5,-6){\makebox(0,0)[b]{\scriptsize \mbox{$F_2$}}}
\put(8,-6){\makebox(0,0)[b]{\scriptsize \mbox{$F_3$}}}
\put(17,-6){\makebox(0,0)[b]{\scriptsize \mbox{$F_{N-1}$}}}
\put(20,-6){\makebox(0,0)[b]{\scriptsize \mbox{$C_{N}$}}}
\put(2,5){\makebox(0,0)[b]{\scriptsize \mbox{$C_1$}}}
\put(5,5){\makebox(0,0)[b]{\scriptsize \mbox{$F_2$}}}
\put(8,5){\makebox(0,0)[b]{\scriptsize \mbox{$F_3$}}}
\put(17,5){\makebox(0,0)[b]{\scriptsize \mbox{$F_{N-1}$}}}
\put(20,5){\makebox(0,0)[b]{\scriptsize \mbox{$C_{N}$}}}
\put(11,0){\makebox(0,0)[l]{$\dots$}}
\end{picture}
}\hspace{5mm}  . \label{variational_lambda}  \\
\nonumber
\end{align}
For $\lambda'$, we similarly define another set of tensors with
\begin{align}
 {Y_n}(\mu_{n+1}\mu'_{n+1} | \mu_{n}\mu'_{n})  \equiv  \sum_{s_n} F_n(s_n|\mu'_n \mu'_{n+1}) F_n (s_n|\mu_n \mu_{n+1}) \, ,
\end{align}
for $1<n<N$, and the boundary tensors
\begin{align}
{Y_1}(\mu_{2}\mu'_{2}) &\equiv \sum_{s_1} C_1(s_1|\mu'_{2}) C_1 (s_1|\mu_{2}) \, ,\\ 
{Y_N}(\mu_{N}\mu'_{N}) &\equiv \sum_{s_N} C_N(s_N|\mu'_N)  C_N (s_N|\mu_N) \, .
\end{align}
As in Eq. (\ref{eq_defX}), these $Y$-tensors can be diagrammatically illustrated as 
\begin{align}
Y_1 =  \hspace{4mm}
{\setlength\unitlength{1.6mm}
\begin{picture}(2,6)(0,-0.3)
\put(0,-2.5){\line(0,1){5}}
\put(-0.7,-3.6){\mbox{$\square$}}
\put(-0.7,2.4){\mbox{$\square$}}
\linethickness{1.5pt}
\put(2,-3){\line(-1,0){1.5}}
\put(2,3){\line(-1,0){1.5}}
\put(-0.9,3.7){\makebox(0,0)[b]{\scriptsize \mbox{$C_1$}}}
\put(-0.9,-4.9){\makebox(0,0)[b]{\scriptsize \mbox{$C_1$}}}
\end{picture}
}\quad , \quad 
Y_n =  {\setlength\unitlength{1.6mm}
\begin{picture}(4,6)(0,-0.3)
\put(2,-2.6){\line(0,1){5.2}}
\put(1.5,-3.6){\mbox{$\circ$}}
\put(1.5,2.4){\mbox{$\circ$}}
\linethickness{1.5pt}
\put(4,-3){\line(-1,0){1.6}}
\put(4,3){\line(-1,0){1.6}}
\put(0,-3){\line(11,0){1.6}}
\put(0,3){\line(1,0){1.6}}
\put(2,3.7){\makebox(0,0)[b]{\scriptsize \mbox{$F_n$}}}
\put(2,-4.9){\makebox(0,0)[b]{\scriptsize \mbox{$F_n$}}}
\end{picture}
}\quad , \quad
Y_N={\setlength\unitlength{1.6mm}
\begin{picture}(2,6)(0,-0.3)
\put(2,-2.5){\line(0,1){5}}
\put(1.3,-3.6){\mbox{$\square$}}
\put(1.3,2.4){\mbox{$\square$}}
\linethickness{1.5pt}
\put(0,-3){\line(1,0){1.5}}
\put(0,3){\line(1,0){1.5}}
\put(3.5,3.7){\makebox(0,0)[b]{\scriptsize \mbox{$C_N$}}}
\put(3.5,-4.9){\makebox(0,0)[b]{\scriptsize \mbox{$C_N$}}}
\end{picture}
}\quad.
 \\ \nonumber
\end{align}
We then obtain 
\begin{align}
 & \lambda'=  \sum_{\{s,\mu,\mu'\}} \prod_{n=1}^N Y_n = 
{\setlength\unitlength{1.6mm}
\begin{picture}(20,6)(0,-0.3)
\put(2,-2.5){\line(0,1){5}}
\put(5,-2.6){\line(0,1){5.2}}
\put(8,-2.6){\line(0,1){5.2}}
\put(17,-2.6){\line(0,1){5.2}}
\put(20,-2.5){\line(0,1){5}}
\multiput(4.4,-3.6)(3.1,0){2}{\mbox{$\circ$}}
\multiput(4.4,2.4)(3.1,0){2}{\mbox{$\circ$}}
\put(16.5,-3.6){\mbox{$\circ$}}
\put(16.5,2.4){\mbox{$\circ$}}
\put(1.3,-3.6){\mbox{$\square$}}
\put(1.3,2.4){\mbox{$\square$}}
\put(19.3,-3.6){\mbox{$\square$}}
\put(19.3,2.4){\mbox{$\square$}}
\linethickness{1.5pt}
\put(10.0,-3){\line(-1,0){1.6}}
\put(7.7,-3){\line(-1,0){2.4}}
\put(4.7,-3){\line(-1,0){2.2}}
\put(19.5,-3){\line(-1,0){2.1}}
\put(16.7,-3){\line(-1,0){1.8}}
\put(10.0,3){\line(-1,0){1.6}}
\put(4.7,3){\line(-1,0){2.2}}
\put(7.7,3){\line(-1,0){2.4}}
\put(19.5,3){\line(-1,0){2.1}}
\put(16.7,3){\line(-1,0){1.8}}
\put(1,-5.1){\makebox(0,0)[b]{\scriptsize \mbox{$C_1$}}}
\put(5,-4.9){\makebox(0,0)[b]{\scriptsize \mbox{$F_2$}}}
\put(8,-4.9){\makebox(0,0)[b]{\scriptsize \mbox{$F_3$}}}
\put(17,-4.9){\makebox(0,0)[b]{\scriptsize \mbox{$F_{N-1}$}}}
\put(21.3,-5.1){\makebox(0,0)[b]{\scriptsize \mbox{$C_{N}$}}}
\put(1,3.7){\makebox(0,0)[b]{\scriptsize \mbox{$C_1$}}}
\put(5,3.6){\makebox(0,0)[b]{\scriptsize \mbox{$F_2$}}}
\put(8,3.6){\makebox(0,0)[b]{\scriptsize \mbox{$F_3$}}}
\put(17,3.6){\makebox(0,0)[b]{\scriptsize \mbox{$F_{N-1}$}}}
\put(21.3,3.7){\makebox(0,0)[b]{\scriptsize \mbox{$C_{N}$}}}
\put(11,0){\makebox(0,0)[l]{$\dots$}}
\end{picture}
}\hspace{5mm} . \\
\nonumber
\end{align}
Note that $X_n$ and $Y_n$ correspond to double-layer transfer matrices in the MPS language.

As in Eqs. (\ref{eq_MPS_inf}) and (\ref{eq_MPS_fin}), we can now extract $\lambda$ and $\lambda'$ from the renormalized transfer matrices $X_n$ and $Y_n$.
In the bulk limit $N\to \infty$, moreover, the tensor $F_n$ is expected to become position independent, implying that $\lambda$ and $\lambda'$ can be respectively extracted from the maximum eigenvalue problems of the renormalized transfer matrices $X$ and $Y$, where we have omitted the site index $n$.
In order to construct the variational states for $X$ and $Y$, moreover, we can use the same arguments as Eq. (\ref{eq_MPS_inf}) - Eq. (\ref{variationalMPS}).
As depicted in Fig. \ref{fig_mps_ctm}(a), for instance, the maximum eigenvector of $X$ is represented as  $|\Phi_X\rangle =  \lim_{N\to \infty} X^{N/2-1} |\Phi_X^{(0)}\rangle$, where $|\Phi_X^{(0)}\rangle$ is a boundary state that does not orthogonal to $|\Phi_X\rangle$.
This boundary state corresponds to the $X_N$ tensor in Eq. (\ref{eq_defX}).\cite{ket_phi}
Clearly, we can do the same thing for the $Y$ tensor.

Taking account of the lattice structure in Fig. \ref{fig_mps_ctm}(a), we decompose $|\Phi_{X}\rangle$ into three pieces, as depicted in Fig. \ref{fig_mps_ctm}(b).
We then introduce new $m$-state renormalized spin variables $\nu$ and $\nu'$ originating from $\{s\}$ spins in the row direction and then write the variational states for $X$ as  
\begin{align}
|\tilde{\Phi}_X\rangle & = \sum_{\nu \nu'} C(\mu'|\nu') G(t|\nu \nu') C(\mu|\nu) = 
\hspace{3mm}{\setlength\unitlength{1.6mm}
\begin{picture}(4,4)(0,-0.3)
\put(0,0){\line(1,0){1.6}}
\put(1.4,-3.6){\mbox{$\square$}}
\put(1.4,2.4){\mbox{$\square$}}
\put(1.5,-0.6){\mbox{$\circ$}}
\linethickness{1.5pt}
\put(2,-2.5){\line(0,1){2.15}}
\put(2,2.5){\line(0,-1){2.15}}
\put(0,-3){\line(1,0){1.5}}
\put(0,3){\line(1,0){1.5}}
\put(3,3.8){\makebox(0,0)[b]{\scriptsize \mbox{$C$}}}
\put(3,-5.0){\makebox(0,0)[b]{\scriptsize \mbox{$C$}}}
\put(3.4,-0.5){\makebox(0,0)[b]{\scriptsize \mbox{$G$}}}
\put(-1,0){\makebox(0,0)[c]{\scriptsize \mbox{$t$}}}
\put(-1,-3){\makebox(0,0)[c]{\scriptsize \mbox{$\mu$}}}
\put(-1,3){\makebox(0,0)[c]{\scriptsize \mbox{$\mu'$}}}
\end{picture}
}\,  .
\end{align}
For the $Y$ tensor, we also have 
\begin{align}
|\tilde{\Phi}_Y \rangle &= \sum_{\nu} C(\mu'|\nu)  C(\mu|\nu) =\hspace{3mm} {\setlength\unitlength{1.6mm}
\begin{picture}(2,4)(0,-0.3)
\put(1.3,-2.6){\mbox{$\square$}}
\put(1.3,1.4){\mbox{$\square$}}
\linethickness{1.5pt}
\put(2,-1.5){\line(0,1){3}}
\put(0,-2){\line(1,0){1.5}}
\put(0,2){\line(1,0){1.5}}
\put(3.2,2.4){\makebox(0,0)[b]{\scriptsize \mbox{$C$}}}
\put(3.2,-3.6){\makebox(0,0)[b]{\scriptsize \mbox{$C$}}}
\put(-1,-2){\makebox(0,0)[c]{\scriptsize \mbox{$\mu$}}}
\put(-1,2){\makebox(0,0)[c]{\scriptsize \mbox{$\mu'$}}}
\end{picture}
}\quad .
\end{align}
Here,the tensor (elements) of $C$ and $G$ are explicitly defined by
\begin{align}
C\equiv C(\mu|\nu) = 
{\setlength\unitlength{1.8mm}
\begin{picture}(2,4)(0,-0.3)
\put(2.4,-1.5){\mbox{$\square$}}
\linethickness{1.5pt}
\put(3,-0.5){\line(0,1){1.5}}
\put(2.5,-1){\line(-1,0){1.5}}
\put(0,-1){\makebox(0,0)[l]{\scriptsize \mbox{$\mu$}}}
\put(3,1.5){\makebox(0,0)[b]{\scriptsize \mbox{$\nu$} }}
\end{picture}
}\hspace{5mm} , \quad  
G\equiv G(t|\nu \nu') = 
{\setlength\unitlength{1.8mm}
\begin{picture}(2,4)(0,-0.3)
\put(2.6,0){\line(-1,0){1.6}}
\put(2.5,-0.6){\mbox{$\circ$}}
\linethickness{1.5pt}
\put(3,-2){\line(0,1){1.6}}
\put(3,2){\line(0,-1){1.7}}
\put(0,0){\makebox(0,0)[l]{\scriptsize \mbox{$t$}}}
\put(3,2.8){\makebox(0,0)[b]{\scriptsize \mbox{$\nu'$} }}
\put(3,-3.2){\makebox(0,0)[b]{\scriptsize \mbox{$\nu$} }}
\end{picture}
}\hspace{5mm} , \quad \label{eq_ctm_step3} \\ 
\nonumber
\end{align}
where the thick vertical lines  correspond to the $\nu$ or $\nu'$ spins and the thick horizontal line originates from the $\mu$ spin.\cite{C_leg}
The dimension of $C$ is $m \times m$  and that of $G$ is $2\times m\times m$.
Note that $C$ is Baxter's CTM.

Let us write the maximum eigenvalues of $X$ and $Y$ as $\xi$ and $\eta$, respectively.
Using the above variational states, we then obtain the following expressions, 
\begin{align}
\xi  &\equiv  \frac{\langle\tilde{\Phi}_X|  X|\tilde{\Phi}_X\rangle}{\langle\tilde{\Phi}_X|\tilde{\Phi}_X\rangle}
=\frac{\sum (C G C)(F W F)(C G C)}{\sum (CGC)(CGC)} \nonumber \\
&=\hspace{5mm}
{\setlength\unitlength{1.8mm}
\begin{picture}(6,4)(0,-0.3)
\put(5.6,0){\line(-1,0){5.2}}
\put(3,-2.7){\line(0,1){5.4}}
\put(5.4,2.5){\mbox{$\square$}}
\put(5.4,-3.5){\mbox{$\square$}}
\put(-0.6,2.5){\mbox{$\square$}}
\put(-0.6,-3.5){\mbox{$\square$}}
\put(-0.5,-0.6){\mbox{$\circ$}}
\put(5.5,-0.6){\mbox{$\circ$}}
\put(2.5,2.5){\mbox{$\circ$}}
\put(2.5,-3.6){\mbox{$\circ$}}
\linethickness{1.5pt}
\put(6,-2.5){\line(0,1){2.2}}
\put(0,-2.5){\line(0,1){2.2}}
\put(6,2.5){\line(0,-1){2.2}}
\put(0,2.5){\line(0,-1){2.2}}
\put(5.5,-3){\line(-1,0){2.2}}
\put(5.5,3){\line(-1,0){2.2}}
\put(0.5,-3){\line(2,0){2.2}}
\put(0.5,3){\line(21,0){2.2}}
\put(7.4,3.4){\makebox(0,0)[b]{\scriptsize \mbox{$C$}}}
\put(7.4,-4.6){\makebox(0,0)[b]{\scriptsize \mbox{$C$}}}
\put(7.4,-0.5){\makebox(0,0)[b]{\scriptsize \mbox{$G$}}}
\put(3,3.4){\makebox(0,0)[b]{\scriptsize \mbox{$F$}}}
\put(-1.4,3.4){\makebox(0,0)[b]{\scriptsize \mbox{$C$}}}
\put(-1.4,-4.6){\makebox(0,0)[b]{\scriptsize \mbox{$C$}}}
\put(-1.4,-0.5){\makebox(0,0)[b]{\scriptsize \mbox{$G$}}}
\put(3,-4.6){\makebox(0,0)[b]{\scriptsize \mbox{$F$}}}
\put(3.7,-1.5){\makebox(0,0)[b]{\scriptsize \mbox{$W$}}}
\end{picture}
}\hspace{8mm} \Big/
\hspace{6mm}
{\setlength\unitlength{1.8mm}
\begin{picture}(4,4)(0,-0.3)
\put(4.6,0){\line(-1,0){3.3}}
\put(4.4,2.5){\mbox{$\square$}}
\put(4.4,-3.5){\mbox{$\square$}}
\put(0.4,2.5){\mbox{$\square$}}
\put(0.4,-3.5){\mbox{$\square$}}
\put(0.5,-0.6){\mbox{$\circ$}}
\put(4.5,-0.6){\mbox{$\circ$}}
\linethickness{1.5pt}
\put(5,-2.5){\line(0,1){2.2}}
\put(1,-2.5){\line(0,1){2.2}}
\put(5,2.5){\line(0,-1){2.2}}
\put(1,2.5){\line(0,-1){2.2}}
\put(4.6,-3){\line(-1,0){3.1}}
\put(4.6,3){\line(-1,0){3.1}}
\put(6.4,3.4){\makebox(0,0)[b]{\scriptsize \mbox{$C$}}}
\put(6.4,-4.6){\makebox(0,0)[b]{\scriptsize \mbox{$C$}}}
\put(6.4,-0.5){\makebox(0,0)[b]{\scriptsize \mbox{$G$}}}
\put(-0.4,3.4){\makebox(0,0)[b]{\scriptsize \mbox{$C$}}}
\put(-0.4,-4.6){\makebox(0,0)[b]{\scriptsize \mbox{$C$}}}
\put(-0.4,-0.5){\makebox(0,0)[b]{\scriptsize \mbox{$G$}}}
\end{picture}
}\label{eq_xi}
\hspace{8mm}, \\
\nonumber \\
\eta &\equiv  \frac{\langle\tilde{\Phi}_Y|Y|\tilde{\Phi}_Y\rangle}{\langle\tilde{\Phi}_Y|\tilde{\Phi}_Y\rangle}
=\frac{\sum (C C)(F F)(C C) }{\sum (CC)(CC)} \nonumber \\ 
&= \hspace{5mm}
{\setlength\unitlength{1.8mm}
\begin{picture}(6,4)(0,-0.3)
\put(3,-1.6){\line(0,1){3.3}}
\put(5.4,1.5){\mbox{$\square$}}
\put(5.4,-2.5){\mbox{$\square$}}
\put(-0.6,1.5){\mbox{$\square$}}
\put(-0.6,-2.5){\mbox{$\square$}}
\put(2.5,1.5){\mbox{$\circ$}}
\put(2.5,-2.5){\mbox{$\circ$}}
\linethickness{1.5pt}
\put(6,-1.5){\line(0,1){3.1}}
\put(0,-1.5){\line(0,1){3.1}}
\put(5.5,-2){\line(-1,0){2.2}}
\put(5.5,2){\line(-1,0){2.2}}
\put(0.5,-2){\line(1,0){2.2}}
\put(0.5,2){\line(1,0){2.2}}
\put(7,2.6){\makebox(0,0)[b]{\scriptsize \mbox{$C$}}}
\put(7,-3.6){\makebox(0,0)[b]{\scriptsize \mbox{$C$}}}
\put(-1,2.6){\makebox(0,0)[b]{\scriptsize \mbox{$C$}}}
\put(-1,-3.6){\makebox(0,0)[b]{\scriptsize \mbox{$C$}}}
\put(3,2.6){\makebox(0,0)[b]{\scriptsize \mbox{$F$}}}
\put(3,-3.6){\makebox(0,0)[b]{\scriptsize \mbox{$F$}}}
\end{picture}
}\hspace{5mm} \Big/
\hspace{5mm}
{\setlength\unitlength{1.8mm}
\begin{picture}(4,4)(0,-0.3)
\linethickness{1.5pt}
\put(5,-1.5){\line(0,1){3.1}}
\put(1,-1.5){\line(0,1){3.1}}
\put(4.4,1.5){\mbox{$\square$}}
\put(4.4,-2.5){\mbox{$\square$}}
\put(0.4,1.5){\mbox{$\square$}}
\put(0.4,-2.5){\mbox{$\square$}}
\put(4.6,-2){\line(-1,0){3.1}}
\put(4.6,2){\line(-1,0){3.1}}
\put(6,2.6){\makebox(0,0)[b]{\scriptsize \mbox{$C$}}}
\put(6,-3.8){\makebox(0,0)[b]{\scriptsize \mbox{$C$}}}
\put(0,2.6){\makebox(0,0)[b]{\scriptsize \mbox{$C$}}}
\put(0,-3.8){\makebox(0,0)[b]{\scriptsize \mbox{$C$}}}
\end{picture}
}\hspace{6mm} ,
\label{eq_eta}
 \\ \nonumber
\end{align}
where we have suppressed the leg-indices of tensors for simplicity.
However, the connectivity of the tensor legs is straightforwardly reproduced from the diagrammatic representation.
Using $\xi$ and $\eta$, we finally obtain that $\lambda \approx \xi^N$ and $\lambda' \approx \eta^N$ and thus  $\Lambda \approx (\xi/\eta)^N$.

In Eqs. (\ref{eq_xi}) and (\ref{eq_eta}),  the variational parameters are installed as tensor elements of $F, G, C$. 
If we take the variation of $\xi$ and $\eta$ with respect $C$ and $G$, we then have variational equations
\begin{align}
 (FWF)(CGC) & = \xi (CGC) \label{veq_column1}\, ,\\
 (FF)(CC) & = \eta (CC) \label{veq_column2}\, ,
\end{align}
The diagram representations of Eqs.(\ref{veq_column1}) and (\ref{veq_column2}) are respectively given by
\begin{align}
{\setlength\unitlength{1.8mm}
\begin{picture}(6,4)(0,-0.3)
\put(5.6,0){\line(-1,0){4.1}}
\put(3,-2.7){\line(0,1){5.4}}
\put(5.4,2.5){\mbox{$\square$}}
\put(5.4,-3.5){\mbox{$\square$}}
\put(5.5,-0.6){\mbox{$\circ$}}
\put(2.5,2.5){\mbox{$\circ$}}
\put(2.5,-3.6){\mbox{$\circ$}}
\linethickness{1.5pt}
\put(6,-2.5){\line(0,1){2.1}}
\put(6,2.5){\line(0,-1){2.2}}
\put(5.6,-3){\line(-1,0){2.3}}
\put(5.6,3){\line(-1,0){2.3}}
\put(2.6,-3){\line(-1,0){1.1}}
\put(2.6,3){\line(-1,0){1.1}}
\put(7.4,3.6){\makebox(0,0)[b]{\scriptsize \mbox{$C$}}}
\put(7.4,-4.8){\makebox(0,0)[b]{\scriptsize \mbox{$C$}}}
\put(7.4,-0.5){\makebox(0,0)[b]{\scriptsize \mbox{$G$}}}
\put(3,3.6){\makebox(0,0)[b]{\scriptsize \mbox{$F$}}}
\put(3,-4.8){\makebox(0,0)[b]{\scriptsize \mbox{$F$}}}
\put(3.7,-1.5){\makebox(0,0)[b]{\scriptsize \mbox{$W$}}}
\end{picture}
}\hspace{6mm} = \xi \times
{\setlength\unitlength{1.8mm}
\begin{picture}(4,4)(2,-0.3)
\put(4.6,0){\line(-1,0){1.6}}
\put(4.4,2.5){\mbox{$\square$}}
\put(4.4,-3.5){\mbox{$\square$}}
\put(4.5,-0.6){\mbox{$\circ$}}
\linethickness{1.5pt}
\put(5,-2.5){\line(0,1){2.1}}
\put(5,2.5){\line(0,-1){2.2}}
\put(4.6,-3){\line(-1,0){1.5}}
\put(4.6,3){\line(-1,0){1.5}}
\put(6.4,3.6){\makebox(0,0)[b]{\scriptsize \mbox{$C$}}}
\put(6.4,-4.8){\makebox(0,0)[b]{\scriptsize \mbox{$C$}}}
\put(6.4,-0.5){\makebox(0,0)[b]{\scriptsize \mbox{$G$}}}
\end{picture}
} \qquad ,
\label{geq_FWF}
 \\
\nonumber 
\end{align}
\begin{align}
{\setlength\unitlength{1.8mm}
\begin{picture}(6,4)(0,-0.3)
\put(3,-1.6){\line(0,1){3.3}}
\put(5.4,1.5){\mbox{$\square$}}
\put(5.4,-2.5){\mbox{$\square$}}
\put(2.5,1.5){\mbox{$\circ$}}
\put(2.5,-2.5){\mbox{$\circ$}}
\linethickness{1.5pt}
\put(6,-1.5){\line(0,1){3.1}}
\put(5.6,-2){\line(-1,0){2.3}}
\put(5.6,2){\line(-1,0){2.3}}
\put(2.6,-2){\line(-1,0){1.3}}
\put(2.6,2){\line(-1,0){1.3}}
\put(7,2.6){\makebox(0,0)[b]{\scriptsize \mbox{$C$}}}
\put(7,-3.8){\makebox(0,0)[b]{\scriptsize \mbox{$C$}}}
\put(3,2.6){\makebox(0,0)[b]{\scriptsize \mbox{$F$}}}
\put(3,-3.8){\makebox(0,0)[b]{\scriptsize \mbox{$F$}}}
\end{picture}
}\hspace{5mm} = \eta \times
{\setlength\unitlength{1.8mm}
\begin{picture}(4,4)(2,-0.3)
\linethickness{1.5pt}
\put(5,-1.5){\line(0,1){3.1}}
\put(4.4,1.5){\mbox{$\square$}}
\put(4.4,-2.5){\mbox{$\square$}}
\put(4.6,-2){\line(-1,0){1.5}}
\put(4.6,2){\line(-1,0){1.5}}
\put(6,2.6){\makebox(0,0)[b]{\scriptsize \mbox{$C$}}}
\put(6,-3.8){\makebox(0,0)[b]{\scriptsize \mbox{$C$}}}
\end{picture}
}\hspace{6mm} ,
\label{geq_FF}
 \\ \nonumber
\end{align}
For $\xi/\eta$, we similarly take the variation with respect to $F$ and $C$ to obtain
\begin{align}
 (GWG)(CFC) & = \xi' (CFC)  \, ,\label{veq_row1}\\
 (GG)(CC) & = \eta' (CC) \, ,\label{veq_row2}
\end{align}
with $\xi \eta' = \xi' \eta$. 
The corresponding diagrams are 
\begin{align}
{\setlength\unitlength{1.8mm}
\begin{picture}(6,3)(0,-1.5)
\put(5.65,0){\line(-1,0){5.3}}
\put(3,-2.7){\line(0,1){4.2}}
\put(5.4,-3.5){\mbox{$\square$}}
\put(-0.6,-3.5){\mbox{$\square$}}
\put(-0.5,-0.6){\mbox{$\circ$}}
\put(5.5,-0.6){\mbox{$\circ$}}
\put(2.5,-3.6){\mbox{$\circ$}}
\linethickness{1.5pt}
\put(6,-2.5){\line(0,1){2.1}}
\put(0,-2.5){\line(0,1){2.1}}
\put(6,1.6){\line(0,-1){1.4}}
\put(0,1.6){\line(0,-1){1.4}}
\put(5.6,-3){\line(-1,0){2.3}}
\put(0.5,-3){\line(1,0){2.2}}
\put(7.4,-4.8){\makebox(0,0)[b]{\scriptsize \mbox{$C$}}}
\put(7.4,-0.5){\makebox(0,0)[b]{\scriptsize \mbox{$G$}}}
\put(-1.4,-4.8){\makebox(0,0)[b]{\scriptsize \mbox{$C$}}}
\put(-1.4,-0.5){\makebox(0,0)[b]{\scriptsize \mbox{$G$}}}
\put(3,-4.7){\makebox(0,0)[b]{\scriptsize \mbox{$F$}}}
\put(3.7,-1.5){\makebox(0,0)[b]{\scriptsize \mbox{$W$}}}
\end{picture}
}\hspace{6mm} = \xi'\times
\hspace{4mm}
{\setlength\unitlength{1.8mm}
\begin{picture}(6,4)(1,-1.5)
\put(3,-1.7){\line(0,1){1.7}}
\put(5.4,-2.5){\mbox{$\square$}}
\put(-0.6,-2.5){\mbox{$\square$}}
\put(2.5,-2.6){\mbox{$\circ$}}
\linethickness{1.5pt}
\put(6,-1.5){\line(0,1){1.5}}
\put(0,-1.5){\line(0,1){1.5}}
\put(5.6,-2){\line(-1,0){2.3}}
\put(0.5,-2){\line(1,0){2.2}}
\put(6,-3.6){\makebox(0,0)[b]{\scriptsize \mbox{$C$}}}
\put(0,-3.6){\makebox(0,0)[b]{\scriptsize \mbox{$C$}}}
\put(3,-3.5){\makebox(0,0)[b]{\scriptsize \mbox{$F$}}}
\end{picture}
}\hspace{6mm}, 
\label{geq_GWG}
\\
\nonumber \\
\hspace{4mm}
{\setlength\unitlength{1.8mm}
\begin{picture}(4,4)(0,-1.5)
\put(4.6,0){\line(-1,0){3.3}}
\put(4.4,-3.5){\mbox{$\square$}}
\put(0.4,-3.5){\mbox{$\square$}}
\put(0.5,-0.6){\mbox{$\circ$}}
\put(4.5,-0.6){\mbox{$\circ$}}
\linethickness{1.5pt}
\put(5,-2.5){\line(0,1){2.1}}
\put(1,-2.5){\line(0,1){2.1}}
\put(5,1.5){\line(0,-1){1.3}}
\put(1,1.5){\line(0,-1){1.3}}
\put(4.6,-3){\line(-1,0){3.1}}
\put(6.4,-4.8){\makebox(0,0)[b]{\scriptsize \mbox{$C$}}}
\put(6.4,-0.5){\makebox(0,0)[b]{\scriptsize \mbox{$G$}}}
\put(-0.4,-4.8){\makebox(0,0)[b]{\scriptsize \mbox{$C$}}}
\put(-0.4,-0.5){\makebox(0,0)[b]{\scriptsize \mbox{$G$}}}
\end{picture}
}
\hspace{6mm} = \eta' \times
{\setlength\unitlength{1.8mm}
\begin{picture}(4,4)(0,-1.5)
\put(4.4,-2.5){\mbox{$\square$}}
\put(0.4,-2.5){\mbox{$\square$}}
\linethickness{1.5pt}
\put(5,-1.5){\line(0,1){1.5}}
\put(1,-1.5){\line(0,1){1.5}}
\put(4.5,-2){\line(-1,0){3.1}}
\put(6,-3.6){\makebox(0,0)[b]{\scriptsize \mbox{$C$}}}
\put(0,-3.6){\makebox(0,0)[b]{\scriptsize \mbox{$C$}}}
\end{picture}
}\hspace{6mm}. \label{geq_GG}\\
\nonumber 
\end{align}
All tensor elements should be consistently optimized to enjoy a set of variational equations above.
Then, we finally arrive at the partition function per unit vertex as
\begin{align}
\kappa \equiv \xi/\eta = \xi'/\eta' \, ,
\end{align}
with which the partition function for the $N\times M$ lattice is written as $ \Lambda^M \approx (\xi/\eta)^{NM}$.
Note that if the Boltzmann weight $W$ has the 90$^\circ$ rotation symmetry, it turns out that $F = G$ and $C$ is a real symmetric matrix, which gives $\xi = \xi'$ and $\eta= \eta'$.

In the above variational equations, an important point is that, with help of CTM, the variational equations for the original MPS of Eq. (\ref{variationalMPS}) were converted to the symmetric form in the row and column directions.
As in Eqs. (\ref{geq_FWF}) and (\ref{geq_GWG}), for instance,  $FWF$ and $GWG$ can be illustrated as the renormalized version of the row-to-row and column-to-column transfer matrices, and $CGC$ and $CFC$ play the role of the corresponding eigenvectors.
This provides an essential view for the entanglement structure embedded in the variational problem based on the MPS framework.
Here, we note that in principle, the above relation for $GWG$ and $CFC$ can be respectively translated to the renormalized Hamiltonian and the ground-state wavefunction for a 1D quantum system, as will be discussed in \S \ref{Sec_4}.

\section{CTMRG}
\label{Sec_3}
In \S\ref{Sec_2}, the CTM was introduced as a corner tensor connecting the renormalized spins in the row and column directions in the variational states for the renormalized transfer matrices $X$ and  $Y$. 
For constructing a solution of the variational equations of (\ref{veq_column1}), (\ref{veq_column2}), (\ref{veq_row1}), and (\ref{veq_row2}), it is not so efficient to directly deal with the eigenvalue problems of the renormalized row-to-row or column-to-column transfer matrices.
Taking account of the physical origin of $F$, $G$ and $C$ tensors, we can systematically formulate recursive relations for CTMs, which lead us to a real-space-RG-like algorithm, i.e. corner-transfer-matrix renormalization group (CTMRG).

In the CTMRG, the variational principle for the partition function is reformulated with the use of SVD for the CTM.
We then demonstrate that the fixed point of the CTMRG satisfies the same variational equation as  (\ref{veq_column1}), (\ref{veq_column2}), (\ref{veq_row1}) and (\ref{veq_row2}) that are based on the MPS formulation.
From the entanglement point of view,  an important point is that the singular value spectrum of the CTM is essentially equivalent to that of the reduced density matrix for the bipartitioned half-infinite worldsheet of the system. 
Accordingly, the CTMRG satisfies the area-law of EE consistently with the MPS description of the 1D quantum system and enables us to obtain very accurate numerical results in the off-critical regime.

\subsection{Corner Transfer Matrix}


As in Eqs. (\ref{eq_ctm_step1}), (\ref{eq_ctm_step2}) and (\ref{eq_ctm_step3}),  the CTM originates from the boundary corner tensors in the MPS for the row-to-row transfer matrix.
Here, we  directly introduce the CTM, taking account of its physical meaning of transferring spins between the row and column directions.
Let us assume a $2N\times 2N$ square-lattice vertex model with the fixed boundary condition. 
Then, each matrix element of the CTM corresponds to a partition function of the quadrant of the lattice with certain edge-spin configurations in the row and column directions.
For example, the CTM for the lower right quadrant is explicitly defined as  
\begin{align}
C^{(N)}(\{t\}|\{s\}) &= \sum_{\{s',t'\}}\prod_{i,j} W_{ij} =\quad
 {\setlength\unitlength{1.8mm}
\begin{picture}(10,6)(0,-0.3)
\put(1,-4){\line(1,0){9}}
\put(1,-2){\line(1,0){9}}
\put(1,0){\line(1,0){9}}
\put(1,2){\line(1,0){9}}
\put(1,4){\line(1,0){9}}
\put(2,5){\line(0,-1){9}}
\put(4,5){\line(0,-1){9}}
\put(6,5){\line(0,-1){9}}
\put(8,5){\line(0,-1){9}}
\put(10,5){\line(0,-1){9}}
\put(0,4){\makebox(0,0)[l]{\scriptsize \mbox{$t_1$}}}
\put(0,2){\makebox(0,0)[l]{\scriptsize \mbox{$t_2$}}}
\put(0,-2){\makebox(0,0)[b]{\scriptsize \mbox{$\vdots$}}}
\put(0,-4){\makebox(0,0)[l]{\scriptsize \mbox{$t_N$}}}
\put(2,5.5){\makebox(0,0)[c]{\scriptsize \mbox{$s_1$}}}
\put(4,5.5){\makebox(0,0)[c]{\scriptsize \mbox{$s_2$}}}
\put(8,5.5){\makebox(0,0)[r]{\scriptsize \mbox{$\cdots$}}}
\put(10,5.5){\makebox(0,0)[c]{\scriptsize \mbox{$s_N$}}}
\end{picture}
}\hspace{6mm}\\
\nonumber 
\end{align}
where $\prod_{i,j}$ denotes the product of the vertices in the quadrant and  $\sum_{\{s', t'\}}$ represents configuration sum for the legs of all the inside vertices.
Also, spin configurations along the upper and left edges are indicated by $\{s, t\}$,  so that the matrix size of $C^{(N)}$ is $2^N \times 2^N$.
Note that the boundary condition along the bottom and right edges are assumed to be fixed.
We also define the finite-size version of $F$ and $G$ as
\begin{align}
F^{(N)}(s |\{t\}\{t'\}) = \sum_{\{s' \}}\prod_{j} W_{j} =
\quad {\setlength\unitlength{1.8mm}
\begin{picture}(10,6)(0,-0.3)
\put(1,-4){\line(1,0){2}}
\put(1,-2){\line(1,0){2}}
\put(1,0){\line(1,0){2}}
\put(1,2){\line(1,0){2}}
\put(1,4){\line(1,0){2}}
\put(2,5){\line(0,-1){9}}
\put(-0.2,4){\makebox(0,0)[l]{\scriptsize \mbox{$t_1$}}}
\put(-0.2,2){\makebox(0,0)[l]{\scriptsize \mbox{$t_2$}}}
\put(-0.2,-2){\makebox(0,0)[b]{\scriptsize \mbox{$\vdots$}}}
\put(-0.2,-4){\makebox(0,0)[l]{\scriptsize \mbox{$t_N$}}}
\put(3.3,4){\makebox(0,0)[l]{\scriptsize \mbox{$t'_1$}}}
\put(3.3,2){\makebox(0,0)[l]{\scriptsize \mbox{$t'_2$}}}
\put(3.8,-2){\makebox(0,0)[b]{\scriptsize \mbox{$\vdots$}}}
\put(3.3,-4){\makebox(0,0)[l]{\scriptsize \mbox{$t'_N$}}}
\put(2,5.5){\makebox(0,0)[c]{\scriptsize \mbox{$s$}}}
\end{picture}
} ,
\\ \nonumber \\
G^{(N)}(t |\{s\} \{s'\}) = \sum_{\{t' \}}\prod_{i} W_{i} =
\quad {\setlength\unitlength{1.8mm}
\begin{picture}(10,3)(0,-0.3)
\put(1,0){\line(1,0){9}}
\put(2,1){\line(0,-1){2}}
\put(4,1){\line(0,-1){2}}
\put(6,1){\line(0,-1){2}}
\put(8,1){\line(0,-1){2}}
\put(10,1){\line(0,-1){2}}
\put(2,-2){\makebox(0,0)[c]{\scriptsize \mbox{$s_1$}}}
\put(4,-2){\makebox(0,0)[c]{\scriptsize \mbox{$s_2$}}}
\put(7,-2){\makebox(0,0)[c]{\scriptsize \mbox{$\cdots$}}}
\put(10,-2){\makebox(0,0)[c]{\scriptsize \mbox{$s_N$}}}
\put(2,2){\makebox(0,0)[c]{\scriptsize \mbox{$s'_1$}}}
\put(4,2){\makebox(0,0)[c]{\scriptsize \mbox{$s'_2$}}}
\put(7,2){\makebox(0,0)[c]{\scriptsize \mbox{$\cdots$}}}
\put(10,2){\makebox(0,0)[c]{\scriptsize \mbox{$s'_N$}}}
\put(0,0){\makebox(0,0)[c]{\scriptsize \mbox{$t$}}}
\end{picture}
}\hspace{6mm}. 
\\ \nonumber
\end{align}
As in Figs. \ref{fig_mps}(b) and \ref{fig_mps_ctm}(b),   regarding $\{s\}$ and $\{t\}$ as collective spin variables $\nu$ and $\mu$ respectively, we then introduce  
\begin{align}
\quad C^{(N)} = 
{\setlength\unitlength{1.8mm}
\begin{picture}(4,2)(0,-0.3)
\put(2.4,-1.5){\mbox{$\square$}}
\linethickness{1.5pt}
\put(3,-0.5){\line(0,1){1.5}}
\put(2.6,-1){\line(-1,0){1.5}}
\put(0,-1){\makebox(0,0)[l]{\scriptsize \mbox{$\mu$}}}
\put(3,1.5){\makebox(0,0)[b]{\scriptsize \mbox{$\nu$}}}
\end{picture}
} \, , \quad
 F^{(N)}=
\hspace{3mm}
{\setlength\unitlength{1.8mm}
\begin{picture}(5,2)(0,-0.3)
\put(2,-0.6){\line(0,1){1.5}}
\put(1.5,-1.5){\mbox{$\circ$}}
\linethickness{1.5pt}
\put(3.8,-1){\line(-1,0){1.5}}
\put(0.1,-1){\line(1,0){1.6}}
\put(-1.5,-1){\makebox(0,0)[l]{\scriptsize \mbox{$\mu$}}}
\put(4.2,-1){\makebox(0,0)[l]{\scriptsize \mbox{$\mu'$}}}
\put(2,1.5){\makebox(0,0)[b]{\scriptsize \mbox{$s$}}}
\end{picture}
}\; , \quad 
G^{(N)} = 
{\setlength\unitlength{1.8mm}
\begin{picture}(4,2)(0,-0.3)
\put(2.6,0){\line(-1,0){1.4}}
\put(2.5,-0.6){\mbox{$\circ$}}
\linethickness{1.5pt}
\put(3,-2){\line(0,1){1.6}}
\put(3,2){\line(0,-1){1.7}}
\put(0,0){\makebox(0,0)[l]{\scriptsize \mbox{$t$}}}
\put(3.2,2.3){\makebox(0,0)[b]{\scriptsize \mbox{$\nu'$} }}
\put(3.2,-3.2){\makebox(0,0)[b]{\scriptsize \mbox{$\nu$} }}
\end{picture}
}\,. \quad 
\label{eq:matrices_for_N}
\end{align}
Then, an important implication to solve the variational equations is that recursive relations  of  $C^{(N)}$, $F^{(N)}$ and $G^{(N)}$ between $N$ and $N+1$ are straightforwardly constructed.
With the diagrammatic representation, we draw the recursion relations as follows,
\begin{align}
 F^{(N+1)}(s'|t\mu t'\mu') & = 
\hspace{3mm}
{\setlength\unitlength{1.8mm}
\begin{picture}(6,5)(0,-0.3)
\put(2,-0.6){\line(0,1){4.6}}
\put(4,2){\line(-1,0){4}}
\put(1.5,-1.5){\mbox{$\circ$}}
\linethickness{1.5pt}
\put(4,-1){\line(-1,0){1.7}}
\put(0.0,-1){\line(1,0){1.6}}
\put(-1,2){\makebox(0,0)[c]{\scriptsize \mbox{$t$}}}
\put(4.7,2){\makebox(0,0)[c]{\scriptsize \mbox{$t'$}}}
\put(-1,-1){\makebox(0,0)[c]{\scriptsize \mbox{$\mu$}}}
\put(4.7,-1){\makebox(0,0)[c]{\scriptsize \mbox{$\mu'$}}}
\put(2,4.6){\makebox(0,0)[c]{\scriptsize \mbox{$s'$}}}
\put(2.8,-2.4){\makebox(0,0)[c]{\scriptsize \mbox{$F^{(N)}$} }}
\put(1.4,2.7){\makebox(0,0)[c]{\scriptsize \mbox{$W$} }}
\end{picture}
}\; ,  \label{eq_recursionF} \\
\nonumber \\
 G^{(N+1)}(t|s\nu s'\nu') &= 
{\setlength\unitlength{1.8mm}
\begin{picture}(7,3)(0,-0.3)
\put(5.65,0){\line(-1,0){4.6}}
\put(3,-2.0){\line(0,1){4}}
\put(5.5,-0.6){\mbox{$\circ$}}
\linethickness{1.5pt}
\put(6,-2){\line(0,1){1.6}}
\put(6,2){\line(0,-1){1.7}}
\put(0,0){\makebox(0,0)[l]{\scriptsize \mbox{$t$}}}
\put(6,2.8){\makebox(0,0)[b]{\scriptsize \mbox{$\nu'$} }}
\put(6,-3.2){\makebox(0,0)[b]{\scriptsize \mbox{$\nu$} }}
\put(3,2.8){\makebox(0,0)[b]{\scriptsize \mbox{$s'$} }}
\put(3,-3.2){\makebox(0,0)[b]{\scriptsize \mbox{$s$} }}
\put(2,0.9){\makebox(0,0)[c]{\scriptsize \mbox{$W$} }}
\put(8,0){\makebox(0,0)[c]{\scriptsize \mbox{$G^{(N)}$} }}
\end{picture}
}\qquad , \label{eq_recursionG}\\ 
\nonumber \\
 C^{(N+1)}(t\mu|s'\nu')& = 
{\setlength\unitlength{1.8mm}
\begin{picture}(8,5)(0,-0.3)
\put(5.65,1){\line(-1,0){4.6}}
\put(3,-1.6){\line(0,1){4.6}}
\put(5.5,0.4){\mbox{$\circ$}}
\put(2.5,-2.5){\mbox{$\circ$}}
\put(5.5,-2.5){\mbox{$\square$}}
\linethickness{1.5pt}
\put(6,-1.5){\line(0,1){2.1}}
\put(6,1.3){\line(0,1){1.7}}
\put(5.7,-2){\line(-1,0){2.4}}
\put(2.7,-2){\line(-1,0){1.6}}
\put(0.2,1){\makebox(0,0)[c]{\scriptsize \mbox{$t$}}}
\put(6,3.8){\makebox(0,0)[c]{\scriptsize \mbox{$\nu'$} }}
\put(0.2,-2){\makebox(0,0)[c]{\scriptsize \mbox{$\mu$} }}
\put(3.6,3.8){\makebox(0,0)[c]{\scriptsize \mbox{$s'$} }}
\put(2.2,1.9){\makebox(0,0)[c]{\scriptsize \mbox{$W$} }}
\put(3,-3.2){\makebox(0,0)[c]{\scriptsize \mbox{$F^{(N)}$} }}
\put(8,-3.2){\makebox(0,0)[c]{\scriptsize \mbox{$C^{(N)}$} }}
\put(8,1){\makebox(0,0)[c]{\scriptsize \mbox{$G^{(N)}$} }}
\end{picture}
}\quad . \label{eq_recursionC}\\
\nonumber
\end{align}
Starting from $N=1$, thus, we can systematically construct $C^{(N)}$, $F^{(N)}$ and $G^{(N)}$ with iterative computations.
In principle,  one can expect that $C^{(N)}$, $F^{(N)}$ and $G^{(N)}$ approaches the bulk tensors satisfying the variational equations in the $N\to \infty$ limit.
However, bond dimensions of tensors are doubled in each recursion step by bundling $s \otimes \mu \to \mu$ and  $t \otimes \nu \to \nu$.
For practical computations, we need to truncate the tensor dimension by $m (\ll 2^N) $, which specifies the total number of the variational parameters.

\subsection{Recursion relation of CTMRG}

So far, we have basically used  the tensor element representation and taken contraction of them by specifying their indices to be summed. 
Here, let us introduce the matrix notation of tensors for later convenience.
For example, ${\sf C}^{(N)}$ and ${\sf F}^{(N)}(s)$ denote the matrices whose elements are respectively defined by
\begin{align}
\left [{\sf C}^{(N)}\right]_{\nu, \mu}  = C^{(N)}(\mu|\nu) \, ,\quad \left [{\sf F}^{(N)}(s)\right]_{\mu', \mu}  = F^{(N)}(s|\mu \mu')\, ,  
\end{align}
where the matrix size is assumed to be $m\times m$.
Here, note that the bare spin index $s$ in ${\sf F}^{(N)}(s)$ is not regarded as a matrix index.
Similarly, ${\sf C}^{(N+1)}$ and ${\sf F}^{N+1}(s) $ are defined by 
\begin{align}
\left [{\sf C}^{(N+1)}\right]_{s \nu, t\mu} & = C^{(N+1)}(t \mu|s\nu) \, ,\\
\left [{\sf F}^{(N+1)}(s)\right]_{t' \mu', t\mu} & = F^{(N+1)}(s| t \mu t'\mu') \, ,
\end{align}
whose matrix size is $2m\times 2m$.
In the diagrammatic representation, the index flow of matrix-matrix multiplication is assumed to be in the {\it anticlockwise direction around the center of the system}.
If the vertex weight has 90$^\circ$-rotational symmetry, ${\sf C}^{(N)}$ is real symmetric and the partition function for the $2N\times 2N$ lattice is compactly written by $Z^{(N)} = \tr[  {{\sf C}^{(N)}}^4 ] $.
In connection with the quantum system, we assume the parity symmetry of the vertex weight, implying that the row-to-row(or column-to-column) transfer matrix is symmetric, but we do not assume  the 90$^\circ$ rotational symmetry for generality.

For truncating the increased matrix dimension of  ${\sf C}^{(N+1)}$, we now employ SVD,  
\begin{align}
{\sf C}^{(N+1)} = {\sf V}^{(N+1)} {\sf \Omega}^{(N+1)} {{\sf U}^{(N+1)}}^\dagger \, ,
\label{eq_svd_ctm}
\end{align}
where  $\sf \Omega$ denotes the diagonal matrix whose diagonal entries are the $2m$ singular values, and ${\sf U}^{(N+1)}$ and ${\sf V}^{(N+1)}$ are the corresponding singular vectors satisfying the orthonormal relation
\begin{align}
{{\sf U}^{(N+1)}}^\dagger {\sf U}^{(N+1)} = {\sf I} \, , \quad {{\sf V}^{(N+1)}}^\dagger {\sf V}^{(N+1)} = {\sf I}\, .
\label{eq_norm_N+1}
\end{align}
Note that the matrix size of ${\sf U}^{(N+1)}$ and ${\sf V}^{(N+1)}$ is also $2m\times 2m$, and  ${\sf I}$ is the identity matrix of $2m$.
Instead of the SVD, we may use the ``reduced density matrices"
\begin{align}
{\sf \rho}^{(N+1)} & = {\sf C}^{(N+1)\dagger}   {\sf C}^{(N+1)}    {\sf C}^{(N+1)\dagger}   {\sf C}^{(N+1)} \, , \\
{\sf \rho}'^{(N+1)} & = {\sf C}^{(N+1)}   {\sf C}^{(N+1)\dagger}    {\sf C}^{(N+1)}   {\sf C}^{(N+1)\dagger} \,, 
\end{align}
where the four CTMs respectively correspond to the four quadrants of the system.
We then diagonalize
\begin{align}
{\sf \rho}^{(N+1)}  & =   {\sf U}^{(N+1)} {{\sf \Omega}^{(N+1)}}^4 {\sf U}^{(N+1)\dagger}  \, ,\nonumber \\
{\sf \rho}'^{(N+1)}  & =  {\sf V}^{(N+1)} {{\sf \Omega}^{(N+1)}}^4 {\sf V}^{(N+1)\dagger} \, ,
\end{align}
which also provide 
\begin{align}
Z^{(N+1)} = \Tr \rho^{(N+1)} =\Tr \rho'^{(N+1)}  = \Tr {{\sf \Omega}^{(N+1)}}^4 \, .
\label{Z_ctm4}
\end{align}

Here, we should remark that these reduced density matrices are basically equivalent to those for the half-bipartitioned ground-state wavefunction of the corresponding 1D quantum system if $N$ is sufficiently larger than the correlation length of the system.
This is because the ground state wavefunction can be represented as a half-infinite world sheet generated by the Suzuki-Trotter decomposition [See also Fig. \ref{itebd_lattice}(d)]. 
More explicitly, regarding the product of two CTMs as a wavefunction 
\begin{align}
\Psi^{(N+1)}(s\nu|s'\nu') \sim \sum_{t \mu}{{ C}^{(N+1)}}(t\mu|s\nu) { C}^{(N+1)}(t\mu|s'\nu')\, ,
\label{wf_classical_quantum}
\end{align}
we have $\rho'^{(N+1)}(s\nu|s'\nu') = \sum_{s''\nu''}{\Psi^{(N+1)}}^*(s\nu|s''\nu'')  \Psi^{(N+1)}(s''\nu''|s'\nu)$ in the manner consistent with the standard definition of the reduced density matrix for 1D quantum systems\cite{imaginary_d} [See Eqs. (\ref{offo_even}) and (\ref{psi_LR})].
From these, it follows that $S_{\rm EE}\equiv - \Tr {\sf \Omega}^4 \log {\sf \Omega}^4$ and $-\log {\sf \Omega}^4$ respectively correspond to the EE and the entanglement spectrum, where we have omitted a superscript of $(N+1)$ for ${\sf \Omega}$.
The property of the entanglement in the CTM/MPS formulation is discussed later in \S \ref{sec_4_5}.

In Eq. (\ref{eq_svd_ctm}), we assume that the singular values in ${\sf \Omega}$ are aligned in the decreasing order, $\omega_1 \ge \omega_2 \ge \cdots \omega_m \ge \omega_{m+1} \cdots \omega_{2m}(\ge 0)$.
If decay of the spectrum is fast,  then, we can well approximate the CTM by retaining up to $\omega_1, \cdots, \omega_m$ and the corresponding singular vectors in  ${\sf U}^{(N+1)}$ and ${\sf V}^{(N+1)}$.
Note that this truncation gives rise to the best approximation of Eq. (\ref{Z_ctm4}) within the truncated number of singular values.
Thus, the CTMRG can be interpreted as {\it a direct variational approximation in the sense of maximizing the partition function (\ref{Z_ctm4})}.\cite{CTMRG2}
From an algorithmic viewpoint, an essential point is that the matrix size of ${\sf U}^{(N+1)}$ and ${\sf V}^{(N+1)}$ of the retained singular vectors is reduced to $2m \times m$, which can be served as renormalization transformation matrices.
On the above basis, we consider the renormalized representation of the matrices,  
\begin{align}
\tilde{\sf F}^{(N+1)}(s) = {\sf U}^{(N+1)\dagger} {\sf F}^{(N+1)}(s) {\sf U}^{(N+1)}\, , \\
\tilde{\sf G}^{(N+1)}(t) =  {\sf V}^{(N+1)\dagger} {\sf G}^{(N+1)}(t) {\sf V}^{(N+1)}  .
\end{align}

In order to extract recursion relations for the renormalized matrices, moreover, we further rewrite the matrices in Eq.~(\ref{eq:matrices_for_N}) for the system size $N$ in the CTM diagonal representation,
\begin{align}
&{\sf C}^{(N)} =   {\sf V}^{(N)} \sf{ \Omega}^{(N)}{\sf U}^{(N)\dagger}\, ,\\
&{\sf U}^{(N)\dagger}  {\sf U}^{(N)}={\sf I},\quad {\sf V}^{(N)\dagger}{\sf V}^{(N)}={\sf I}\, , \label{eq_norm_N}
\end{align}
and
\begin{align}
\tilde{\sf F}^{(N)}(s)= {\sf U}^{(N)\dagger}{\sf F}^{(N)}(s) {\sf U}^{(N)}\, ,\nonumber \\
\tilde{\sf G}^{(N)}(t)= {\sf V}^{(N)\dagger}{\sf G}^{(N)}(t) {\sf V}^{(N)}\, ,\nonumber
\end{align}
where the matrix size is $m\times m$ by definition.
We also introduce the RG transformation matrices in the form of the $m\times m$ block matrix as follows,
\begin{align}
\left[ \tilde{\sf U}^{(N+1)\dagger}(t) \right]_{\xi,\mu} & =\;
{\setlength\unitlength{1.8mm}
\begin{picture}(5,3)(0,-0.3)
\put(2.1,-1){\line(0,1){3}}
\linethickness{1.5pt}
\put(1,-1){\line(1,0){3}}
\put(2.5,-1){\makebox(0,0)[c]{\mbox{$\blacktriangleright$}}}
\put(1,2){\makebox(0,0)[c]{\scriptsize \mbox{$t$}}}
\put(0,-1){\makebox(0,0)[c]{\scriptsize \mbox{$\mu$} }}
\put(5,-1){\makebox(0,0)[c]{\scriptsize \mbox{$\xi$} }}
\end{picture}
}\;
\equiv  \left[ {\sf U}^{(N+1)\dagger}{\sf U}^{(N)} \right]_{\xi,t\mu} , \label{eq_UUgap}
\\
\left[ { \tilde{\sf V}^{(N+1)} } (s)\right]_{\nu,\xi}  & = \; 
{\setlength\unitlength{1.8mm}
\begin{picture}(5,3)(0,-0.3)
\put(2.9,-1){\line(0,1){3}}
\linethickness{1.5pt}
\put(1,-1){\line(1,0){3}}
\put(2.5,-1){\makebox(0,0)[c]{\mbox{$\blacktriangleleft$}}}
\put(4,2){\makebox(0,0)[c]{\scriptsize \mbox{$s$}}}
\put(5,-1){\makebox(0,0)[c]{\scriptsize \mbox{$\nu$} }}
\put(0,-1){\makebox(0,0)[c]{\scriptsize \mbox{$\xi$} }}
\end{picture}
}\;
\equiv  \left[ { {\sf V}^{(N)\dagger} {\sf V}^{(N+1)}} \right]_{s \nu,\xi}  \quad , 
\end{align}
where the solid triangle indicates the direction of the renormalized leg index $\xi$ that is connected to the singular values ${\sf \Omega}^{(N+1)}$.
Here, note that, for example, ${\sf U}^{(N)}$ in Eq. (\ref{eq_UUgap}) is respectively multiplied to ${\sf U}^{(N+1)\dagger}$ from the right for each $t$.
Thus $\mu$, $\nu$ and $\xi$ run from 1 to $m$ and the matrix size of $\tilde{\sf U}^{(N+1)}(t)$ and $\tilde{\sf V}^{(N+1)}(s)$ for a given $t$ or $s$ is $m\times m$.
From the modern MPS point of view, the above transformations by ${\sf U}^{(N)}$ and ${\sf V}^{(N)}$ is nothing but a gauge transformation for the tensor legs  of the auxiliary degrees of freedom.

\begin{figure}[tb]
\includegraphics[width = 8 cm]{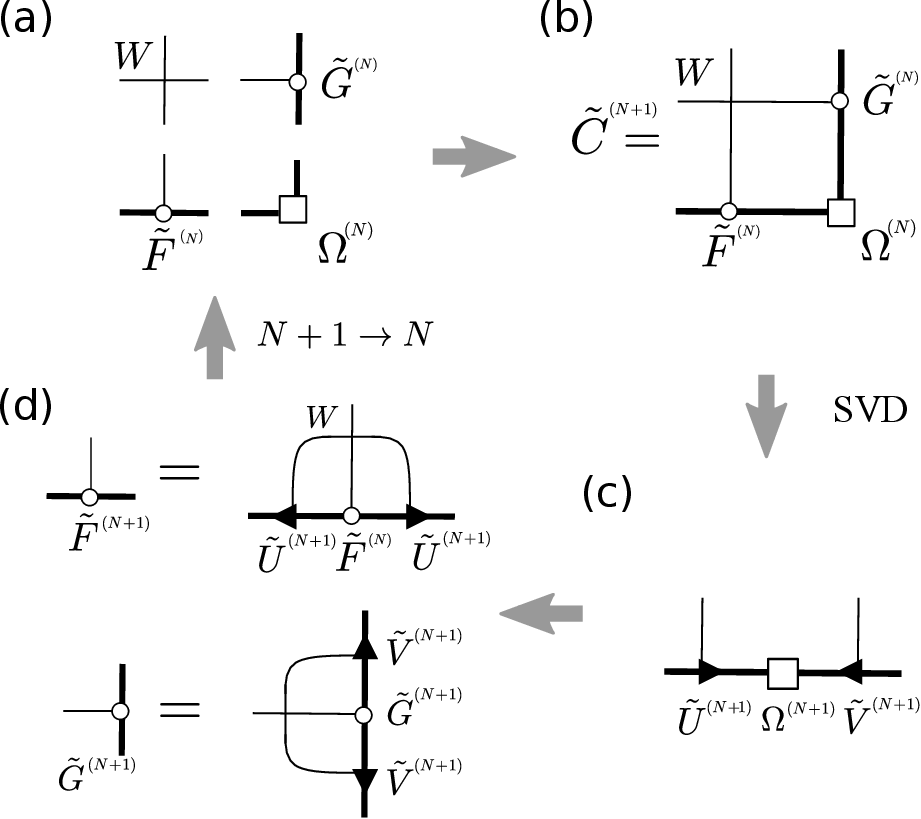}
\caption{Diagrammatic representation of the CTMRG algorithm. $W$ is the vertex weight of Eq. (\ref{vertex_weight}).
(a)  $\Omega^{(N)}$ denotes the spectrum of the CTM after $N$ iterations. 
Also, $\tilde{F}^{(N)}$ and $\tilde{G}^{(N)}$ represent 3-leg tensors originating from the MPS for the row-to-row or column-to-column transfer matrices, respectively.
(b) Contracting $W$, $\Omega^{(N)}$, $\tilde{F}^{(N)}$ and $\tilde{G}^{(N)}$, we calculate an extended CTM $\tilde{C}^{(N+1)}$.
(c) Perform SVD of $\tilde{C}^{(N+1)}$ (or equivalently diagonalization of the reduced density matrix) to obtain the new singular values $\Omega^{(N+1)}$ and the corresponding singular vectors $\tilde{U}^{(N+1)}$ and $\tilde{V}^{(N+1)}$.
d) Retaining the half of $\tilde{U}^{(N+1)}$ and $\tilde{V}^{(N+1)}$, we perform renormalization transformation for $\tilde{F}^{(N+1)}$ and $\tilde{G}^{(N+1)}$.
Then, we go back to (a) with replacing $N+1\to N$.
}
\label{CTMRG_loop}
\end{figure}

We can now reconstruct the recursive relations of Eqs. (\ref{eq_recursionF}), (\ref{eq_recursionG}) and (\ref{eq_recursionC}) for the renormalized matrices in the CTM diagonal representation.
We summarize the practical CTMRG algorithm in Fig. \ref{CTMRG_loop}.
Given $\tilde{F}^{(N)}$,  $\tilde{G}^{(N)}$ and $\Omega^{(N)}$, the extension of the CTM $\tilde{C}^{(N+1)}$ and its SVD to generate the transformation matrices $\tilde{ U}^{(N+1)}$ and $\tilde{ V}^{(N+1)}$ respectively correspond to Fig. \ref{CTMRG_loop}(b) and (c).
Also, renormalization of the extended $\tilde{F}^{(N+1)}$ and $\tilde{G}^{(N+1)}$ tensors are depicted as  Fig. \ref{CTMRG_loop}(d).
In the matrix notation, the corresponding recursion relation is explicitly written down as, 
\begin{align}
\tilde{\sf F}^{(N+1)}(s') &= \sum_{stt'} W(tst's')\tilde{\sf U}^{(N+1)\dagger}(t') \tilde{\sf F}^{(N)}(s) \tilde{\sf  U}^{(N+1)}(t) \, ,   \\
\tilde{\sf G}^{(N+1)}(t) &= \sum_{ss't'} W(tst's')  \tilde{\sf V}^{(N+1)\dagger}(s') \tilde{\sf G}^{(N)}(t') \tilde{\sf V}^{(N+1)}(s) \, ,  \label{ctmrg_recursion_G} \\
 {\sf \Omega }^{(N+1)} &= \nonumber \\
 \sum_{ss'tt'} W(ts&t's')  \tilde{\sf V}^{(N+1)\dagger}(s')  \tilde{\sf G}^{(N)}(t'){\sf \Omega}^{(N)} \tilde{\sf F}^{(N)}(s) \tilde{\sf U}^{(N+1)}(t)\, ,
\label{SVD_RCTM}
\end{align}
in which the matrix dimension is self-consistently maintained to be $m$. 
Also, it follows from Eqs. (\ref{eq_norm_N+1}) and (\ref{eq_norm_N}) that $\sum_t \tilde{\sf U}^{(N+1)^\dagger}(t) \tilde{\sf U}^{(N+1)}(t) = \sum_s \tilde{\sf V}^{(N+1)\dagger}(s) \tilde{\sf V}^{(N+1)}(s) = {\sf I}$ with ${\sf I}$ being the $m$-dimensional identity matrix, so that Eq. (\ref{SVD_RCTM}) corresponds to the SVD for the extended CTM with the truncated bases. 
Thus these recursive relations enable us to iteratively increase system sizes of them by replacing $N+1 \to N$ with keeping the matrix dimension.

\subsection{Fixed point and variational equations}

After a sufficient number of CTMRG iterations, the matrices converge to the bulk ones.
We then drop the labels of $N$ and $N+1$ in the recursion relations:  
 $\sf{\Omega}^{(N)} \to \sf{\Omega}$ ($\tilde{\sf C}^{(N)} \to \tilde{\sf C}$), $\tilde{\sf F}^{(N)} \to \tilde{\sf F}$ and  $\tilde{\sf G}^{(N)} \to \tilde{\sf G}$, as well as the RG transformation matrices $\tilde{\sf U}^{(N)} \to \tilde{\sf U}$ and $\tilde{\sf V}^{(N)} \to \tilde{\sf V}$, which lead us to a couple of self-consistent equations. 
 In the matrix notation, we explicitly have  
\begin{align}
&\tilde{\sf F}(s') = \alpha   \sum_{s tt'} W(tst's') \tilde{\sf U}^\dagger(t') \tilde{\sf F}(s)  \tilde{\sf U}(t)    \label{eq_self_F} \, ,   \\
&\tilde{\sf G}(t) = \beta    \sum_{ss't'} W(tst's') \tilde{\sf V}^\dagger(s')  \tilde{\sf G}(t') \tilde{\sf V}(s)     \label{eq_self_G} \, ,  \\
&{\sf \Omega }= \gamma  \sum_{ss'tt'} W(tst's') \tilde{\sf V}^\dagger(s')  \tilde{\sf G}(t')\Omega \tilde{\sf F}(s) \tilde{\sf U}(t)  \, , \label{eq_self_omega} \\
&\sum_t \tilde{\sf U}^\dagger(t) \tilde{\sf U}(t) = {\sf I} \, , \quad  \sum_s \tilde{\sf V}^\dagger(s) \tilde{\sf V}(s)= {\sf I}\, ,  \label{eq_self_norm}
\end{align}
where $\alpha$, $\beta$ and $\gamma$ are certain normalization constants.
These equations describe  the fixed point of the CTMRG in the bulk limit.

We demonstrate that the above fixed-point equations satisfy the variational equations (\ref{veq_column1}), (\ref{veq_column2}), (\ref{veq_row1}) and (\ref{veq_row2}), which consist of 7 equations in the unit of $m \times m$ matrix. 
In order to eliminate $\tilde{\sf U}$ and $\tilde{\sf V}$ from the fixed point equations (\ref{eq_self_F})-(\ref{eq_self_norm}), we examine  the similarity transformation for $\tilde{\sf U}$ and $\tilde{\sf V}$,  
\begin{align}
\tilde{\sf U}(t) = {\sf \Omega} \tilde{\sf G}(t) {\sf \Omega}^{-1}\,, \quad \tilde{\sf V}(s) = {\sf \Omega} \tilde{\sf F}(s) {\sf \Omega}^{-1} \, .
\label{UV_MPS}
\end{align}
Then, Eqs. (\ref{eq_self_norm}) turn out
\begin{align}
\sum_t \tilde{\sf G}(t){\sf \Omega} {\sf \Omega} \tilde{\sf G}(t) = {\sf \Omega} {\sf \Omega}  \, , \quad  
\sum_s \tilde{\sf F}(s){\sf \Omega \Omega}\tilde{\sf  F}(s) = {\sf \Omega \Omega } \,\label{eq_self_v}
\end{align}
which are equivalent to Eqs. (\ref{veq_column2}) and (\ref{veq_row2}) with $\eta=1 = \eta'=1$.
Here, note that  $\tilde{\sf G}^\dagger(t)=\tilde{\sf G}(t)$ and $\tilde{\sf F}^\dagger(s)=\tilde{\sf F}(s)$, and the orthonormal condition of the singular vectors naturally results in $\eta=1 = \eta'=1$.
Substituting Eq. (\ref{UV_MPS}) into Eqs. (\ref{eq_self_F}) and (\ref{eq_self_G}), we obtain
\begin{align}
\sum_{stt'} W(tst's') \tilde{\sf G}(t){\sf \Omega} \tilde{\sf F}(s) {\sf \Omega} \tilde{\sf G}(t') &= \kappa {\sf \Omega}\tilde{\sf F}(s') {\sf \Omega}  \, , \label{ctmrg_fix1}\\ 
\sum_{ss't'} W(tst's') \tilde{\sf F}(s'){\sf \Omega } \tilde{\sf G}(t') {\sf \Omega} \tilde{\sf F}(s) &= \kappa {\sf \Omega} \tilde{\sf G}(t) \sf {\Omega } \,,
\label{ctmrg_fix2}
\end{align}
which are respectively equivalent to the variational equations (\ref{veq_column1}) and (\ref{veq_row1}) with  $\alpha=\beta=\kappa$.
Finally, using Eq. (\ref{ctmrg_fix1}), we can reduce Eq. (\ref{eq_self_omega}) to be Eq. (\ref{eq_self_v}).
We can then see that these fixed point equations of CTMRG correspond to Eqs. (\ref{geq_FWF}), (\ref{geq_FF}), (\ref{geq_GWG}) and (\ref{geq_GG}) within the gauge transformation.
Thus, the fixed point of the CTMRG is attributed to the variational equations based on the MPS formulation.
In other words, the two variational principles, i.e. the variational principle for the row-to-row transfer matrix and the low-rank approximation of the CTM through the SVD, lead to the same equations.
This is the reason why the MPS algorithms are able to provide stable and accurate numerical results for 2D classical models as well as 1D quantum systems.

\subsection{MPS revisited}

In the CTMRG, $\tilde{\sf U}$ and $\tilde{\sf V}$ are computed as the singular vectors for the CTM.  
At the fixed point of the CTMRG, nevertheless, we already see that Eq. (\ref{UV_MPS})  directly relate $\tilde{\sf V}(s)$ to 
 $\tilde{\sf F}(s)$ in the MPS for the row-to-row transfer matrix.
Substituting $ {\sf F}(s) =  {\sf \Omega}^{-1} \tilde{\sf V}(s) {\sf \Omega} =  {\sf \Omega} \tilde{\sf V}^\dagger (s) {\sf \Omega}^{-1}  $ to the original MPS form of Eq. (\ref{variationalMPS}),  we can reconstruct other MPS representations of the variational state as follows,
\begin{align}
|\Psi\rangle 
 & = \cdots\tilde{\sf V} (s_{n+1}) \tilde {\sf V} (s_n) \tilde{\sf V} (s_{n-1}) \cdots \, , \label{leftMPS} \\
 & = \cdots \tilde{\sf V}^\dagger (s_{n+1}) \tilde{\sf V}^\dagger (s_n) \tilde{\sf V}^\dagger (s_{n-1}) \cdots \, ,\label{rightMPS}\\
 & = \cdots \tilde{\sf V} (s_{n+2}) \tilde{\sf V} (s_{n+1}) {\sf \Omega}^2 \tilde{\sf V}^\dagger (s_n) \tilde{\sf V}^\dagger (s_{n-1})  \cdots \, ,
\label{mixedMPS}
\end{align}
which are respectively called left canonical, right canonical and mixed canonical forms.\cite{site_matrix}
A crude derivation of the MPS from the CTM was also discussed in Ref. [\citen{Ueda2010}]. 
Note that for the 1D quantum system, ${\sf \Omega}^2$ in Eq. (\ref{mixedMPS}) corresponds to the singular values for the bipartitioned wavefunction.
From Eq. (\ref{UV_MPS}) combined with the fact of $\tilde{\sf F}(s)$ being the symmetric matrix, moreover, it follows that the relation
\begin{align}
 \tilde{\sf V} (s) {\sf \Omega}^2 =  {\sf \Omega}^2 \tilde{\sf V}^\dagger (s) 
\label{VOOV}
\end{align}
holds for the transformation matrix.
This implies that the location of ${\sf \Omega}$ can be shifted in Eq. (\ref{mixedMPS}).\cite{PWFRG}
These MPS representations extracted from the RG transformation matrix play an essential role for 1D quantum systems where the background transfer matrix and ${\sf F}$ matrix are not known apriori.
In particular, the DMRG is formulated based on the mixed canonical form, without explicitly referring to $F$ and $C$ tensors.
Details of the DMRG for 1D quantum systems will be explained in the next section, where Eq. (\ref{VOOV}) can be generalized to the position dependent form for a finite-size system.

It is also worth noticing that relation of Eqs. (\ref{UV_MPS}) and (\ref{VOOV}) can be used for acceleration of MPS-based algorithms.
As seen so far, convergence of the CTMRG type algorithms basically follows from the power method with respect to the renormalized transfer matrices.
By imposing the uniform bulk relation of Eqs. (\ref{UV_MPS}) and (\ref{VOOV}) to MPS tensors, an accelerated algorithm can be formulated in Ref.~[\citen{Fishman2018}].
This acceleration is also possible for the variational MPS algorithm for 1D quantum systems, which is dubbed as VUMPS\cite{Zauner-Stauber2018}.
This algorithm is useful for dealing with critical systems where convergence of the power method becomes slow down.

\subsection{accuracy of CTMRG/MPS and entanglement}
\label{sec_4_5}

Accuracy of the CTMRG is attributed to the low-rank approximation of SVD for the CTM or the reduced density matrix and thus is controlled by a cutoff number of retained basis (bond dimension) $m$.\cite{cutoffm}
The intrinsic reason why the CTMRG has succeeded in precisely calculating thermodynamic quantities is that the spectrum ${\sf \Omega}$ at the fixed point is discrete and its decay is very rapid except at the critical point.
In the qualitative level, this is because the CTM involving the sum with respect to the quadrant of the infinite system would more directly reflect the bulk information, in contrast to the row-to-row transfer matrix which generally contains continuous excitation spectrum.
As mentioned for Eq. (\ref{wf_classical_quantum}), moreover, the product of two CTMs corresponds to the ground-state wavefunction for 1D quantum systems through the Suzuki-Trotter decomposition
(See also \S \ref{Sec_4}).
We also note that the row-to-row transfer matrix and the corresponding Hamiltonian for the integrable model have simultaneous eigenstates.
Thus, the situation of ${\sf \Omega}$ basically holds for MPS approaches such as DMRG and infinite TEBD for 1D quantum systems.

From the entanglement viewpoint,  an essential point is that the EE, say  $S_{\rm EE} $, for a certain bipartitioning of the ground-state wavefunction of quantum systems in the off critical regime generally exhibits so-called area law; it is asymptotically proportional to the area of the interface between system and reservoir parts,  $S_{\rm EE} \sim \ell^{d-1} $, where $\ell$ is the linear dimension of a system part and $d$ is the spatial dimension of the ground-state wavefunction.
In particular,  we have $S_{\rm EE} \sim {\rm const} $ for $\ell \gg 1$ for a 1D quantum system.
Since the upper bound of EE represented by the CTM/MPS with the bond dimension is given by $\sim \log m$, the MPS/CTM with a large but finite $m$ is capable of well approximating the bulk state.
Actually, the rapid convergence and decay of $\sf \Omega$ in the CTMRG for the corresponding 2D classical system ensures $S_{\rm EE} = -\Tr {\sf \Omega}^4 \log {\sf \Omega}^4 \sim {\rm const}$ consistent with the area law of EE.

For quantitative analysis of the CTMRG accuracy, a primal problem is how the distribution of the CTM spectrum(or equivalently the reduced-density-matrix spectrum) behaves at the fixed point.
Then, the exact CTM spectra calculated for integrable models such as Ising model and eight-vertex model\cite{Baxter1976,Baxter1977} provide essential insight for general cases.
The asymptotic behavior of the spectrum was extracted in Ref. [\citen{OHA}] to show the universal form
\begin{align}
\omega_m \sim e^{-\alpha (\log m)^2 }
\label{eq_asym_ev}
\end{align}
where $\alpha$ is a nonuniversal dimensionless constant reflecting a distance from the critical point.
The same asymptotic formula was also derived in the context of grand-canonical density matrices\cite{GarnetChan2002} and  perturbed CFTs\cite{Lefevre2008}.
Although Eq. (\ref{eq_asym_ev}) seems to show relatively slow decay with increasing $m$, a practical  point is that the coefficient $\alpha$ is usually moderately large except at the critical point, which ensures very accurate CTMRG results within  $m\sim$ a few hundred and the consistency with the area law of EE.
For example,  $\alpha$ of the Ising model is of order of unity even at $|T-T_c |/J \simeq 0.01$, so that the CTMRG yields numerically exact results with $m\sim 100$ ($e^{-(\log 100)^2}\sim 6\times 10^{-10}$).

As the system approaches the critical point, the decay rate $\alpha$ of the CTM spectrum also becomes small and  the log-correction to the area law of EE emerges reflecting the critical fluctuation.
In principle, the exact spectrum at the critical point is not normalizable in the bulk limit, for which Eq. (\ref{eq_asym_ev}) breaks down.
If the CTMRG with a finite $m$ is applied to the critical system, however, CTMRG iterations finally converge to a certain approximated value of the partition function $\kappa$.
This finite $m$ fixed point of the CTM  was firstly investigated in the framework of the low-temperature series expansion and the crossover to the mean-field-like behavior was concluded as $T \to T_c$\cite{Tsang1979}.
This suggests that an effective correlation length $\xi_{\rm e}(m)$ specified by a finite $m$  was brought into fixed-point tensors of the CTMRG, which is often mentioned as ``finite-$m$ effect" or ``finite-entanglement effect".

In the CTMRG framework,  $\xi_{\rm e}(m)$ can be numerically evaluated by the largest and second-largest eigenvalues of the effective row-to-row transfer matrix, e.g.  $\tilde{T} \equiv \tilde{G} W \tilde{G}$,  constructed from the fixed-point tensors.
Recall that the iteration number $N$ in the CTMRG is proportional to the system size.
For $N \ll  \xi_{\rm e}(m) $, thus, the finite-size scaling behavior of physical quantities can be observed, while for $N \gg \xi_{\rm e}(m)$, the CTMRG calculation converges to the fixed point characterized by $m$.
With increasing $m$, the accuracy of the finite-$m$ fixed point is gradually improved toward the exact one corresponding to $m\to \infty$.
Note that such analysis of the $m$-dependence of the MPS at the critical point is  recently  termed as ``finite-$m$ scaling'' or ``finite-entanglement scaling". 

The crossover between the finite-size scaling and the finite-entanglement scaling can be described by the two-parameter-scaling ansatz, which was originally proposed in Ref. [\citen{NOK1996}].
According to recent developments in the MPS approach, moreover, the effective correlation length $\xi_{\rm e}$ behaves   
\begin{align}
\xi_{\rm e}(m) \sim m^\theta
\end{align}
for critical systems, where $\theta$ is a new phenomenological exponent characterizing the divergence of $\xi_{\rm e}(m)$ in the context of the finite-$m$ scaling.\cite{Tagliacozzo2008}
Moreover, an argument based on a perturbed CFT combined with the MPS provides
\begin{align}
 \theta = \frac{6}{c\left(1+\sqrt{12/c} \right)}, 
 \label{eq_theta_c}
\end{align}
where $c$ denotes the central charge in the corresponding CFT.\cite{Pollmann2009}
The two-parameter scaling analysis combined with the CTMRG has been successfully applied to exotic phase transitions.\cite{UedaIcosa,UedaBKT,UedaDodeca}.
For 1D critical spin systems, the two-parameter scaling combined with the MPS approaches is also confirmed.\cite{Tagliacozzo2008,Pirvu2012}.
Nevertheless, it is also known that some numerical results suggest a weak violation of Eq. (\ref{eq_theta_c}).\cite{Schmoll2019,UedaBKT}
Further investigations will be required  to check the relation (\ref{eq_theta_c}).

\section{MPS-type algorithms for 1D quantum systems}
\label{Sec_4}

In general,  1D quantum systems can be mapped into the corresponding 2D classical models with the Suzuki-Trotter decomposition\cite{Trotter1959, Suzuki1976}.
Thus, it is expected that the CTM formulation of the variational approximation based on the MPS in \S \ref{Sec_2} can be recast to the ground-state problem of  the 1D quantum systems.
Historically, however, the DMRG was invented as a numerical RG approach to 1D  quantum systems, independently of the CTM-based variational approximation. 
More precisely, in the DMRG, the ground-state wavefunction of the super-block Hamiltonian is directly calculated, and then reduced density matrices for subsystems of the left or right halves are diagonalized to construct the RG transformation matrix, without exploiting the CTM representation.
This is because the 2D classical model generated by the Suzuki-Trotter decomposition is usually a highly anisotropic system with the periodic boundary in the imaginary time direction, where the CTM representation of the ground-state wavefunction is not so convenient.
Here, we discuss how the above difficulty can be bypassed in the formulation of infinite TEBD (iTEBD) and DMRG for 1D quantum systems.

\subsection{1D quantum vs 2D classical} \label{1d2d}

Let us briefly review basic features of the Suzuki-Trotter decomposition for such a typical system as $S=1/2$ Heisenberg spin chain.
The Hamiltonian is written as
\begin{equation}
{\hat H} = {\hat H}_{\rm e} + {\hat H}_{\rm o} \, ,
\label{Heisenberg_H}
\end{equation}
where
\begin{align}
{\hat H}_{\rm e}^{~}  = \sum_{n = {\rm even}}^{~} \hat{h}_{n,n+1}\, , \quad 
{\hat H}_{\rm o}^{~}  = \sum_{n = {\rm odd}}^{~} \hat{h}_{n,n+1}\, , 
\end{align}
with
\begin{align}
\hat{h}_{n,n+1} \equiv J \bm{S}_{n}\cdot \bm{S}_{n + 1} \, .
\label{eq_localH}
\end{align}
Here,  $ \bm{S}$ denotes the $S=1/2$ spin matrices.
We assume the system length $N$ with  $N$ being an even number.
Note that  $[{\hat H}_{\rm e}^{~}  ,{\hat H}_{\rm o}^{~}  ] =0$.
Using the Suzuki-Trotter decomposition to Eq. ({\ref{Heisenberg_H}), 
we can rewrite the partition function at an inverse temperature $\beta \equiv 1 / k_{\rm B}^{~} T$ as 
\begin{equation}
Z( \beta ) = {\rm Tr} \, e^{- \beta {\hat H}}_{~} \simeq {\rm Tr} \left[ e^{-\epsilon {\hat H}_{\rm e}}  e^{-\epsilon {\hat H}_{\rm o}} \right ]^M \, ,
\label{Suzuki-Trotter}
\end{equation}
with $\epsilon = \beta / M $ $(M \gg 1)$.
As shown in Fig. \ref{itebd_lattice}(a), Eq. (\ref{Suzuki-Trotter}) defines the 45$^\circ$-rotated vertex model with the local Boltzmann weight 
\begin{align}
U_{n,n+1}= e^{- \epsilon J \bm{S}_n \cdot \bm{S}_{n+1}} 
=
\hspace{3mm}{\setlength\unitlength{2mm}
\begin{picture}(6,3)(0,-0.3)
\put(1,-2){\line(1,1){4}}
\put(1,2){\line(1,-1){4}}
\put(0.3,-2.7){\makebox(0,0)[c]{\scriptsize \mbox{$s_n$}}}
\put(6,-2.7){\makebox(0,0)[c]{\scriptsize \mbox{$s_{n+1}$}}}
\put(0.3,2.5){\makebox(0,0)[c]{\scriptsize \mbox{$s'_n$}}}
\put(6,2.5){\makebox(0,0)[c]{\scriptsize \mbox{$s'_{n+1}$}}}
\end{picture}
}  \label{vertex_ST} \, ,\\
\nonumber
\end{align}
which is nothing but the local imaginary time evolution operator of a tiny time step $\epsilon$ in the $S^z$-diagonal representation.
In this section, we will omit the leg indices of $U_{n,n+1}$ for legibility.

In principle, it is possible to perform the CTMRG for this $45^\circ$-rotated lattice. 
In the CTMRG, however, the periodic boundary in the imaginary time direction is difficult to handle and thus the infinite trotter number limit($M\to \infty$) is not well controllable.
Instead, the quantum transfer matrix approach with a finite $M$ is often used,\cite{Suzuki1987,WangXiang1997,Shibata1997,Shibata2003,Okunishi1999}  but the double extrapolations with respect to $\beta\to \infty$ and $M\to \infty$ is required.
Also, the asymmetric quantum transfer matrix implies that an efficient treatment of dual biorthonormal bases is required for a stable computation\cite{Huang2012}.
These difficulties basically originate from the fact that the 1D quantum system corresponds to the highly anisotropic limit of the 2D classical model generated by Eq. (\ref{Suzuki-Trotter}).
Thus, we had better directly deal with the ground-state wavefunction of the Hamiltonian to avoid the above cumbersome extrapolations. 
In addition,  the local vertex of Eq. (\ref{vertex_ST}) is a 4-leg tensor acting on two spins, which should be contrasted to the vertex weight of Eq. (\ref{vertex_weight}) that only has a single connection in the column and row directions.
This implies that an appropriate adjustment of the MPS algorithm in the previous sections is needed for practical calculations.

\subsection{iTEBD}

As mentioned above, a direct evaluation of $Z(\beta)$ with the CTM accompanies some technical difficulties.
We explain the iTEBD for 1D quantum systems\cite{iTEBD}, which is a compact updating algorithm of local tensors based on the mixed canonical MPS with an implicit use of the background CTM structure.
The basic idea is to represent the ground-state wavefunction as  $|\Psi\rangle = \lim_{\beta\to \infty} e^{-\beta \hat{H} } |\Psi_0\rangle$, through the imaginary-time evolution based on the Trotter decomposition 
\begin{align}
 e^{-\tau \hat{H} }\simeq  \left[ e^{-\epsilon {\hat H}_{\rm e}}    e^{-\epsilon {\hat H}_{\rm o}} \right ]^M \, ,
\label{checker_board_decomp}
\end{align}
where $\tau \equiv \epsilon M$ with a small $\epsilon$ fixed denotes a discretized imaginary time.
Starting from an initial state $|\Psi_0\rangle$, then, we can systematically construct a final state at a sufficiently large $\tau$ ($M \gg 1 $), which provides a good approximation of the ground-sate wavefunction.


\begin{figure}
\begin{center} \includegraphics[width = 8.5 cm]{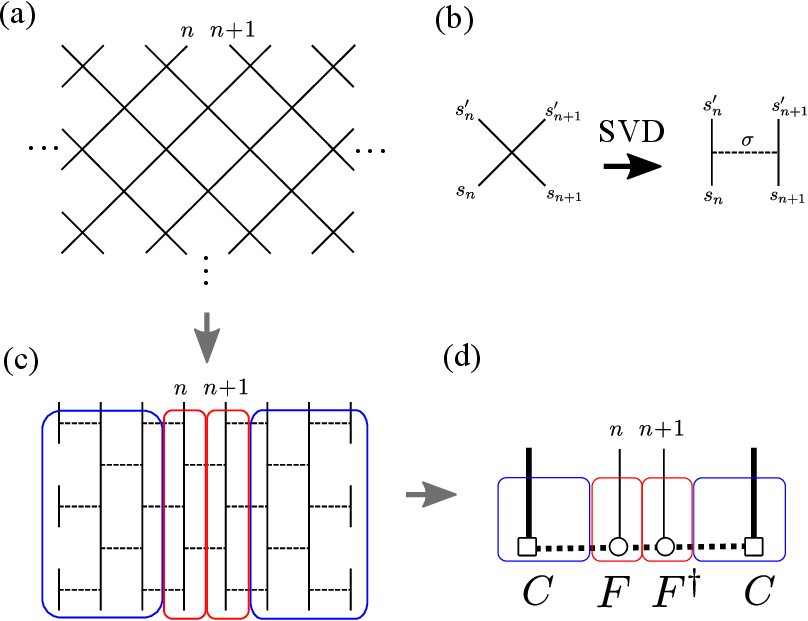} \end{center}
\caption{(Color online) The MPS and CTM structure behind the $S=1/2$ Heisenberg chain.
(a) The 45$^\circ$-rotated vertex model generated by the Suzuki-Trotter decomposition. 
The half-infinite world-sheet of this vertex model corresponds to the ground-state wavefunction. 
(b) The diagram of SVD for the local vertex weight $U_{n,n+1}$, which is given by a contraction of two 3-leg vertices with the horizontal dotted line representing an effective spin $\sigma$. 
(c) A brick-wall lattice converted from the panel (a). 
(d) The renormalized tensors $F$ and $C$ respectively correspond to the red and blue lines in the panel (c). 
The thick horizontal broken lines of $C$ and $F$ represent the effective spins corresponding to bunches of the horizontal dotted lines of $\sigma$ spins in (c), while the thick vertical line of $C$ indicates the renormalized spin indices of the vertical lines corresponding to bunches of physical spin legs in (c). }
\label{itebd_lattice}
\end{figure}

In analogy with the maximal eigenvector of the transfer matrix, the ground-state wavefunction in the bulk limit can be represented as the half-infinite world-sheet of the 45$^\circ$-rotated vertex model generated by the Suzuki-Trotter decomposition.
In Fig. \ref{itebd_lattice} (a),  certain two adjacent sites in the bulk wavefunction are labeled by $n$ and $n+1$.
In order to illustrate the hidden CTM structure of the wavefunction, it is helpful to use SVD of the local vertex weight, 
\begin{eqnarray}
U_{n,n+1}
= \sum_{\sigma=1}^{4} W(\sigma, s'_{n} s_{n}) \Sigma_\sigma W(\sigma, s'_{n+1} s_{n+1} )
\label{eq_Uto3leg}
\end{eqnarray}
where $\Sigma_\sigma$ denotes singular values and $W$ represents the corresponding singular vectors.
Note that $U_{n,n+1}$ is a real symmetric matrix with respect to $s_ns_{n}'$ and $s_{n+1}s'_{n+1}$ due to the  bond parity of $\hat{h}_{n,n+1}$. 
Then, the  singular vectors attached with the weight of square root of the singular values, i.e. $ \sqrt{\Sigma_\sigma }  W(\sigma, s'_{n} s_{n} ) $  can be regarded as a 3-leg vertex weight, as depicted in Fig. \ref{itebd_lattice} (b), where the horizontal dotted line represents 4 state index of $\sigma$.
Using this 3-leg vertex, we formally convert the 45$^\circ$-rotated vertex model into the brick-wall-lattice model in Fig. \ref{itebd_lattice}(c).

The next step is to decompose the brick-wall lattice in Fig. \ref{itebd_lattice}(c) into four pieces surrounded by red and blue lines, reflecting the two-sublattice structure.   
As in Fig. \ref{itebd_lattice} (d), we then construct two $F$ tensors sandwiched by two $C$ tensors, bundling the vertical lines and horizontal dotted lines of the tensors into the thick lines that represent the renormalized indices of corrective spin degrees of freedom.
Of course, the two $F$ tensors reflect the two-sublattice version of MPS and $C$ corresponds to the CTM.

For later convenience, we turn to the matrix representation with the CTM-diagonal basis, which was already introduced in the previous section.
Assuming $n$=even, we may write the wavefunction matrix as 
\begin{align}
{\sf \Psi}_\mathrm{e}( s_{n} s_{n+1}) &= {\sf \Omega} {\sf F}^\dagger(s_{n+1}) {\sf F}(s_n) {\sf \Omega}  
 = \hspace{3mm}{\setlength\unitlength{2mm}
\begin{picture}(8,3)(0,-0.3)
\put(2,2.3){\line(0,-1){2.0}}
\put(5,2.3){\line(0,-1){2.0}}
\put(1.5,-0.5){\mbox{$\circ$}}
\put(4.5,-0.5){\mbox{$\circ$}}
\put(-0.5,-0.5){\mbox{$\square$}}
\put(6.3,-0.5){\mbox{$\square$}}
\linethickness{1.5pt}
\put(0.1,2.3){\line(0,-1){1.8}}
\put(6.85,2.3){\line(0,-1){1.8}}
\multiput(0.6,0)(0.7,0){2}{\line(1,0){0.3}}
\multiput(2.2,0)(0.7,0){4}{\line(1,0){0.3}}
\multiput(5.33,0)(0.7,0){2}{\line(1,0){0.3}}
\put(2,-1){\makebox(0,0)[c]{\scriptsize \mbox{$ F$}}}
\put(5,-1){\makebox(0,0)[c]{\scriptsize \mbox{$ { F^\dagger}$ } }}
\put(2,3){\makebox(0,0)[c]{\scriptsize \mbox{$s_n$}}}
\put(5,3){\makebox(0,0)[c]{\scriptsize \mbox{$s_{n+1}$}}}
\put(0,-1){\makebox(0,0)[c]{\scriptsize \mbox{${ \Omega}$}}}
\put(7,-1){\makebox(0,0)[c]{\scriptsize \mbox{${ \Omega}$}}}
\end{picture}
}\; , 
\label{offo_even}
\end{align}
where  ${\sf F}(s_n)$ and ${\sf \Omega}$ respectively denote matrix representations of the $F$-tensor and the singular values of the CTM.\cite{site_matrix,comment_tilde}
In the diagram of Eq.~(\ref{offo_even}),  ${\sf \Omega}$ is the singular values of the CTM, which joints the horizontal dotted line with the vertical one. 
This representation of the wavefunction clearly corresponds to Eq. (\ref{wf_classical_quantum}) for the CTMRG within a tiny deviation due to the difference of the boundary conditions.
Note that the bond-parity symmetry of $\hat{h}_{n,n+1}$ yields ${\sf F}^\dagger(s_{n+1})$ at the $n+1$th site.

As mentioned before, a direct application of the CTM to the Suzuki-Trotter decomposed system may not be so convenient.
The iTEBD provides an easy-to-use iterative algorithm of evaluating Eq. (\ref{offo_even}) with no explicit use of  the 3-leg vertex weight and the CTM.   
In order to improve ${\sf F}$ and ${\sf \Omega}$ starting from certain initial tensors, we directly operate the 4-leg weight $U_{n,n+1}$ to ${\sf \Psi}_\mathrm{e}$, 
\begin{align}
\bar{\sf \Psi}_\mathrm{e}(s'_ns'_{n+1} ) & =\!\! \sum_{s_n,s_{n+1}} U_{n,n+1} {\sf \Omega} {\sf F}^\dagger(s_{n+1}) {\sf  F}(s_n)  {\sf \Omega} \nonumber \\
& = \hspace{1mm}{\setlength\unitlength{2mm}
\begin{picture}(8,5)(0,-0.3)
\put(2,0){\line(1,1){3}}
\put(2,3){\line(1,-1){3}}
\put(2,0){\line(0,-1){1.7}}
\put(5,0){\line(0,-1){1.7}}
\put(1.5,-2.5){\mbox{$\circ$}}
\put(4.5,-2.5){\mbox{$\circ$}}
\put(-0.5,-2.5){\mbox{$\square$}}
\put(6.3,-2.5){\mbox{$\square$}}
\linethickness{1.5pt}
\put(0.1,0){\line(0,-1){1.6}}
\put(6.85,0){\line(0,-1){1.6}}
\multiput(0.6,-2)(0.7,0){2}{\line(1,0){0.3}}
\multiput(2.2,-2)(0.7,0){4}{\line(1,0){0.3}}
\multiput(5.33,-2)(0.7,0){2}{\line(1,0){0.3}}
\put(2,-3){\makebox(0,0)[c]{\scriptsize \mbox{$ F$}}}
\put(5,-3){\makebox(0,0)[c]{\scriptsize \mbox{$ { F}^\dagger $ } }}
\put(1.3,2.5){\makebox(0,0)[c]{\scriptsize \mbox{$s'_n$}}}
\put(6.5,2.5){\makebox(0,0)[c]{\scriptsize \mbox{$s'_{n+1}$}}}
\put(0,-3){\makebox(0,0)[c]{\scriptsize \mbox{${ \Omega}$}}}
\put(7,-3){\makebox(0,0)[c]{\scriptsize \mbox{${ \Omega}$}}}
\end{picture}
} \, , \label{eq_U_Psi_e} \\
\nonumber
\end{align}
which is slightly improved toward the ground-state wavefunction.
In the context of the 3-leg vertex weight, this process increases the number of the vertices included in the $F$ tensor by one in the imaginary-time direction.  

In order to extract an improved $F$ tensor from Eq. (\ref{eq_U_Psi_e}),  we regard $\bar{\sf \Psi}_\mathrm{e}(s'_n s'_{n+1} )$ as a symmetric matrix of $2m\times 2m$ and then perform SVD,
\begin{align}
\bar{\sf \Psi}_\mathrm{e}(s'_{n} s'_{n+1}) = \bar{\sf V}(s'_{n+1}) \bar{\sf \Lambda} \bar{\sf V}^\dagger(s'_{n})\, ,
\label{svd_psi_e}
\end{align}
where  $\bar{\sf \Lambda}$  and  $\bar{\sf V}$ respectively denote the singular values and the corresponding singular vectors.
We then retain the larger half of $\bar{\sf \Lambda}$  and the corresponding  $\bar{\sf V}$ to maintain the matrix dimensions.
However,  Eq. (\ref{svd_psi_e}) has the sublattice structure shifted from Eq. (\ref{offo_even}), reflecting the 45$^\circ$-rotated square (or brick wall) lattice due to the Suzuki-Trotter decomposition.  
For the purpose of restoring the improved $\bar{F}$,  an important point is that Eq. (\ref{svd_psi_e}) is in the mixed canonical form of the wavefunction;
On the basis of the bulk relations of Eq. (\ref{UV_MPS}), we can extract $\bar{F}$ tensor with   
\begin{align}
\bar{\sf F}^\dagger (s'_{n})  = \bar{\sf \Omega}^{} \bar{\sf V}^\dagger (s'_{n}) {\sf \Omega}^{-1} \, ,
\label{itebd_FV}
\end{align}
or equivalently $\bar{\sf F} (s'_{n+1})  = {\sf \Omega}^{-1} \bar{\sf V} (s'_{n+1}) \bar{\sf \Omega}^{}$,  where $ \bar{\Omega} \equiv \sqrt{\bar{\Lambda}}$.
Here, note that the lhs of Eq.(\ref{itebd_FV}) is not $ \bar{\sf F}$ but $\bar{\sf F}^\dagger $, because the sublattice of $\bar{\sf F}$ is shifted. 
At the same time,  the sublattice associated with $\bar{\sf \Omega}$ has been also shifted by one site in the spatial direction.

In order to proceed to the next step of optimization for the $F$ and $\Omega$ tensors, we should take account of the shift of the sublattice structure. 
With moving $n \to n+1$, we next consider $\bar{\sf \Psi}_\mathrm{o}$ constructed from $\bar{\sf F}$ and $\bar{\sf \Omega}$ for another sublattice, 
\begin{align}
\bar{\sf \Psi}_\mathrm{o}(s'_{n+1} s'_{n+2} )  =\bar{\sf \Omega}\, \bar{\sf F}^\dagger (s'_{n+2}) \bar{\sf F} (s'_{n+1})\bar{\sf \Omega} 
\label{mps_psi_o}\, 
\end{align}
Then, we operate $U_{n+1,n+2}$ to $\bar{\sf \Psi}_\mathrm{o}$,  
\begin{align}
\bar{ \bar{ {\sf \Psi} }}_{\rm o}(s''_{n+1} s''_{n+2})  & \equiv \sum_{s'_{n+1},s'_{n+2}} U_{n+1,n+2} \bar{\sf \Omega} \bar{\sf F}^\dagger (s'_{n+2})  \bar{\sf  F}(s'_{n+1}) \bar{\sf \Omega} \nonumber \\
& = 
\hspace{3mm}{\setlength\unitlength{2mm}
\begin{picture}(8,5)(0,-0.3)
\put(2,0){\line(1,1){3}}
\put(2,3){\line(1,-1){3}}
\put(2,0){\line(0,-1){1.7}}
\put(5,0){\line(0,-1){1.7}}
\put(1.5,-2.5){\mbox{$\circ$}}
\put(4.5,-2.5){\mbox{$\circ$}}
\put(-0.5,-2.5){\mbox{$\square$}}
\put(6.3,-2.5){\mbox{$\square$}}
\linethickness{1.5pt}
\put(0.1,0){\line(0,-1){1.6}}
\put(6.85,0){\line(0,-1){1.6}}
\multiput(0.6,-2)(0.7,0){2}{\line(1,0){0.3}}
\multiput(2.2,-2)(0.7,0){4}{\line(1,0){0.3}}
\multiput(5.33,-2)(0.7,0){2}{\line(1,0){0.3}}
\put(2,-3.2){\makebox(0,0)[c]{\scriptsize \mbox{$\bar{ F} $}}}
\put(5,-3.2){\makebox(0,0)[c]{\scriptsize \mbox{$ \bar{ F}^\dagger $ } }}
\put(1.3,2.5){\makebox(0,0)[c]{\scriptsize \mbox{$s''_{n+1}$}}}
\put(7,2.5){\makebox(0,0)[c]{\scriptsize \mbox{$s''_{n+2}$}}}
\put(0,-3.2){\makebox(0,0)[c]{\scriptsize \mbox{$\bar{ \Omega}$}}}
\put(7,-3.2){\makebox(0,0)[c]{\scriptsize \mbox{$\bar{ \Omega}$}}}
\end{picture}
} \, , \label{eq_U_Psi_o}\\ 
\nonumber 
\end{align}
and perform SVD,
\begin{align}
\bar{ \bar{ {\sf \Psi} }}_{\rm o}(s''_{n+1} s''_{n+2}) = \bar{\bar{ {\sf V} }}(s''_{n+2}) \bar{\bar{ {\sf \Lambda} }} \bar{\bar{ {\sf V} }}^\dagger (s''_{n+1}) \,. \label{eq_SVD_odd}
\end{align}
Similarly to  Eq. (\ref{itebd_FV}),  we then extract an improved $F$-tensor from Eq. (\ref{eq_SVD_odd}) as
\begin{align}
\bar{\bar{ {\sf F} }}^\dagger(s''_{n+1})  = \bar{\bar{ {\sf \Omega} }}^{} \bar{\bar{ {\sf V} }}^\dagger (s''_{n+1}) {\bar{ {\sf \Omega} }}^{-1} 
\end{align}
with $\bar{\bar{\Lambda}} = \bar{\bar{\sf \Omega}}^2 $.
Here, note that $\bar{\bar{ {\sf F} }}(s''_{n+1})$ and $\bar{\bar{\sf \Omega}}$ respectively have the same sublattice structures as ${\sf F}(s_{n+1})$ and $ {\sf \Omega}$ in Eq. (\ref{offo_even}).
Thus, we have established a closed loop of the iterative update of the local tensor $F$ and CTM spectrum $\Omega$ with replacing $\bar{\bar{ {\sf F} }} \to  {\sf F}$ and $\bar{\bar{ {\sf \Omega} }} \to {\sf \Omega}$.

Repeating the above recursive processes from certain initial tensors,  we can iteratively calculate the uniform MPS representation of the ground-state.
Here, note that the operation of $U$ in Eqs. (\ref{eq_U_Psi_e}) and (\ref{eq_U_Psi_o}) increases the system size in the imaginary-time direction, while the insertion of ${\sf F}$ matrix at the center of the wavefunction in Eqs.(\ref{offo_even}) and (\ref{mps_psi_o}) corresponds to extending the size in the spatial direction.
These processes ensure convergence of iTEBD iterations to the bulk ground state.
The error in the iTEBD comes from the cutoff dimension $m$ and the Trotter discretization $\epsilon$.

We comment on the relation to Vidal's notation of iTEBD\cite{iTEBD}, which is explicitly given by 
\begin{align}
{\sf \Gamma}(s_n) \equiv {\sf \Omega}^{-1} {\sf F}(s_n) {\sf \Omega}^{-1}  = {\sf V}^\dagger (s_n) {\sf \Omega}^{-2} 
\end{align}
with $\Lambda= \Omega^2$. 
Thus, the iTEBD algorithm based on the $\Gamma-\Lambda$ representation is equivalent to the CTM-based formulation above, although the original iTEBD was formulated in the context of the Schmidt decomposition for the bulk wavefunction.

\subsection{infinite-size DMRG}

As discussed in \S \ref{1d2d},  numerical RG approaches for the  2D classical system by the Suzuki-Trotter decomposition accompany subtle problems due to the Trotter error and the periodic boundary condition in the imaginary-time direction.
In particular, it seems difficult to directly construct the recursive relation for the ground-state wavefunction without a priori knowledge of such a local tensor as $F(s|\mu\mu')$ in the background classical system.
The DMRG is a milestone algorithm to eventually generate the MPS in the mixed canonical form of Eq. (\ref{mixedMPS})  with the striking use of SVD for the ground-sate wavefunction directly calculated from the total (superblock) Hamiltonian.
Moreover, various considerations about the DMRG mechanism opened the door to further developments in the numerical RG assisted by quantum information theory.

\begin{figure}[bt]
\begin{center} 
\includegraphics[width = 4.8 cm]{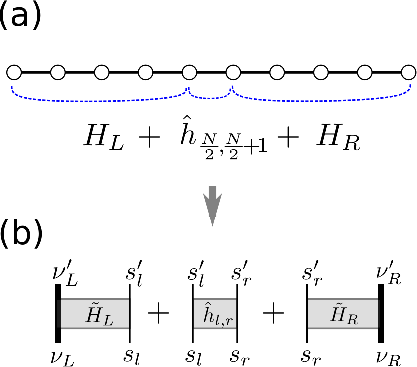}
\includegraphics[width = 3.7 cm]{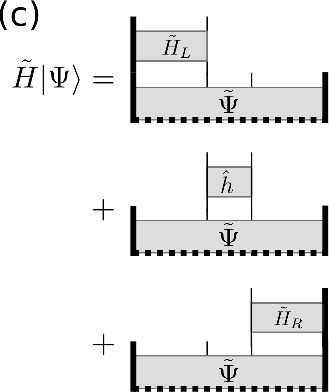}
\end{center}
\caption{(Color online) The left- and right-block Hamiltonians in the DMRG.
(a) The total chain is divided into three pieces: 
The center bond Hamiltonian, $\hat{h}_{\frac{N}{2},\frac{N}{2}+1}$,and the left and right block Hamiltonians, $H_L$ and $H_R$.
(b) The block Hamiltonian $H_L(H_R)$ is renormalized into the $\tilde{H}_L$($\tilde{H}_R$) with the renormalized spin indices $\nu_L$ and $ \nu_L'$($\nu_R$ and $\nu_R'$).
Also, we write $\hat{h}_{l,r}$ with $l\equiv \frac{N}{2}$ and $r\equiv \frac{N}{2}+1$.
(c) Operation of the super-block Hamiltonian to a wavefunction can be also decomposed into three parts, as in Eq. (\ref{HPsi}).
}
\label{fig_DMRG}
\end{figure}

In this subsection, we discuss the DMRG for 1D quantum systems, under the light of the background 2D classical system.  
As an example, we consider the $S=1/2$ Heisenberg spin chain containing $N$(=even) spins with the open boundaries.
The local Hamiltonian was defined by Eq. (\ref{eq_localH}).
As shown in Fig. \ref{fig_DMRG},  the entire Hamiltonian (superblock Hamiltonian) in the DMRG is bipartitioned into left and right block Hamiltonians, 
\begin{align}
{H}_L(s'_{\frac{N}{2}} \nu'_L| s_{\frac{N}{2}} \nu_L)  & =  \sum_{n=1}^{\frac{N}{2}-1}  \hat{h}_{n,n+1},\,  \label{eq_iH_left}\\
{H}_R(s'_{\frac{N}{2}+1} \nu'_R| s_{\frac{N}{2}+1} \nu_R) &= \sum_{n=\frac{N}{2}+1}^{N-1} \hat{h}_{n,n+1} \label{eq_iH_right} \, ,
\end{align}
where $\nu_L$ and $\nu_R$ respectively represent the block spin variables containing $n=1, \cdots, \frac{N}{2}-1 $ and $n= \frac{N}{2}+1, \cdots, N$. 
Note that dimension of $\nu_{L/R}$ is $2^{\frac{N}{2}-1}$, but is usually truncated by a certain number $m (\ll 2^{\frac{N}{2}-1} )$, which is of order of a few hundred in practical computations.
In the following, we put a tilde symbol on  a tensor (or matrix), if it is written in certain renormalized bases with $\nu_{L/R} = 1, \cdots, m$.  
We also introduce $s_l \equiv   s_{\frac{N}{2}}$ and $ s_r \equiv s_{\frac{N}{2}+1}$ for convenience.
Then, the superblock Hamiltonian is written as
\begin{align}
\tilde{H} = \tilde{H}_L + \hat{h}_{l,r} + \tilde{H}_R \, ,
\label{superblock}
\end{align}
where  $\tilde{H}_L $ and  $\tilde{H}_R $ are  the renormalized version of Eqs. (\ref{eq_iH_left}) and (\ref{eq_iH_right}). 
Note that $\tilde{H}_{L/R}$ might be associated with the $\tilde{G}$-tensor in the CTM approach.
However, we would like to remark that the superblock Hamiltonian is represented as the sum of the local Hamiltonians, while the effective transfer matrix in the classical system is given by the product of the local weights.
This difference is technically important to formulate the iterative algorithm for the quantum system.
Also, we will return to this subject in the context of matrix product operator in subsection \ref{sec_MPO}.

As mentioned before,  a direct construction of the recursion relation for the wavefunction is difficult.
In the DMRG, thus, the super-block Hamiltonian is directly diagonalized,
\begin{align}
\tilde{H} |\tilde{\Psi}\rangle = \tilde{E}_0 |\tilde{\Psi} \rangle
\label{diag_superBH}
\end{align}
where  $\tilde{E}_0$ is the ground-state energy  and $|\tilde{\Psi}\rangle$ is the corresponding wavefunction. 
Of course, the computational cost of Eq. (\ref{diag_superBH}) is rather expensive.
However,  bipartitioning of the system helps us to reduce the computational cost.
An important step is to represent $|\tilde{\Psi}\rangle$ in the matrix form with respect to the $L$ and $R$ subspace.
For this purpose, we write the ground-state wavefunction as 
\begin{equation}
| \tilde{\Psi}  \rangle = \sum_{s_{l},s_{r} }\sum_{\nu_L, \nu_R = 1}^{m}  
 |s_{l} \rangle |  \nu_L \rangle | s_{r}\rangle |\nu_R \rangle \tilde{\Psi}(s_{l}\nu_L| s_{r}\nu_R)  \, ,
\label{psi_LR}
\end{equation}
which is also consistent with Eq. (\ref{wf_classical_quantum}) for the CTMRG.
As in the case of the CTMRG, we employ the matrix representation\cite{site_matrix} as 
\begin{align}
\left[ \tilde{\sf \Psi}(s_{l} s_{r} ) \right]_{ \nu_R, \nu_L } = \tilde{\Psi}( s_{l} \nu_L| s_{r} \nu_R ) \, ,
\end{align}
where the matrix size of the block with a given $s_l$ and $ s_r$ is just $m \times m$ and thus the total dimension of $\tilde{\sf \Psi}$ becomes $2m \times 2m$.
We also write the matrix representation of $\tilde{H}_{L/R}$ as 
\begin{align}\
\left[ \tilde{\sf H}_{L}(s'_l|s_l) \right]_{\nu_L,\nu_L'} = H_L(s'_l \nu'_L|s_l \nu_L)\, , \label{tilde_H_L} \\
\left[ \tilde{\sf H}_{R}(s'_r|s_r) \right]_{\nu'_R,\nu_R} = H_R(s'_r\nu'_R|s_r \nu_R)\, 
\end{align}
where  the order of $\nu_L, \nu_L'$ is inverted from that of $\nu_R, \nu_R'$ in our convention of matrix representation.
Then, the operation of $\tilde{H}|\tilde{\Psi}\rangle $, which is a core step in the Lanczos or modified Lanczos diagonalization, is written as 
\begin{align}
\tilde{H}|\tilde{\Psi}\rangle = &\sum_{s_{l} }  \tilde{\sf \Psi}(s_{l}s'_{r} )\tilde{\sf H}_L(s'_{l}|s_{l} )  
 +  \sum_{s_{l}s_{r} }\hat{h}_{ l,r } \tilde{\sf \Psi}(s_{l}s_{r} )  \nonumber \\
& +  \sum_{s_{r}} \tilde{\sf H}_R(s'_{r}|s_{r} ) \tilde{\sf \Psi}(s'_{l}s_{r} )  \,  ,
\label{HPsi}
\end{align}
which can be diagrammatically illustrated in Fig. \ref{fig_DMRG}(c).
The three terms in Eq. (\ref{HPsi}) can be independently computed without dealing with the full matrix elements of the superblock Hamiltonian, where the computational cost is of the same order as the CTMRG.
This allows us to directly manipulate $\tilde{\sf \Psi}$  as a $2m\times 2m$ matrix in Lanczos diagonalization with such an optimized {\tt DGEMM} routine.
Thus, the dominant computational cost in the DMRG is governed by the number of Lanczos iterations.

Suppose that the ground-state wavefunction matrix $\tilde{\sf \Psi}$ is calculated by the Lanczos method, where the dimension of $\nu_{L,R}$ is assumed to be $m$.
Regarding  $s_{l}\nu_L$ and $s_r  \nu_R$ as matrix indices of $\tilde{\sf \Psi}$, we then perform SVD
\begin{align}
\tilde{\sf \Psi}(s_l  s_r) =\tilde{\sf V}_R(s_r) {\sf \Lambda} \tilde{\sf V}_L^\dagger(s_l) \, ,
\end{align}
where $\Lambda$ denotes the singular values and  $\tilde{\sf V}_{L}(s_{l})$ and $\tilde{\sf V}_{R}(s_{r})$ are the corresponding singular vectors.
If we keep the larger $m$ singular values and arrange the block of $s_l$[$s_r$] in the column direction,  we can use $\tilde{\sf V}_{L}(s_{l})$[$\tilde{\sf V}_{R}(s_{r})]$ as a RG transformation matrix of the dimension $2m \times m$.
Note that $\tilde{\sf V}_L = \tilde{\sf V}_R$ if the system has the parity symmetry.

As in Eqs. (\ref{ctmrg_recursion_G}),  the recursive relations for the left- ant right-block Hamiltonians are respectively given by
\begin{align}
\tilde{\sf H}'_L(s'_{\bar{l} }|s_{\bar{l}})&= \sum_{s_l,s_l'}\tilde{\sf V}_L^\dagger(s_l) [\hat{h}_{l,\bar{l}}+ \tilde{\sf H}_L(s'_l|s_l)]\tilde{\sf V}_L(s'_l)\, \label{dmrg_recursion_l}
,\\
\tilde{\sf H}'_R(s'_{\bar{r} }|s_{\bar{r}})&= \sum_{s_r,s_r'}\tilde{\sf V}_R^\dagger(s'_r) [\hat{h}_{\bar{r},r}+ \tilde{\sf H}_R(s'_r|s_r)]\tilde{\sf V}_R(s_r)\, ,
\label{dmrg_recursion_r}
\end{align}
where $\bar{l}$ and $\bar{r}$ denote spins newly inserted at the center of $H_{L/R}$.
If $\tilde{\sf H}_{L/R}$ contains $\frac{N}{2}$ spins, $\tilde{\sf H}'_{L/R}$ contains $\frac{N}{2}+1$ spins, but its matrix size (including $s_{\bar{l}/\bar{r}}$ and $s'_{\bar{l}/\bar{r}}$) remains at $2m\times 2m$.
Thus, we can formulate the closed loop of  DMRG iteration, returning to Eq. (\ref{superblock}) with replacing  $\tilde{\sf H'}_{L/R} \to \tilde{\sf H}_{L/R}$ and $N+2 \to N$.
Starting from a small $N$, we can recursively construct the effective Hamiltonian and the wavefunction in the bulk limit.

An essential point of the DMRG is that the recursive relation of Eqs. (\ref{dmrg_recursion_l}) and (\ref{dmrg_recursion_r}) is established with no direct reference to $F$ tensor. 
However, once we obtained ${\sf V}_R$ and ${\sf V}_L$ with the SVD of the wavefunction matrix,  we can reconstruct the mixed-canonical MPS from ${\sf V}_R$ and ${\sf V}_L$.
This implies that it is also possible to set up the recursion relation for the wavefunction matrix within the framework of the DMRG, which provides a good initial vector for the Lanczos diagonalization.
Actually, it was demonstrated that the drastic reduction of the number of Lanczos iterations with help of the recursive relation for the wavefunction.\cite{PWFRG,Hieida1997,Mcculloch2008,Ueda2010}

From the viewpoint of the underlying classical system,  we can expect that  the relation of  $\Lambda \sim \Omega^2$, as seen in Eqs. (\ref{svd_psi_e}) or (\ref{eq_SVD_odd}).
Equivalently,  we deduce $\Lambda ^2 \sim \Omega^4$ for the spectrum of the reduced density matrix $\rho\equiv {\sf \Psi}^\dagger {\sf \Psi}$.
In particular, the relation was exactly established for integrable models\cite{Peschel_Kaulke_Legeza1999, OHA}, where the Hamiltonian and the transfer matrix have a simultaneous eigenstate with help of the Yang-Baxter relation\cite{BaxterBook}.
However, we should recall that the SVD approach to the bipartitioned wavefunction in the DMRG was introduced independently of the CTM variation.
In order to obtain an well-approximated wavefunction $|\tilde{\Psi}\rangle$  within the truncated number of basis, S.R. White consider the minimization problem of the norm distance $\delta = \left| |\Psi\rangle - |\tilde{\Psi}\rangle \right|^2$.\cite{White1992,White1993}
In the matrix representation, this problem is equivalent to minimizing the Frobenius norm of
\begin{align}
\delta = \big|\big| {\sf \Psi} - \tilde{\sf \Psi} \big|\big| \, ,
\end{align}
for which the low-rank approximation of SVD provides the explicit solution.
This corresponds to the variational approximation of Eq.(\ref{Z_ctm4}) that maximizes the partition function from the view point of the 2D classical model.

Meanwhile, the SVD for the bipartioned wavefunction is also equivalent to the Schmidt decomposition in the quantum information terminology.
This fact attracts much interest to the DMRG from the quantum information side. 
Recently, the logarithm of the reduced-density-matrix spectrum, that is  $-\log \Lambda^2$, is called ``entanglement spectrum" and has been extensively used as a quantitative measure of a nonlocal correlation in the ground state.
Moreover, \"{O}stlund and Rommer revealed that the recursive use of SVD in the DMRG generates the MPS without passing through the background classical model\cite{Ostlund1995,Rommer1997}. 
They also formulated a direct variational algorithm based on the MPS, where the VBS state for the AKLT chain played the role of  benchmark model of the MPS description.
The VBS state can be exactly interpreted as a nontrivial alignment of Bell pairs of auxiliary spins.
Accordingly, the MPS description  in Refs. [\citen{Ostlund1995}] certainly triggered fusion of the DMRG and the concept of entanglement.

\subsection{finite-size DMRG}

So far, we have basically assumed the uniform ground state in the bulk limit, where DMRG/CTMRG can be viewed as iterative methods of solution for the self-consistent variational equations.
Another important aspect of the DMRG is on the finite-size algorithm, which established the position-dependent update scheme  for a position-dependent MPS.
This point also fits the construction of MPS by  recursive use of Schmidt decomposition in quantum information. 
Thus, the finite-size DMRG not only enables precise analyses of a wide class of 1D quantum many-body systems, but also accelerated the subsequent development of TN algorithms.
In this sense, we think that the contribution of the finite-size DMRG is more essential for the TN than the infinite-size DMRG.

Let us consider the $S=1/2$ Heisenberg spin chain of $N$ sites with open boundaries again.
We also divide the total Hamiltonian into three pieces,
$
H = H^{(n)}_L  + \hat{h}_{n,n+1}+ H^{(N-n)}_R 
$, 
but this time assume length of the left and right blocks to be $n$ and $N-n$  respectively.
The block Hamiltonians are explicitly defined as
\begin{align}
H^{(n)}_L = \sum_{i=1}^{n} \hat{h}_{i,i+1}\, , \quad {\rm and} \quad
H^{(N-n)}_R = \sum_{i=n+1}^{N-1} \hat{h}_{i,i+1} \, .
\end{align}
where superscripts $(n)$ and $(N-n)$ indicating the block sizes. 
Further, we introduce the renormalized version of the superblock Hamiltonian as 
\begin{align}
\tilde{H}^{(n)} = \tilde{H}_L^{(n)}+ \hat{h}_{n,n+1} + \tilde{H}_R^{(N-n)}
\label{finiteH}
\end{align} with the cutoff dimension $m$.
In a practical set up of the finite-size DMRG, $\tilde{H}^{(n)}_L$  and $\tilde{H}^{(N-n)}_R $ for all of $1 < n < N-1$ are usually retained in computer memory space.

The $n$-dependent representation of the block Hamiltonians yields the $n$-dependent wavefunctions.
Using the matrix notation, we write the wavefunction in the mixed canonical form at site $n$ as 
\begin{align}
\tilde{\sf \Psi}^{(n)}(s_n s_{n+1}) 
& = \tilde{\sf V}_R^{(n+1)}(s_{n+1}) {\sf \Lambda}^{(n)} \tilde{\sf V}_L^{(n)\dagger}(s_n)  \nonumber \\
& ={\setlength\unitlength{1.8mm}
\begin{picture}(5,3)(0,-0.3)
\put(1.3,-1){\line(0,1){3}}
\put(6.7,-1){\line(0,1){3}}
\put(3.4,-1.5){$\square$}
\linethickness{1.5pt}
\put(0,-1){\line(1,0){3.5}}
\put(8,-1){\line(-1,0){3.5}}
\put(1.7,-1){\makebox(0,0)[c]{\mbox{$\blacktriangleright$}}}
\put(4.3,-2.5){\makebox(0,0)[c]{\scriptsize \mbox{$\Lambda^{(n)}$} }}
\put(6.2,-1){\makebox(0,0)[c]{\mbox{$\blacktriangleleft$}}}
\put(7.7,-2.5){\makebox(0,0)[c]{\scriptsize \mbox{$\tilde{V}_R^{(n+1)}$} }}
\put(1.7,-2.5){\makebox(0,0)[c]{\scriptsize \mbox{$\tilde{V}_L^{(n)}$} }}
\end{picture}
}\; \label{finite_mix} \\
\nonumber
\end{align}
where ${\sf \Lambda}^{(n)}$ denotes singular values and  $\tilde{\sf V}_L^{(n)}(s_n)$ and $\tilde{\sf V}_R^{(n+1)}(s_{n+1})$ are the corresponding singular vectors.
Using  the recursive relations (\ref{dmrg_recursion_l}) or (\ref{dmrg_recursion_r}), then,  we can iteratively update $\tilde{H}_L^{(n)}$  or  $\tilde{H}_R^{(N-n)}$ with shifting $n$ from left to right or right to left.

In order to illustrate the MPS structure generated by the finite-size DMRG, it is instructive to trace the iterative construction of the renormalized block Hamiltonians.
For instance, the expectation value of the renormalized Hamiltonian (\ref{finiteH}) with respect to Eq. (\ref{finite_mix}), that is $ \langle \tilde{\Psi}^{(n)} |\tilde{ H }^{(n)}|\tilde{\Psi}^{(n)} \rangle $,  can be restored into the following from,
\begin{align}
\langle \Psi^{(n)} |H | \Psi^{(n)} \rangle  ={\setlength\unitlength{1.8mm}
\begin{picture}(25,7)(0,-0.3)
\multiput(2.0,-3)(2.0,0){5}{\line(0,1){3}}
\multiput(13.8,-3)(2.0,0){5}{\line(0,1){3}}
\multiput(2.0,2.5)(2.0,0){5}{\line(0,1){3}}
\multiput(13.8,2.5)(2.0,0){5}{\line(0,1){3}}
\put(0,0){\line(1,0){24}}
\put(0,2.5){\line(1,0){24}}
\put(2,0){\oval(4,6)[lb]}
\put(2,2.5){\oval(4,6)[lt]}
\put(0,0){\line(0,1){2.5}}
\put(22,0){\oval(4,6)[rb]}
\put(22,2.5){\oval(4,6)[rt]}
\put(24,0){\line(0,1){2.5}}
\put(11.4,-3.5){$\square$}
\put(11.4,5){$\square$}
\linethickness{1.5pt}
\put(2,-3){\line(1,0){9.5}}
\put(2,5.5){\line(1,0){9.5}}
\put(22,-3){\line(-1,0){9.5}}
\put(22,5.5){\line(-1,0){9.5}}
\multiput(2.4,-3)(2.0,0){5}{\makebox(0,0)[c]{\mbox{$\blacktriangleright$}}}
\multiput(2.4,5.5)(2.0,0){5}{\makebox(0,0)[c]{\mbox{$\blacktriangleright$}}}
\put(12.3,1.3){\makebox(0,0)[c]{\mbox{${H}$} }}
\put(12.3,-4.5){\makebox(0,0)[c]{\scriptsize \mbox{$\Lambda^{\!(n)}$} }}
\multiput(13.5,-3)(2.0,0){5}{\makebox(0,0)[c]{\mbox{$\blacktriangleleft$}}}
\multiput(13.5,5.5)(2.0,0){5}{\makebox(0,0)[c]{\mbox{$\blacktriangleleft$}}}
\put(15.0,-4.5){\makebox(0,0)[c]{\scriptsize \mbox{$\tilde{V}_R^{(n+1)}$} }}
\put(22.5,-4.5){\makebox(0,0)[c]{\scriptsize \mbox{$\tilde{V}_R^{(N-1)}$} }}
\put(9.9,-4.5){\makebox(0,0)[c]{\scriptsize \mbox{$\tilde{V}_L^{(n)}$} }}
\put(1.9,-4.5){\makebox(0,0)[c]{\scriptsize \mbox{$\tilde{V}_L^{(2)}$} }}
\put(5,-4.5){\makebox(0,0)[c]{\scriptsize \mbox{$\cdots$} }}
\put(19,-4.5){\makebox(0,0)[c]{\scriptsize \mbox{$\cdots$} }}
\end{picture}
}\; ,  \label{DMRG_globalH}\\
\nonumber
\end{align} 
which is nothing but the expectation value of the total Hamiltonian with the MPS wavefunction defined as 
\begin{align}
\Psi^{(n)}(s_1, \cdots s_n, &s_{n+1}, \cdots s_N)  = \nonumber \\
 &\tilde{\sf V}_{R}^{(N-1)} \cdots \tilde{\sf V}_R^{(n+1)}{\sf \Lambda}^{(n)} \tilde{\sf V}_{L}^{(n)\dagger} \cdots \tilde{\sf V}_L^{(2)\dagger} \, .
 \label{DMRG_MPS_PSI}
\end{align}
This demonstrates that the DMRG is interpreted as a variational method for the MPS wavefunction.
Nevertheless, it is difficult to directly deal with the total Hamiltonian $H$ and $\Psi$ with no compression of the Hilbert space.
Instead, the DMRG handles the eigenvalue problem of the renormalized super-block Hamiltonian, which is depicted as   
\begin{align}
\tilde{H}^{(n)}  = \; {\setlength\unitlength{1.8mm}
\begin{picture}(25,7)(0,-0.3)
\multiput(2.0,-3)(2.0,0){4}{\line(0,1){3}}
\multiput(15.8,-3)(2.0,0){4}{\line(0,1){3}}
\multiput(2.0,2.5)(2.0,0){4}{\line(0,1){3}}
\multiput(15.8,2.5)(2.0,0){4}{\line(0,1){3}}
\put(10.5,-1.5){\line(0,1){1.5}}
\put(13.3,-1.5){\line(0,1){1.5}}
\put(10.5,2.5){\line(0,1){1.5}}
\put(13.3,2.5){\line(0,1){1.5}}
\put(0,0){\line(1,0){24}}
\put(0,2.5){\line(1,0){24}}
\put(2,0){\oval(4,6)[lb]}
\put(2,2.5){\oval(4,6)[lt]}
\put(0,0){\line(0,1){2.5}}
\put(22,0){\oval(4,6)[rb]}
\put(22,2.5){\oval(4,6)[rt]}
\put(24,0){\line(0,1){2.5}}
\linethickness{1.5pt}
\put(2,-3){\line(1,0){7.5}}
\put(14.5,-3){\line(1,0){7.5}}
\put(2,5.5){\line(1,0){7.5}}
\put(14,5.5){\line(1,0){7.5}}
\multiput(2.4,-3)(2.0,0){4}{\makebox(0,0)[c]{\mbox{$\blacktriangleright$}}}
\multiput(2.4,5.5)(2.0,0){4}{\makebox(0,0)[c]{\mbox{$\blacktriangleright$}}}
\put(12.3,1.3){\makebox(0,0)[c]{\mbox{${H}$} }}
\multiput(15.5,-3)(2.0,0){4}{\makebox(0,0)[c]{\mbox{$\blacktriangleleft$}}}
\multiput(15.5,5.5)(2.0,0){4}{\makebox(0,0)[c]{\mbox{$\blacktriangleleft$}}}
\put(16.5,-4.5){\makebox(0,0)[c]{\scriptsize \mbox{$\tilde{V}_R^{(n+2)}$} }}
\put(23,-4.5){\makebox(0,0)[c]{\scriptsize \mbox{$\tilde{V}_R^{(N-1)}$} }}
\put(8.9,-4.5){\makebox(0,0)[c]{\scriptsize \mbox{$\tilde{V}_L^{(n-1)}$} }}
\put(1.9,-4.5){\makebox(0,0)[c]{\scriptsize \mbox{$\tilde{V}_L^{(2)}$} }}
\put(5,-4.5){\makebox(0,0)[c]{\scriptsize \mbox{$\cdots$} }}
\put(19.5,-4.5){\makebox(0,0)[c]{\scriptsize \mbox{$\cdots$} }}
\end{picture}
}\; . \label{DMRG_MPS_H} \\
\nonumber
\end{align}
and  the wavefunction of Eq. (\ref{finite_mix}).
Clearly, the diagram of  Eq. (\ref{finite_mix}) can be plugged into the ``hole'' of the above $\tilde{H}^{(n)}$, which gives $\langle \tilde{\Psi}^{(n)} |\tilde{H}^{(n)} | \tilde{\Psi}^{(n)} \rangle = \langle {\Psi}^{(n)} |{H} |{\Psi}^{(n)} \rangle $. 
The DMRG calculation gradually optimizes each tensor/matrix embedded in Eq. (\ref{DMRG_MPS_PSI}) with the $n$-dependent update scheme, and finally converges to the global fixed point satisfying the variational condition for Eq. (\ref{DMRG_globalH}),

 After convergence of the DMRG calculation, then,  Eq. (\ref{finite_mix}) with various $n$ gives the $n$ dependent expressions of the fixed-point wavefunction.
Comparing  $\Psi^{(n)}$ with $\Psi^{(n+1)}$, thus, we obtain the finite-size generalization  of Eq.({\ref{VOOV}),
\begin{align}
{\sf \Lambda}^{(n+1)} \tilde{\sf V}_L^{(n+1)\dagger}(s_{n+1})   =  \tilde{\sf V}_R^{(n+1)}(s_{n+1}) {\sf \Lambda}^{(n)}
\end{align}
with the ``boundary condition" ${\sf \Lambda}^{(1)} = {\sf \Lambda}^{(N-1)} = I$.
Moreover, this relation gives rise to the one-site shift operation of $\tilde{\sf \Psi}^{(n)}$, which is explicitly written as
\begin{align}
\tilde{\sf \Psi}^{(n+1)}(s_{n+1} s_{n+2}) = \sum_{s_n} \tilde{\sf V}^{(n+2)}_R(s_{n+2}) \tilde{\sf \Psi}^{(n)}(s_n s_{n+1}) \tilde{\sf V}_L^{(n)}(s_{n})\, .
\end{align}
This shifting operation for $\tilde{\sf \Psi}^{(n)}$ can be used for preparing a good initial wavefunction for the Lanczos diagonalization in the finite-size DMRG\cite{White1996}.

As already mentioned, the MPS representation of the wavefunction (\ref{DMRG_MPS_PSI}) in the finite-size DMRG is obtained  with successive use of SVD, which is also equivalent to the Schmidt decomposition of the wavefunction.
On the other hand,  Eq. (\ref{DMRG_MPS_PSI}) can be straightforwardly converted to the MPS of Eq. (\ref{variationalMPS}) based on the transfer-matrix formulation, through the relation of 
$ {\sf F}^{(n)} ={{\sf \Lambda}^{(n)}}^{-1/2} {\sf V}_R^{(n)} {{\sf \Lambda}^{(n-1)}}^{1/2} = {{\sf \Lambda}^{(n)}}^{1/2} {\sf V}_L^{(n-1)\dagger} {{\sf \Lambda}^{(n-1)}}^{-1/2}$ 
(Recall ${\sf \Omega} = {\sf \Lambda}^{1/2} $).
Thus, an important aspect of  the finite-size DMRG is that it explicitly bridges the above two aspects of quantum information  and transfer matrix formulation through the position-dependent MPS.
In addition, we point out that the expression of Eq.  (\ref{DMRG_MPS_H}) provides a prototype of the variational approach based on TPS or PEPS in a higher dimension.

From the practical point of view, long-range and/or nonuniform interactions up to moderate chain length can be handled by the finite-size DMRG within a realistic computational cost, thanks to the stable position-dependent update scheme.
Thus, the application range of the finite-size DMRG  was expanded to wide variety of quantum systems such as finite-size 2D quantum system\cite{Liang1994}, bosonic systems\cite{Jeckelmann1998}, dynamical quantities\cite{Jeckelmann2002},  momentum space\cite{Xiang1996}, random systems\cite{Hida1996}, quantum Hall systems\cite{Shibata2001}, quantum chemistry\cite{White1999},  numerical RG for the Kondo impurity model~\cite{Hofstetter2000}, and so on.  
We think that  these features of the finite-size DMRG stimulated us to further developments in the TN algorithm.

\subsection{matrix product operator}
\label{sec_MPO}

We explained the MPS on the basis of the variational state for the transfer matrix which is defined as a product of local vertex weights, and then extend it to the 1D quantum system through the Suzuki-Trotter decomposition.
In the context of the MPS,  the row-to-row transfer-matrix in the form of Eq. (\ref{eq_def_rtrtm}) is theoretically more tractable than the Hamiltonian which transfers the spin indices in the diagonal direction as depicted in Eq. (\ref{vertex_ST}).
 The matrix-product operator (MPO) was introduced as a systematical method bridging a gap between the Hamiltonian and the transfer matrix formalism without using the Suzuki-Trotter decomposition.\cite{McCulloch2007}

For the purpose of rewriting the sum of local operators, say $\hat{A}_n$, into a product of local operators, the basic idea of MPO is to use an auxiliary matrix having the nilpotent property.
As a simple example, let us consider
\begin{align}
M= \prod_n (1 + S^- \hat{A}_n)
\label{MPO_naive}
\end{align}
where $S^- \equiv S^x + i S^y$ with  $\vec{S}$ being the $S=1/2$ spin matrices.
Note that $S^-$ acts on the auxiliary Hilbert space distinct from the physical degrees of freedom in the system.
As well known,  $S^- $ is nilpotent, i.e. $(S^- )^2 = 0$.
Expanding $M$, then, we can easily obtain  $M = 1+ S^- (\sum_n \hat{A}_n)$, which contains the sum of $\hat{A}_n$ as $(2,1)$ element of the auxiliary matrix.

For the two-body interaction, we need to extend the size of the auxiliary space and introduce two kind of nilpotent matrices
\begin{align}
\tau= \begin{pmatrix}
0 &0 & 0 \\
1 &0 & 0 \\
0 &1 & 0 \\
 \end{pmatrix} \, ,
\quad 
\sigma= \begin{pmatrix}
0 &0 & 0 \\
0 &0 & 0 \\
1 &0 & 0 \\
 \end{pmatrix} 
\end{align}
which satisfy $\tau^2 = \sigma$,  $\tau^3 = 0$ and $\sigma^2=0$.
As in the case of Eq. (\ref{MPO_naive}), we expand $\prod_n (1 + \tau \hat{A}_n)$.
However, this generates two-body interactions $\hat{A}_n \hat{A}_{n'}$ for the all-to-all pairs with respect to $n$ and $n'$ in the chain.
In order to restrict the spatial range of the interactions, we further introduce a projection matrix 
\begin{align}
P= \begin{pmatrix}
1 &0 & 0 \\
0 &0 & 0 \\
0 &0 & 1 \\
 \end{pmatrix} \, ,
\end{align}
which follows $P \tau^2 P = P\sigma P =\sigma $, $P\tau P=0$,  $P^2=P$.
We then obtain 
\begin{align}
M &=\prod_n ( P + \sigma \hat{A}_n + \tau \hat{B}_n ) \nonumber \\
&= P + \sigma (\sum_n \hat{A}_n + \hat{B}_n \hat{B}_{n+1}), 
\label{MPO_tau}
\end{align}
where $\hat{A}_n$ and  $\hat{B}_n $ denote local operators for the physical degrees of freedom at site $n$. 
If $\hat{A}_n = \Gamma \hat{S}_n^x$ and $\hat{B}_n= J^{1/2} \hat{S}_n^z$,   the coefficient of $\sigma$ matrix in $M$ gives the Hamiltonian of the transverse field Ising model  with the coupling constants, $\Gamma$ and $J$.
By construction, a generalization to the Heisenberg interaction is straightforward.
In addition, we note that the above MPO formulation is generalized to 1D quantum systems with spatially extended interactions by replacing 
\begin{align}
P \to P_\lambda = 
\begin{pmatrix}
1 &0 & 0 \\
0 &\lambda & 0 \\
0 &0 & 1 \\
\end{pmatrix}
\end{align}
with $0 \le \lambda \le 1$.\cite{Pirvu2010}

In Eq. (\ref{MPO_tau}), $ \hat{A}_n $ and $ \hat{B}_n $ are local operators associated with the physical degrees of freedom at site $n$, while  $\tau$ and $\sigma$  act on the auxiliary  Hilbert space, which corresponds to the horizontal line of the row-to-row transfer matrix.
In other words,  $M$ has the same operator structure as the row-to-row transfer matrix for the classical vertex model.
This implies that the Hamiltonian can be generally converted into the row-to-row transfer matrix form defined on the square lattice of the usual direction(not 45$^\circ$-rotated direction).
Although the auxiliary degree of freedom gives rise to extra numerical cost, the MPO representation is useful for systematically dealing with 1D quantum many-body systems in the MPS formulation.

In this subsection, we have explained the MPO in the context of a way of converting the Hamiltonian to the transfer-matrix form.
Nevertheless, we should note that  recently, the term ``MPO" is used in a wider context in the TN literature; 
the row-to-row transfer matrix for the 2D classical model itself is often termed as ``MPO".
This situation may sound slightly confusing in the statistical mechanics side.
However, they are basically the equivalent object in the level of mathematical formulation. 

\subsection{time-dependent algorithms}

The MPS approach has also a significant relevance to numerical simulations of the real-time dynamics of quantum many-body systems.
To our knowledge, the first trial was carried out for a point contact problem of two free fermionic chains with the infinite-size DMRG.\cite{Cazalilla2002} 
In this approach, however, the renormalized Hamiltonian is not changed during the time evolution, so that the time range where numerical accuracy remains good is relatively short. 
More efficient MPS approaches for general setups were formulated as adaptive updating schemes for real-time evolution operators generated by Suzuki-Trotter decomposition with small time slices.\cite{WhiteFeiguin2004,Daley2004}

In the adaptive updating approaches, we assume the MPS form of time-dependent wavefunctions at  $t$ and $t+\Delta t$, which are explicitly related by the time evolution operator with a short time interval $\Delta t$, 
\begin{align}
|\psi(t+\Delta t)\rangle = \exp(-i\Delta t \hat{H})|\psi(t)\rangle \, .
\end{align}
As in Eq. (\ref{checker_board_decomp}), $ \exp(-i\Delta t \hat{H})$ can be decomposed into $ \exp(-i\Delta t \hat{H}) \sim \exp(-i\Delta t \hat{H}_{\rm e})\exp(-i\Delta t \hat{H}_{\rm o})$ with help of Trotter discretization. 
Given MPS representation of  $|\psi(t)\rangle$,  we can sequentially generate $\exp(-i\Delta t \hat{H}_{\rm e})|\psi(t)\rangle \to |\bar{\psi}(t)\rangle$ and $\exp(-i\Delta t \hat{H}_{\rm o})|\bar{\psi}(t)\rangle \to |\psi(t+\Delta t)\rangle $, where $|\bar{\psi}\rangle$ denotes the wavefunction at a mid time step due to the Trotter discretization. 
Then, an advantageous point is that $ \exp(-i\Delta t \hat{H}_{\rm e/o}) $ are represented as a product of local two-sites operators and $|\bar{\psi}(t)\rangle$ and $|\psi(t+\Delta t)\rangle $ can be straightforwardly reconstructed into the MPS form with use of SVD. 
Practically, the finite-size algorithm of DMRG or the TEBD algorithm can be directly applied to this problem.
Also, the Lindblad-type time evolution operator can be constructed as an MPS defined on the doubled Hilbert space of the density operator\cite{Ferstraete_MPDO2004,Zwolak2004}

From the variational point of view, the optimized tensors in the time-dependent algorithms do not satisfy the static variation problem, unlike the DMRG for the ground state.
Instead, the time-dependent algorithms can be viewed as a variational algorithm of minimizing an extended cost function,
\begin{align}
\delta \equiv |\psi(t+\Delta t) - \exp(-i\Delta t \hat{H})\psi(t) |
\label{eq_TDVP}
\end{align}
 at each time step. 
Note that the definition of the cost function is not unique.
For instance, another cost function, 
 \begin{align}
\delta  \equiv  & \; |\exp(i\Delta t \hat{H}_{\rm o})\psi(t+\Delta t) - \bar{\psi}(t) | \nonumber \\
 & + | \bar{\psi}(t) -  \exp(-i\Delta t \hat{H}_{\rm e}) \psi(t) |\, ,
\label{eq_leastaction}
\end{align}
originating from a least action principle also leads to a similar adoptive time evolution scheme based on the MPS.\cite{Ueda_condmat}
For the case of Eq. (\ref{eq_leastaction}), both of forward and backward iterations with respect to the time-evolution direction are required for the convergence of tensor optimization.
More recently, an improved formulation of the real-time evolution algorithm combined with the MPS was also proposed\cite{TDVP2011,TDVP2016} on the basis of the traditional time-dependent variational principle attributed to Dirac and Frenkel\cite{Dirac_1930,Frenkel1934}.
The algorithm based on the time-dependent variational principle enables us to certainly extend the time range where numerical accuracy is maintained, in comparison with the previous ones.

The MPS-based time-evolution algorithms provide practical tools for the real-time simulation of 1D quantum systems and have been extensively used for investigating real-time dynamics of ultra-cold atom systems as well as condensed matter problems. 
In this review, however, we will skip about applications and recent developments in the time-dependent algorithm, and just refer the reader to Ref. [\citen{Paeckel_TD2019}].

\section{Tensor product state and projected entangled pair state}
\label{sec_TPS}

The success of the DMRG and the CTMRG in 1D quantum and 2D classical systems prompts us to generalize the algorithms to higher dimensions.
A straightforward extension of the MPS-type variational state for the transfer matrix of the 3D classical model is a tensor product state (TPS)\cite{tensorproduct} consisting of local tensors which carry a set of leg indices reflecting the lattice structure [See Fig. \ref{l2ltm}(a)].
Then, the 3D version of the variational argument in \S\ref{Sec_2} naturally leads us to the corner transfer tensor\cite{BaxterBook}, which is illustrated in Fig. \ref{CTTRG}.

On the basis of the corner transfer tensor, the first challenge of the TN approach to the higher dimension was made for the cubic-lattice Ising model\cite{CTTRG}. 
As in Fig. \ref{CTTRG}(b), we can formulate the recursion relation for the corner transfer tensor, in cooperation with 6 pieces of supplemental tensors corresponding to the plane and edge structures of the lattice.
In practical numerical calculations, however, the spectra of the reduced density matrices with a certain cut of leg indices exhibit very slow decay even for a temperature moderately away from the critical point, in contrast to the rapid decay in the 2D case.
Whether such a bad situation originated from a problem of the optimization scheme or quality of the variational state was not clear at that time.
Thus, we should have primarily examined the quality of the TPS as a variational state with a direct variational calculation.

\begin{figure}
\begin{center} \includegraphics[width = 7.0 cm]{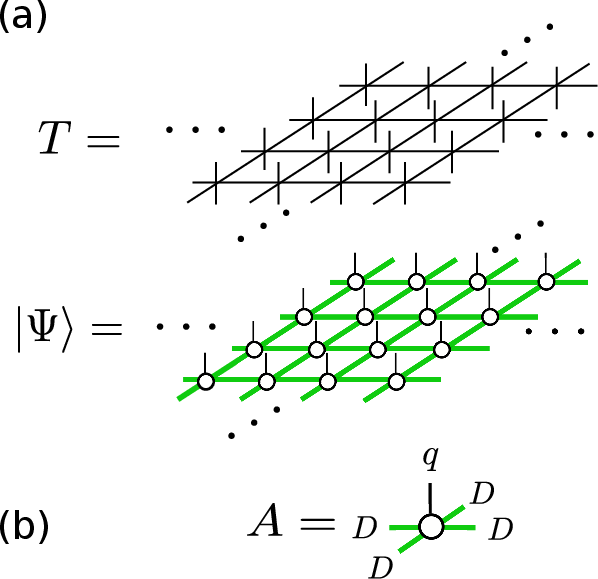} \end{center}\caption{
(Color online) The TPS/PEPS for a cubic lattice vertex model.
(a) The layer-to-layer transfer matrix $T$ and the TPS/PEPS representation of the maximum-eigenvalue eigenvector $|\Psi\rangle$ of $T$, where the circles denote vertices of local variational tensors.
The thin vertical lines indicate the physical degrees of freedom connected to $T$, and  the thick green lines represent auxiliary leg variables carrying the entanglement among lattice sites. 
(b) The local tensor $A$ is defined as a 5-leg tensor with one vertical line of the physical degree of freedom and four thick green lines of auxiliary degrees of freedom. 
The bond dimension of the physical spin is denoted as $q$, whereas that of the thick green lines is written as $D$. 
}   
\label{l2ltm}
\end{figure}

The basic idea of the direct variational approach to the transfer matrix is attributed to the Kramers-Wannier approximation\cite{KW}.
For the 3D Ising case,  the 2D Ising model in an effective magnetic field can be used as a trial state for the maximal eigenvector of the layer-to-layer transfer matrix, where the effective coupling and the effective magnetic field play the role of variational parameters.\cite{KW3D}
Then, an essential point is that the CTMRG can be used for evaluating the norm and expectation values with respect to the trial state that is composed of the effective 2D Ising model.
This mechanism can be viewed as a dimensional reduction by the variational TPS and allows us to directly optimize variational parameters.
Although the Kramers-Wannier approximation for the 3D Ising model contains only two variational parameters, its result is much better than the naive use of the corner-transfer-tensor RG, suggesting that the TPS has sufficient potential to describe  many-body effects in higher dimensions.

For the quantum system, a prototype example of the variational approach is the AKLT model on a honeycomb lattice\cite{AKLT2}, where the ground state is exactly described by the 2D VBS state.
Here, we note that by definition, the 2D VBS state satisfies the variational condition for the AKLT Hamiltonian.
In analogy with the 1D VBS state, the auxiliary spins in the 2D VBS can be regarded as leg indices attached to the physical $S=3/2$ spin, implying that the 2D VBS state is exactly described by a TPS with a finite bond dimension.
 Note that the TPS representation of the VBS state is equivalent to the PEPS, which was introduced in the context of quantum information.
Then, an essential point is that the expectation value of physical quantities is represented as a 2D double-layer vertex model constructed from the local tensors of the TPS/PEPS.[See also Fig. \ref{doublevertex}(a)].
This implies that the CTMRG or DMRG can be used for efficiently evaluating the norm and expectation values of physical quantities for the VBS state.\cite{Hieida1999}
These results prompted us to generalize the TPS/PEPS as a good variational state for 2D quantum lattice systems.

Here, we mention that the TPS/PEPS also satisfies the area law of EE for the 2D gapful ground state.
As discussed in \S \ref{sec_4_5}, the EE for the ground state of $d$D off-critical systems satisfies $S_{\rm EE} \sim \ell^{d-1}$ with $\ell$ being the linear dimension of a system part. 
For $|\Psi\rangle$ in Fig. \ref{l2ltm}(a), then, it is straightforward to see that given a certain system part and its complement,  the number of bonds cut by the interface of the two subsystems is proportional to the perimeter of the system part.
This also implies that the TPS/PEPS is capable of representing the gapful ground state of 2D quantum lattice systems, from the entanglement point of view.

\begin{figure}
\begin{center} \includegraphics[width = 6.0 cm]{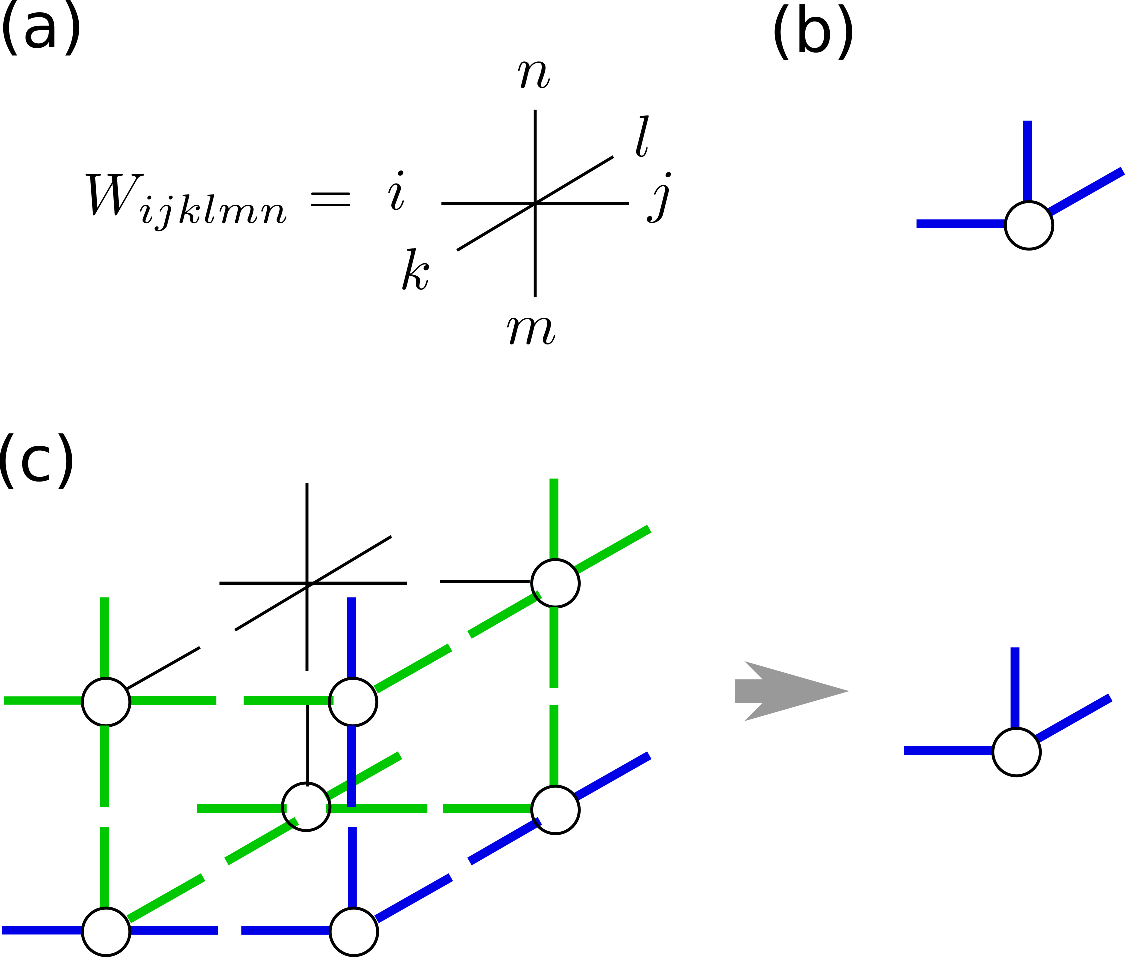} \end{center}\caption{ (Color online) Building block tensors of a cubic lattice system. 
(a) The local Boltzmann weight is represented as a 6-leg vertex tensor.
(b) A corner transfer tensor:  three thick blue legs represent the renormalized spin indices for the plane structure.
(c) The recursion relation can be built up for the corner transfer tensor in combination with the other types of vertex tensors.
The blue and green thick lines respectively represent the leg indices corresponding to the plane and edge structures.}   
\label{CTTRG}
\end{figure}

\begin{figure}
(a)\begin{center} \includegraphics[width =8 cm]{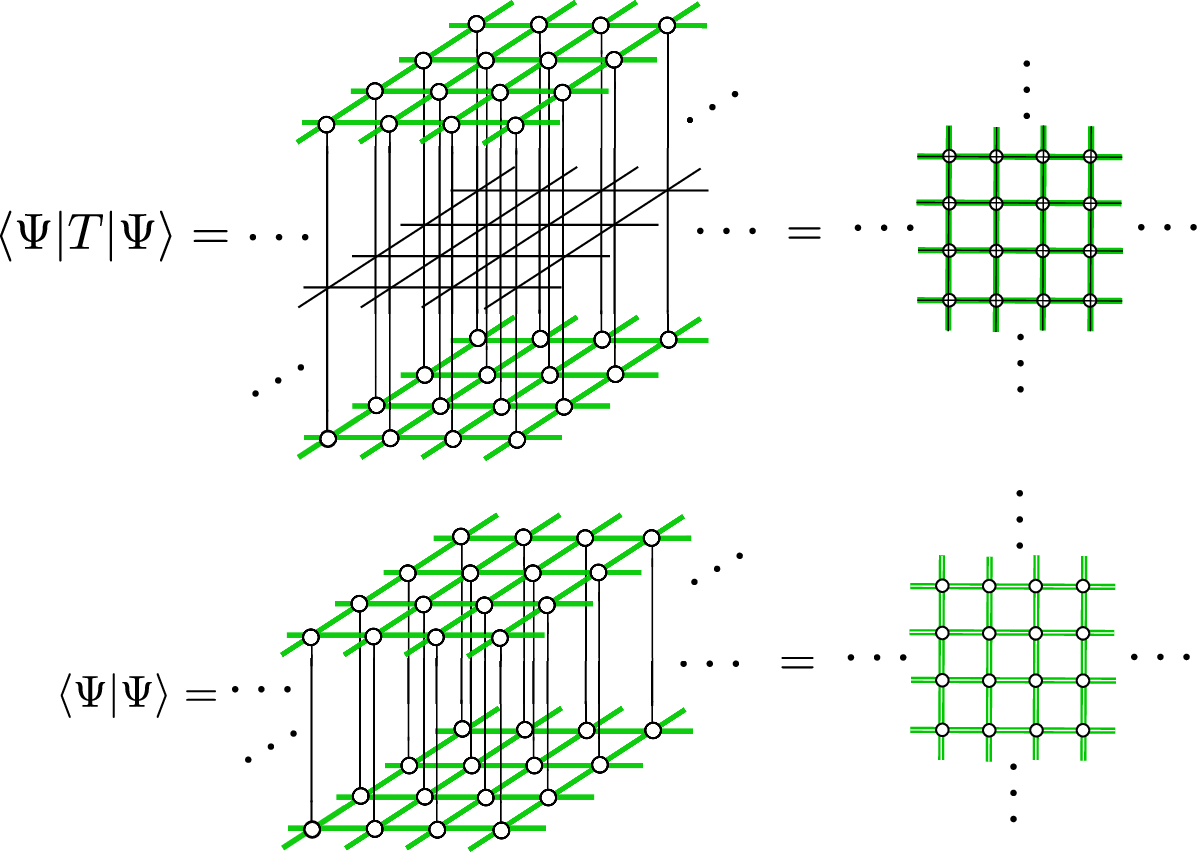} \end{center}
(b)\begin{center} \includegraphics[width = 7.0 cm]{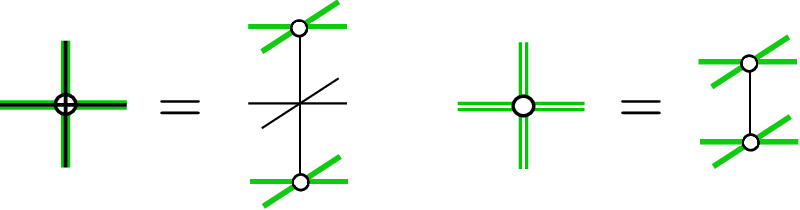} \end{center}
\caption{(Color online) 
Double layer vertex models for the TPS/PEPS.
(a)The double layer representation of $\langle \Psi |T| \Psi \rangle $  and $\langle \Psi| \Psi \rangle $  for the TPS.
The right panels show the top views of the 2D double-layer vertex model.
For the case of PEPS, the transfer matrix $T$ is usually replaced with the infinitesimal (imaginary) time evolution operator.
(b)The definition of the double layer vertices, for which the bond dimensions of the leg variables are  $qD^2$ (left) and $D^2$(right).
}   
\label{doublevertex}
\end{figure}

Let us proceed to details of the TPS-based variational approximation for the 3D vertex model, which is illustrated in Figs. \ref{l2ltm} and \ref{doublevertex}. 
For a layer-to-layer transfer matrix $T$, its largest eigenvalue eigenvector is written as the TPS form\cite{comment_TPS},
\begin{align}
|\Psi\rangle = \sum \prod_n A^{(n)}
\label{eq_TPS}
\end{align}
where $A^{(n)}$ denotes a local tensor depicted in Fig.\ref{l2ltm}(b) and the product with respect to $n$ runs over tensors arranged on the 2D lattice sites in Fig. \ref{l2ltm}(a).
Further, the summation in Eq. (\ref{eq_TPS}) is taken for the auxiliary degrees of freedom depicted as the thick green lines in Fig.\ref{l2ltm}(a).
However, naive use of SVD is not appropriate for the optimization of $A^{(n)}$, unlike the CTMRG/DMRG scheme where the RG transformation matrix due to the singular vectors can be connected to the MPS tensor through Eqs. (\ref{UV_MPS}) or (\ref{VOOV}).
These relations are no longer the case for the 2D TPS, and the reduced density matrix with a certain cut exhibits very slow decay of the spectrum.
Thus, we need a more brute force optimization of the  local tensor based on the direct variation of $\langle \Psi|T|\Psi \rangle/\langle \Psi|\Psi \rangle$ (or $\langle \Psi|\hat{H}|\Psi \rangle/\langle \Psi|\Psi \rangle$  for the 2D quantum case).
Then, an essential point is that $\langle \Psi |T| \Psi \rangle $ and $\langle \Psi | \Psi\rangle$ can be calculated through the CTMRG for the double-layer vertex models depicted in Fig. \ref{doublevertex}.  
Thus, we already have practical numerical tools for optimizing the local tensor so as to maximize the expectation value of $T$.
Here, we explain two typical approaches called ``variational update", and ``full update'' of TPS/PEPS.
We also mention the role of ``simple update'', which is a simple but less accurate updating scheme of local tensors based on the SVD.
In addition, we just notice that there is a hybrid approach of the imaginary time evolution and the RG transformation based on the reduced density matrix with a cut in the imaginary time direction (vertical density matrix ansatz)\cite{VDMA,Maeshima2004}.

\subsection{variational and full updates}

In this subsection, we basically consider a uniform system in the bulk limit, where we may omit the site index $n$ and thus write  $A^{(n)} \to A$.
A solid optimization scheme of the local tensor in the TPS is to use an environment tensor surrounding the tensor to be optimized, which is assumed to be located at the ``center" of the system.
In order to see structure of the environment tensor, we take the variation of the maximum eigenvalue 
\begin{align}
\Lambda \equiv \frac{\langle \Psi|T|\Psi \rangle}{\langle \Psi|\Psi \rangle} \,
\label{tps_variation}
\end{align}
with respect to the local tensor, $A \to A + \delta A$. 
Assuming the translation invariance in the bulk limit, we can write the optimal condition for $A$ in the form of the generalized eigenvalue problem, 
\begin{align}
X A = \Lambda Y A \, ,
\label{tps_gep}
\end{align}
where $X$ and $Y$ respectively represent effective environment tensors defined in Fig.\ref{env_double}.
Note that Eq. (\ref{tps_gep}) is a nonlinear equation for $A$, since the environment tensors $X$ and $Y$ themselves are built from $A$.

\begin{figure}
\begin{center} \includegraphics[width = 6.0 cm]{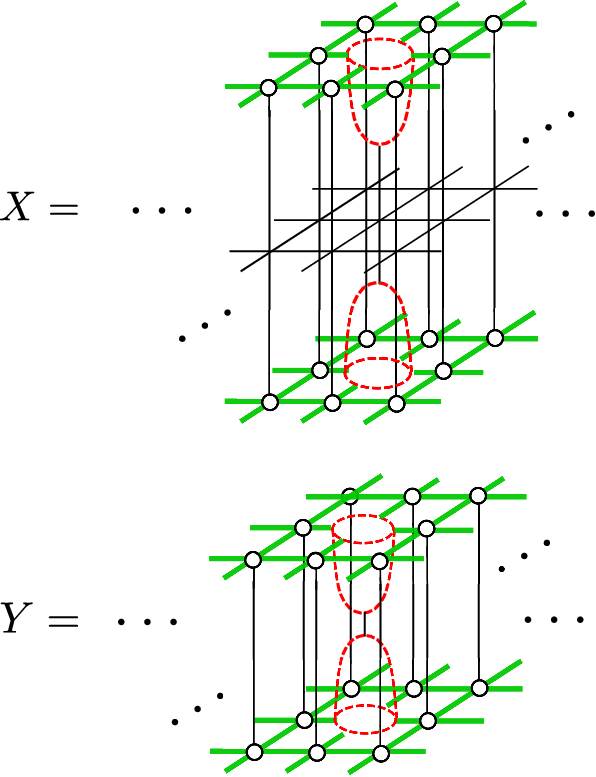} \end{center}
\caption{(Color online) 
The environment tensors  $X $  and $ Y$.
The red cages located at the center of the environments indicate ``holes'' originating from the variation with respect to the local tensor $A$.
These environment tensors can be efficiently calculated by inserting the impurity site after convergence of the CTMRG iterations for the double layer vertex models in Fig. \ref{doublevertex}.  
}   
\label{env_double}
\end{figure}

The basic role of the $X$ tensor is similar to the renormalized Hamiltonian of Eq. (\ref{DMRG_MPS_H} ) in the DMRG.
As mentioned, however, the singular vectors of SVD of the TPS cannot be reduced to the local tensor $A$ in the TPS, unlike the MPS in one dimension [See Eq. (\ref{UV_MPS})].
Thus, we formulate a more direct iterative optimization scheme of $A$ on the basis of  Eq. (\ref{tps_gep});
Given a local tensor of $ A$, we calculate  $X$ and $Y$ with the CTMRG for the effective double-layer vertex models.
In particular,  $X$ and $Y$ can be computed by insertion of the ``impurity" vertex corresponding to the red cages in Fig. \ref{env_double} after a sufficient number of CTMRG iterations for the double-layer vertex models.
Then, we establish the recursive relation,
\begin{align}
 \Lambda^{-1} Y^{-1} X A \to A'\, ,
\label{tps_update1}
\end{align}
where $A'$ provides an improved local tensor for the TPS with an appropriate normalization $\Lambda$. 
For an early stage of recursive iterations, however, this relation usually causes a too drastic change between $A$ and $A'$ to achieve stable optimization.
Thus, we need to introduce an adjustment mechanism of the optimization speed as follows,
\begin{align}
A_{\rm new}=A +\epsilon A' \sim  e^{\epsilon \tilde{K}}A\,,
\label{tps_tensor_update}
\end{align} 
where $\tilde{K}\equiv   \Lambda^{-1} Y^{-1} X$ and $\epsilon$ denotes a controlling parameter of convergence to the maximum-eigenvalue eigenstate.
A typical value is $\epsilon =0.01 \sim 0.1$ in practical situations.
Updating $A\to A_{\rm new}$ in Eq. (\ref{eq_TPS}), we now have the closed form of an iterative variational algorithm for the TPS.
Numerical results for the 3D Ising model demonstrated that the variational update provided systematical improvements of the Kramers-Wannier approximation.\cite{TPVA1,TPS1,TPS2}

If we replace $T$ with a Hamiltonian $\hat{H}$ together with $\epsilon \to -\epsilon$, Eqs.(\ref{tps_update1}) and (\ref{tps_tensor_update}) turn out to be the variational update algorithm for the infinite PEPS.\cite{Corboz2016} 
A particular point for quantum systems is that the Hamiltonian is the sum of two-body interaction terms rather than the product of local vertices. 
Thus, all of the two-body terms in the Hamiltonian must be carefully summed up during CTMRG calculations of the effective double-layer models in Fig. \ref{env_double}.
Here,  one might view  the quantum version of Eq. (\ref{tps_tensor_update}}) with $\tilde{H}= E_g^{-1} Y^{-1}X$ as an effective imaginary time evolution based on a renormalized Hamiltonian.
In contrast to the DMRG for  1D quantum systems, however, $\tilde{H}$ is transformed with the nonorthnormal basis and its Hermiticity is explicitly broken.
Thus, direct use of $\tilde{H}$ is not appropriate usually.

Another  update scheme of the local tensor is based on the imaginary-time evolution combined with the environment tensors, which is often called ``full update" in PEPS literature.
The basic idea is similar to the variational update for the 3D classical system above;
The imaginary time evolution operator $ \exp(-\epsilon \hat{H})$  with a small time slice $\epsilon$ plays a role of the layer-to-layer transfer matrix $T$ as in the case of TPS, since $ \exp(-\epsilon \hat{H})$ can be decomposed into a product of local weights with the sublattice structure through the Suzuki-Trotter decomposition. 
Given a state $|\Psi \rangle$ in the PEPS representation, an improved PEPS $|\Psi'\rangle $ for $\exp(-\epsilon \hat{H})|\Psi\rangle $ is constructed with minimizing the distance defined by 
\begin{align}
\delta  = \big| |\Psi'\rangle - e^{-\epsilon \hat{H}}|\Psi \rangle \big|\, .
\label{tdvariation_peps}
\end{align}
The outcome tensor $A'$ in $|\Psi'\rangle$ is constructed so as to minimize $\delta$ of Eq. (\ref{tdvariation_peps}) for the given input tensor $A$ in $|\Psi \rangle$ at each time step.
Thus, the full update is based on an imaginary-time dependent variation for the PEPS between two adjacent imaginary time slices, as in  Eq. (\ref{eq_TDVP}).  
Repeating updates along the imaginary time evolution, one can finally obtain a good approximation of the ground-state wavefunction corresponding to the $\beta \to \infty$ limit.

An interesting aspect of the optimal condition based on Eq. (\ref{tdvariation_peps}) is that the environment tensors to determine $A'$ are formally very similar to $X$ and $Y$ in Fig. \ref{env_double}.
In the variational update,  however, the tensor $A$ in $|\Psi \rangle$ are iteratively revised through Eq. (\ref{tps_update1}), which is a highly nonlinear equation of the local tensor.
In the full update, by contrast, the input tensor $A$ is already fixed at the previous time step, and thus roles of $A$ and $A'$  in the environment tensors are slightly different from the case of the variational update. (e.g. See Sec.II in Ref. [\citen{Murg2007}], where $A$ and $B$ respectively correspond to the present $A'$ and $A$ in Eq. (\ref{tdvariation_peps}))
This implies that the states in intermediate steps of iterations may be different between the above two algorithms, but the final wavefunctions after convergence in the $\beta\to \infty$ limit should become equivalent in principle.
We also note that the dominant computational cost is on the environment tensor for both update schemes.

Here, we should remark that  in the practical full-update algorithm of infinite-system size PEPS (iPEPS), a simplified version of the environment tensor is often used, \cite{iPEPS,Corboz2010} where the imaginary time evolution operator is applied to only the center sites of the environment block, on the assumption of the translation invariance.
Thus, the environment tensor used in the iPEPS  may not be exactly the same as that in the original full-update algorithm for the finite-size system.
The variational update for TPS/iPEPS achieves certainly accurate results compared with the iPEPS with the simplified environment tensor.\cite{Corboz2016}.

\subsubsection{simple update}

The computational cost of the variational update or the full update for 3D classical or 2D quantum systems is basically proportional to $(q D^2 m)^3$, which is attributed to computation of the environment tensor of $X$ in Fig. \ref{env_double}, where $m$ denotes the number of retained basis in the CTMRG for the double layer vertex model. 
This implies that the TPS/PEPS approaches become expensive  rapidly as the bond dimension $D$ increases.
The simple update is a simple but less accurate updating algorithm of the local tensor, which bypasses a direct computation of the environment tensor.
A background idea may be considered as a higher dimensional analogy of iTEBD for the MPS, where the effect of the environment is partially taken into account through singular values.\cite{Xiang_simple2008,Orus2009}

\begin{figure}
\begin{center} \includegraphics[width = 8.0 cm]{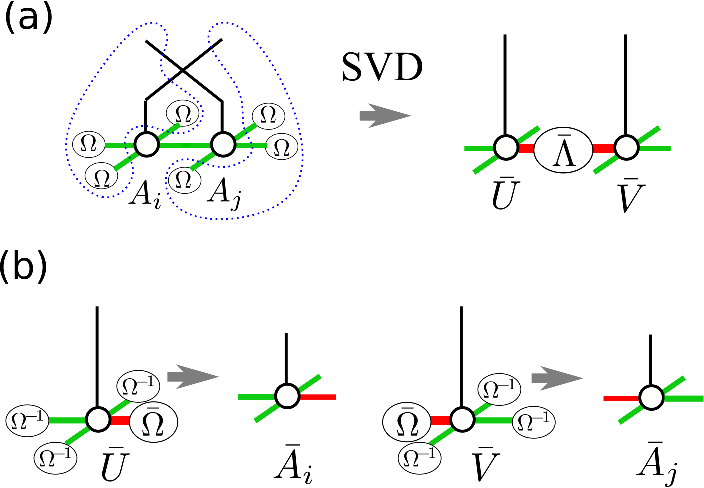} \end{center}
\caption{(Color online) 
A simple update of the local tensors $A_{i}$ and $A_j$ for two adjacent $i$-$j$ sites.
(a) Apply the local imaginary-time-evolution operator $e^{-\epsilon \hat{h}_{i,j}}$ to the local tensors $A_{i,j}$ attached with $\Omega$,  as in Eq. (\ref{offo_even}) for the iTEBD.
Regarding the dangling legs surrounded by the dotted lines as matrix indices, we perform SVD, where $\bar{\Lambda}$ indicate the singular values and $\bar{U}$ and $\bar{V}$ with the green and thick red lines represent the singular vectors.
(b) On the basis of the low-rank approximation of $\bar{\Lambda}$, the improved tensors $\bar{A}_{i,j}$ with the red leg renormalized is constructed from the singular vectors attached with $\bar{\Omega}\equiv \sqrt{\bar{\Lambda}}$. 
We can repeat a similar process for the other remaining green legs.
}   
\label{simple}
\end{figure}

Let us consider the imaginary-time evolution problem for a 2D quantum system.
As depicted in Fig. \ref{simple}(a), we consider two local tensors $A_{i}$ and $A_{j}$ attached with $\Omega$, where $i,j$ indicate two adjacent sites.
Here, this $\Omega$ corresponds to the square root of the singular values and is expected to include some environment effect like Eq. (\ref{offo_even}) of the iTBED.
We then apply the local imaginary-time-evolution operator $e^{-\epsilon \hat{h}_{i,j}}$  to the physical legs of $A_{i,j}$, where  $\hat{h}_{i,j}$ is the local Hamiltonian.
Regarding the dangling legs surrounded by the dotted lines as matrix indices,  we perform SVD to obtain the singular values $\bar{\Lambda}$ and the corresponding singular vectors $\bar{U}$ and $\bar{V}$.
As in Fig. \ref{simple}(b). we then renormalize $\bar{\Omega}\equiv \sqrt{\bar{\Lambda}}$ into the red leg indices and construct new local tensors $\bar{A}_{i,j}$, rearranging the matrix indices of $\bar{U}$ and $\bar{V}$ with $\Omega^{-1}$.
Repeating these processes for all legs of the local tensors, we can finally obtain those effectively absorbing the single-step evolution of $e^{\epsilon \hat{H}}$ in the imaginary time direction.

In the simple update, direct computations of the environment tensors are skipped, so that its numerical cost is very cheap. 
However, the renormalization due to $\bar{\Omega}$ attached to the red legs may have missed some part of environment effects.
In contrast to the MPS for the 1D quantum system, the singular vectors in the simple update for the 2D quantum system no longer generate the translation operation on the 2D lattice, since they do not lead to such relations as Eqs. (\ref{UV_MPS}) and (\ref{VOOV}).
This implies that the result of the simple update does not meet the optimal PEPS for the ground state  and has a certain deviation from the result of the variational or full update, even for the gapful ground state with a short correlation length.
In practical situations, thus, the simple update should be used for preparing initial local tensors for the full update or variational update.
From the theoretical viewpoint, thus, how canonical forms of TPS/PEPS can be recovered from transformation matrices due to SVD is an essential problem, which was actually discussed in Refs. [\citen{Zaletel2020,Haghshenas2019}].
Also, this point may be associated with how to construct the correct transformation matrix in the vertical density matrix approach \cite{VDMA,Maeshima2004}.

Finally, we would like to comment on recent developments in optimization schemes such as gradient method \cite{Gradient2016} and automatic differentiation approach\cite{AutoDiff2019}.
These approaches enable us to achieve accuracy comparable to the variational update, with skipping direct computations of the environment tensors that is the most costly part of numerical computations.
Another interesting development is calculation of low-energy excitations based on the PEPS.\cite{Vanderstraeten2019,Ponsioen2020}

\section{Real-space renormalization groups and tensor networks}

As discussed so far, the CTMRG/DMRG type algorithms provide iterative numerical methods of solution for the variational equation, where the eigenvectors of reduced density matrices play the role of projective transformation with keeping the important information in the bulk states.
Also, the linear system size of the renormalized tensors/matrices basically increases in proportion to the number of iterations.
Thus, they basically seem to share certain characteristics with real-space RGs.
In Kadanoff-Wilson type real-space RG, however, the length scale or the energy scale of the system are changed in the exponential manner with respect to RG steps,  and then a response of the system against such an active control of the length scale combined with the coarse graining allows us to extract information of the critical behavior. 
In the CTMRG/DMRG,  the system-size increase is just linear and the spectrum of the reduced density matrix is not directly related to the low-energy spectrum of the system Hamiltonian.
In particular, the critical behavior described by the fixed point of the CTMRG/DMRG with a finite $m$ is basically of the mean-field type\cite{Tsang1979,NOK1996}.
Thus, in principle, the framework of CTMRG/DMRG is insufficient to extract proper information about the RG flow.
Instead,  we often combine the finite-size or finite-entanglement scaling analysis to investigate critical behaviors, as discussed in \S \ref{sec_4_5}.

The TN approach incorporating the scale transformation was initiated with the tensor renormalization group (TRG) for the Ising model in 2007.\cite{TRG2007}
An advanced point of the TRG from the conventional real-space RGs is on massive use of SVD for vertex-type Boltzmann weights instead of the Hamiltonian.
The approximation based on the dominant singular values and corresponding singular vectors practically works very well.
Moreover,  the TRG  involves the scale transformation equivalent to the conventional real-space RG.
This  suggests that in principle the scaling dimension can be extracted within the framework of the TRG formulation.\cite{Lyu2021}
Although the TRG turns out to belong to a class of tree tensor networks (TTNs) and the resulting accuracy is of the same order as CTMRG for the 2D case, it stimulates the following development of TNs associated with the real-space RG. 
For instance,  the HOTRG, which is assisted by the higher-order SVD\cite{HOSVD1,HOSVD2}, makes it easier to a systematical extension of the TN to higher dimensional systems,\cite{HOTRG2012} and is appreciated for recent TN studies of lattice gauge theories\cite{Carmen_Banuls_LG2020}.

In the context of the critical phenomena, moreover, a significant development was achieved by the tensor network renormalization group (TNR)\cite{TNR2015} or  the multi-scale entanglement renormalization ansatz (MERA) for 1D quantum systems\cite{MERA2007,MERA2008}, both of which succeeded in extracting the numerically exact scaling dimensions beyond the mean-field level.
In particular, the concept of disentangler implemented in these TNs enables us to  filter out short-range entanglements irrelevant to critical behaviors, where the total network structures become capable of representing the log-correction to the area law of EE. 
In this sense,  the concept of the real-space RG presented in the 70's-80's was finally justified by the recent development of the TN in the quantitative level.
Also, the MERA clarified the interesting connection between the TN  and the Ryu-Takayanagi formula in the AdS/CFT\cite{RT_PRL2006,Swingle2012}, which stimulates us to further studies around the nexus of TN physics and quantum gravity.
In the followings, we overview a series of developments in the TN from the TRG to the TNR, focusing on their fixed point structures rather than technical details. 
After that, we will discuss the MERA and related issues.
This is because the TNR is easier to see the connection to the real-space RG, although the MERA was presented earlier than the TNR.

\subsection{tensor renormalization group}

Let us begin with the local Boltzmann weight of the 2D isotropic Ising model in the vertex representation,
\begin{align}
W(s_1\, s_2\, s_3\, s_4)
=
\hspace{3mm}{\setlength\unitlength{2mm}
\begin{picture}(5,3)(0,-0.3)
\put(2,-2){\line(0,1){4}}
\put(0,0){\line(1,0){4}}
\put(-1,0){\makebox(0,0)[c]{\scriptsize \mbox{$s_1$}}}
\put(5,0){\makebox(0,0)[c]{\scriptsize \mbox{$s_{3}$}}}
\put(2,-3){\makebox(0,0)[c]{\scriptsize \mbox{$s_2$}}}
\put(2,3){\makebox(0,0)[c]{\scriptsize \mbox{$s_{4}$}}}
\end{picture}
}  \label{trg_vertex} \quad .\\
\nonumber
\end{align}
In the previous MPS-based formulation,  we have always considered the row-to-row transfer matrix and its maximal eigenvector.
In the TRG, meanwhile, the coarse graining and scale transformations are directly realized with SVD for vertex weights and their contraction.
More precisely, we regard the vertex weight as a matrix that transfers the spins, for example,  in the NE to SW direction and then perform SVD,
\begin{align}
W(s_1\, s_2\, s_3\, s_4)
&= \sum_{\sigma=1}^{4} V(\sigma, s_{1} s_{2}) \Sigma_\sigma V(\sigma, s_{3} s_{4} ) \nonumber \\ 
&=\hspace{3mm}{\setlength\unitlength{2mm}
\begin{picture}(5,5)(0,-0.3)
\put(2,-3){\line(0,1){2}}
\put(0,-1){\line(1,0){2}}
\put(4,1){\line(0,1){2}}
\put(4,1){\line(1,0){2}}
\put(3.25,0.9){\rotatebox{135}{$\blacktriangle$}}
\put(1.25,-0.95){\rotatebox{-45}{$\blacktriangle$}}
\multiput(2.0,-1)(0.25,0.25){9}{\circle*{0.1}}
\put(-1,-1){\makebox(0,0)[c]{\scriptsize \mbox{$s_1$}}}
\put(7,1){\makebox(0,0)[c]{\scriptsize \mbox{$s_{3}$}}}
\put(2,-4){\makebox(0,0)[c]{\scriptsize \mbox{$s_2$}}}
\put(4,4){\makebox(0,0)[c]{\scriptsize \mbox{$s_{4}$}}}
\put(2.5,0.5){\makebox(0,0)[c]{\scriptsize \mbox{$\Sigma$}}}
\put(5.2,2.2){\makebox(0,0)[c]{\scriptsize \mbox{$V$}}}
\put(0.5,-2.5){\makebox(0,0)[c]{\scriptsize \mbox{$V$}}}
\end{picture}
}  \qquad ,  \label{trg_svd}
\\
\nonumber
\end{align} 
where $\Sigma$ and $V$ respectively denote singular values and the corresponding singular vectors.
Note that $\sigma$ can be viewed as a coarse-grained spin variable originating from the two $s$ spins and the solid triangle in Eq. (\ref{trg_svd}) indicates the direction of the coarse graining.
Using $\Sigma$ and $V$, we construct a 3-leg vertex tensor 
\begin{align}
\tilde{V}(\sigma,s_1,s_2) = \sqrt{ \Sigma_\sigma } V(\sigma, s_{1} s_{2}) \, .
\end{align}
Clearly, we can do the same computation in the SE to NW direction of the vertex weight.
We then connect four $\tilde{V}$ tensors and contract them with respect to $s_1, s_2, s_3, s_4$,
\begin{align}
&W'(\sigma_1\, \sigma_2\, \sigma_3\, \sigma_4) \nonumber \\
&= \sum_{s_1,s_2, s_3, s_4} \tilde{V}(\sigma_1, s_{1} s_{2}) \tilde{V}(\sigma_2, s_{2} s_{3} ) \tilde{V}(\sigma_3, s_{3} s_{4} )\tilde{V}(\sigma_4, s_{4} s_{1} )\nonumber \\ 
&=\hspace{3mm}{\setlength\unitlength{2mm}
\begin{picture}(6,5)(0,-0.3)
\put(2,2){\line(1,0){3}}
\put(2,-1){\line(1,0){3}}
\put(2,-1){\line(0,1){3}}
\put(5,-1){\line(0,1){3}}
\put(1.5,-1.0){\rotatebox{135}{\scriptsize $\blacktriangle$}}
\put(4.5,2){\rotatebox{-45}{\scriptsize $\blacktriangle$}}
\put(1.5,1.5){\rotatebox{45}{\scriptsize $\blacktriangle$}}
\put(4.5,-0.5){\rotatebox{-135}{\scriptsize $\blacktriangle$}}
\multiput(0.5,-2.5)(0.25,0.25){7}{\circle*{0.1}}
\multiput(5,2)(0.25,0.25){7}{\circle*{0.1}}
\multiput(5,-1)(0.25,-0.25){7}{\circle*{0.1}}
\multiput(2,2)(-0.25,0.25){7}{\circle*{0.1}}
\put(1.5,0.5){\makebox(0,0)[c]{\scriptsize \mbox{$s_1$}}}
\put(3.5,-1.6){\makebox(0,0)[c]{\scriptsize \mbox{$s_{2}$}}}
\put(5.6,0.5){\makebox(0,0)[c]{\scriptsize \mbox{$s_3$}}}
\put(3.5,2.5){\makebox(0,0)[c]{\scriptsize \mbox{$s_{4}$}}}
\put(0.3,-1.7){\makebox(0,0)[c]{\scriptsize \mbox{$\sigma_1$}}}
\put(6.7,-1.7){\makebox(0,0)[c]{\scriptsize \mbox{$\sigma_2$}}}
\put(0.3,2.7){\makebox(0,0)[c]{\scriptsize \mbox{$\sigma_4$}}}
\put(6.7,2.7){\makebox(0,0)[c]{\scriptsize \mbox{$\sigma_3$}}}
\end{picture}
}  \qquad  \Rightarrow 
\hspace{3mm}{\setlength\unitlength{2mm}
\begin{picture}(6,5)(0,-0.3)
\multiput(1.5,-1.5)(0.25,0.25){7}{\circle*{0.1}}
\multiput(5.1,-1.6)(-0.25,0.25){7}{\circle*{0.1}}
\multiput(5.1,2.1)(-0.25,-0.25){7}{\circle*{0.1}}
\multiput(1.5,2)(0.25,-0.25){7}{\circle*{0.1}}
\put(2.9,-0.1){\scriptsize $\square$}
\put(0.5,-2){\makebox(0,0)[c]{\scriptsize \mbox{$\sigma_1$}}}
\put(6,-2){\makebox(0,0)[c]{\scriptsize \mbox{$\sigma_2$}}}
\put(0.5,2.5){\makebox(0,0)[c]{\scriptsize \mbox{$\sigma_4$}}}
\put(6,2.5){\makebox(0,0)[c]{\scriptsize \mbox{$\sigma_3$}}}
\end{picture}
}
\qquad ,\label{trg_contraction}\\
\nonumber
\end{align} 
which gives a new  4-leg renormalized vertex  $W'$.
Then, this new vertex $W'$ defines a 45$^\circ$ rotated square lattice with the lattice space scaled by $\sqrt{2}$.
Instead, a number of the effective spin degrees of freedom in the renormalized vertex, which is represented as dotted lines,  are increased twice in the contraction step of Eq. (\ref{trg_contraction}). 
As expected,  the small singular values and the corresponding singular vectors should be truncated to maintain the bond dimension of the renormalized legs constant.
In the following, let us write this bond dimension as $\chi$.\cite{cutoffm}
For $W'$,  we can further do the process from Eqs. (\ref{trg_svd}) to (\ref{trg_contraction}) again, which generate a square lattice vertex weight, say $\tilde{W}(\tilde{s}_1 \tilde{s}_2 \tilde{s}_3 \tilde{s}_4)$ in the original lattice orientation, where $\tilde{s}_i$ with $i=1, \cdots, 4$ indicate  renormalized spin indices of $\tilde{W}$.

\begin{figure}[bth]
\begin{center} \includegraphics[width = 8.0 cm]{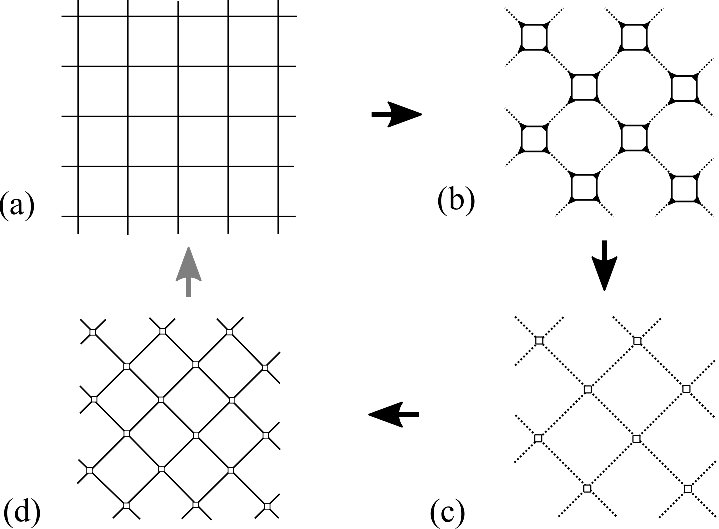} \end{center}
\caption{Iteration process of the TRG. 
(a) A vertex model of a certain length scale
(b) Perform SVD for the vertices in the NE-SW or SE-NW directions. The dotted lines indicate the renormalized spin indices to which $\sqrt{\Sigma}$ is attached.
(c) Contracting solid square lines in (b), we have a $45^{\circ}$ rotated vertex model.
(d) Rescaling  the vertex model in the panel (c) by the scale factor $\sqrt{2}$, we have the effective vertex model in the original scale. Repeating the same process of (a) - (d) for the $45^{\circ}$-rotated vertex model again, we arrive at the effective vertex model scaled by the scale factor $2$.}
\label{trg_iteration}
\end{figure}

Now, we have the closed loop of TRG iteration, which is summarized as Fig. \ref{trg_iteration}.
For a square-lattice vertex model of Fig. \ref{trg_iteration}(a) , we decompose the local vertex tensor, using Eq. (\ref{trg_svd}).
Then, the system is converted into an intermediate lattice model of  Fig. \ref{trg_iteration}(b).
Contracting out the solid lines in Eq. (\ref{trg_contraction}), we obtain a coarse-grained square-lattice vertex model in  Fig. \ref{trg_iteration}(c), where the renormalized vertices have 4 dotted lines of the renormalized spins. 
Rescaling the lattice space in Fig. \ref{trg_iteration}(c) by $\sqrt{2}$, we then have the square-lattice vertex model, as illustrated in Fig. \ref{trg_iteration}(d).
Although the lattice in Fig. \ref{trg_iteration}(d) is $45^\circ$ rotated,  we can do the TRG iteration for Fig. \ref{trg_iteration}(d) again and then arrive at the renormalized square-lattice vertex model of $\tilde{W}(\tilde{s}_1 \tilde{s}_2 \tilde{s}_3 \tilde{s}_4)$ in the same orientation as  Fig. \ref{trg_iteration}(a) with the scale factor $2$ of the lattice space.
Replacing $\tilde{W}(\tilde{s}_1 \tilde{s}_2 \tilde{s}_3 \tilde{s}_4) \to W(s_1 s_2 s_3 s_4)$, we have the recursive relation of the vertex weight in the TRG. 
After $N$ iterations of the above recursion process, we obtain the effective vertex weight containing $2^N \times 2^N$ number of the original vertices, and finally arrived at the fixed point vertex representing the bulk limit.
Then, the partition function of the system with the periodic boundary conditions is calculated as 
\begin{align}
Z= \sum_{s,s'} W(s s' s s')\, .
\end{align}
The accuracy of the TRG algorithm is good,  except for the vicinity of the critical point\cite{TRG2007}.

\begin{figure}[t]
\begin{center} \includegraphics[width = 8.0 cm]{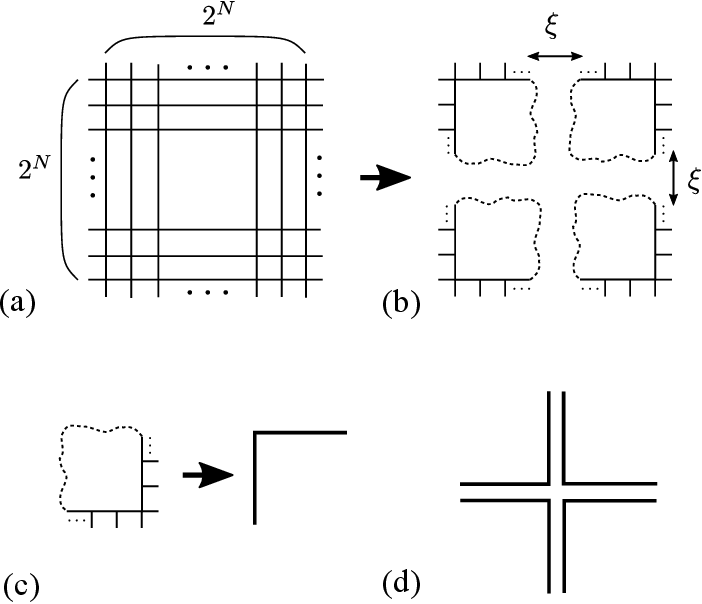} \end{center}
\caption{The fixed point tensor of the TRG.
(a) The effective vertex weight after $N$ iterations of the TRG contains $2^N \times 2^N$ number of the original vertex weights.
(b) If the system is off critical and $2^N \gg \xi$, the entanglement between the two opposite edges of the effective vertex are totally decoupled from each other.
(c) A quadrant in the panel (b) corresponds to the situation of CTM with free boundaries.
(d) Diagrammatic representation of the fixed-point vertex weight, which is called corner double-line tensor(CDL).
Note that the square symbol is not assigned for the CTMs in the CDL tensor.
}
\label{trg_cdl}
\end{figure}

For analyzing accuracy of the TRG algorithm, it is essential to clarify the fixed point structure of the renormalized vertex.
Here, we assume that there is no topological entanglement in the system.
The cycle of the TRG iteration in Fig. \ref{trg_iteration} provides an RG transformation with scale factor 2, implying that the effective vertex after $N$ iterations contains $2^N \times 2^N$ number of bare vertices.
As shown in Fig. \ref{trg_cdl}(a), then,  each leg spin index of the renormalized vertex represents a bunch of $2^N$ legs of the bare vertices.
If the system is away from the critical point and the correlation length $\xi$ is finite,  the TRG iteration basically  converges to the fixed point for $2^N \gg \xi$.
Then, an essential feature of the fixed point vertex, say $W^*(s_1 s_2 s_3 s_4)$,  is that the leg spin indices of the opposite sides of $W^*$, i.e. $s_1$ and $s_3$ or $s_2$ and $s_4$,  are spatially separated away beyond $\xi$, where the entanglement between them is negligible.
The remaining entanglement involved in $W^*$ is attributed to the correlations between the two edges around each corner of the lattice, as depicted in Fig. \ref{trg_cdl}(b).
For the tensor representing each quadrant, the inside boundaries (wavy dashed curves) are sufficiently away from the outer corner.
Thus the situation of the corner tensor in Fig. \ref{trg_cdl}(c) is basically the same as that of the CTM with free boundaries.
Gluing four decoupled CTMs, then, we may represent the fixed-point tensor as 
\begin{align}
W^*(s_1 s_2 s_3 s_4) &= C(\mu_1|\nu_1) C(\nu_2 |\mu_2) C(\mu_3|\nu_3) C(\nu_4|\mu_4) 
 \nonumber \\
&=
\hspace{3mm}{\setlength\unitlength{2mm}
\begin{picture}(5,6)(0,-0.3)
\linethickness{1.0pt}
\put(3,0){\line(0,-1){3}}
\put(3.5,0){\line(0,-1){3}}
\put(3,0){\line(-1,0){3}}
\put(3.5,0){\line(1,0){3}}
\put(3,0.5){\line(-1,0){3}}
\put(3.5,0.5){\line(1,0){3}}
\put(3,0.5){\line(0,1){3}}
\put(3.5,0.5){\line(0,1){3}}
\put(-1,0){\makebox(0,0)[c]{ \mbox{$s_1$}}}
\put(1,1.3){\makebox(0,0)[c]{\scriptsize \mbox{$\mu_4$}}}
\put(1,-1){\makebox(0,0)[c]{\scriptsize \mbox{$\mu_1$}}}
\put(7.5,0){\makebox(0,0)[c]{ \mbox{$s_{3}$}}}
\put(6,1.3){\makebox(0,0)[c]{\scriptsize \mbox{$\mu_3$}}}
\put(6,-1){\makebox(0,0)[c]{\scriptsize \mbox{$\mu_2$}}}
\put(3.3,-4){\makebox(0,0)[c]{ \mbox{$s_2$}}}
\put(2.2,-2.5){\makebox(0,0)[c]{\scriptsize \mbox{$\nu_1$}}}
\put(4.5,-2.5){\makebox(0,0)[c]{\scriptsize \mbox{$\nu_2$}}}
\put(3.3,4.1){\makebox(0,0)[c]{ \mbox{$s_{4}$}}}
\put(2.2,3){\makebox(0,0)[c]{\scriptsize \mbox{$\nu_4$}}}
\put(4.5,3){\makebox(0,0)[c]{\scriptsize \mbox{$\nu_3$}}}
\end{picture}
}  \hspace{10mm} ,\label{trg_cdl_vertex} \\
\nonumber
\end{align}
where the leg spin variables $\{s \}$ in $W^*$ consist of a bunch of the double line indices of $\{\mu\}$ or $\{\nu\}$. 
For instance, $s_1 \equiv  \mu_1 \otimes \mu_4$, $s_2 = \nu_1 \otimes \nu_2$, etc., where if the dimensions of $\{s\}$ is $\chi$, those for $\mu$ and $\nu$ should be $\sqrt{\chi}$. 
Thus, the fixed point vertex tensor $W^*$ is often called corner double line (CDL) tensor.\cite{TRG2007,Gu_TRG2008}
Here, we note that the square symbols are not assigned for the CTMs in the CDL tensor for simplicity.
Also, we  assume that $C$ is a real symmetric matrix below.

An important nature of the CDL tensor is that it automatically satisfies the fixed point condition of the TRG recursion relation.
In order to see this, we consider SVD of the CDL tensor, which can be described in terms of SVD for CTMs in Eq. ({\ref{trg_cdl_vertex}).
Let us write the SVD of the CTM as
\begin{align}
C = U \Omega U^\dagger \, 
\label{trg_ctm_ev}
\end{align}
where the number of singular values in $\Omega$ is assumed to be $\sqrt{\chi}$ and thus matrix size of $U$ is also $\sqrt{\chi} \times \sqrt{\chi}$.
Here, recall that $C$ is assumed to be real symmetric.
 Then, the SVD of $W^*$ tensor [Eq. (\ref{trg_svd})] can be written as 
\begin{align}
W^* = V^* \Sigma^* {V^*}^\dagger 
\end{align}
with $\Sigma^* \equiv \Omega \otimes \Omega$ and $ V^*\equiv U \otimes U$, which can be diagrammatically represented as
\begin{align}
{\setlength\unitlength{2mm}\begin{picture}(8,5)(0,-0.3)
\put(2,-3){\line(0,1){2}}
\put(0,-1){\line(1,0){2}}
\put(4,1){\line(0,1){2}}
\put(4,1){\line(1,0){2}}
\put(3.25,0.9){\rotatebox{135}{$\blacktriangle$}}
\put(1.25,-0.95){\rotatebox{-45}{$\blacktriangle$}}
\multiput(2.0,-1)(0.25,0.25){9}{\circle*{0.1}}
\put(-1,-1){\makebox(0,0)[c]{\scriptsize \mbox{$s_1$}}}
\put(7,1){\makebox(0,0)[c]{\scriptsize \mbox{$s_{3}$}}}
\put(2,-4){\makebox(0,0)[c]{\scriptsize \mbox{$s_2$}}}
\put(4,4){\makebox(0,0)[c]{\scriptsize \mbox{$s_{4}$}}}
\put(2.5,0.5){\makebox(0,0)[c]{\scriptsize \mbox{$\Sigma^*$}}}
\end{picture}
} \Rightarrow \quad
{\setlength\unitlength{2mm}\begin{picture}(8,5)(0,-0.3)
\linethickness{1.0pt}
\put(2.1,-3.3){\line(0,1){2.2}}
\put(-0.3,-0.9){\line(1,0){2.2}}
\put(1.7,-3.3){\line(0,1){2}}
\put(-0.3,-1.3){\line(1,0){2}}
\put(3.9,1.2){\line(0,1){2.1}}
\put(4.1,0.9){\line(1,0){2.2}}
\put(4.3,1.3){\line(0,1){2}}
\put(4.3,1.3){\line(1,0){2}}
\multiput(2.1,-1.1)(0.25,0.25){9}{\circle*{0.1}}
\multiput(1.9,-0.9)(0.25,0.25){9}{\circle*{0.1}}
\put(-1,-1){\makebox(0,0)[c]{\scriptsize \mbox{$s_1$}}}
\put(7,1){\makebox(0,0)[c]{\scriptsize \mbox{$s_{3}$}}}
\put(2,-4){\makebox(0,0)[c]{\scriptsize \mbox{$s_2$}}}
\put(4,4){\makebox(0,0)[c]{\scriptsize \mbox{$s_{4}$}}}
\put(2.4,0.6){\makebox(0,0)[c]{\scriptsize \mbox{$\Omega$}}}
\put(3.5,-0.5){\makebox(0,0)[c]{\scriptsize \mbox{$\Omega$}}}
\put(3.3,2.3){\makebox(0,0)[c]{\scriptsize \mbox{$U$}}}
\put(5.5,-0.0){\makebox(0,0)[c]{\scriptsize \mbox{$U$}}}
\put(2.8,-2.5){\makebox(0,0)[c]{\scriptsize \mbox{$U$}}}
\put(0.5,-0.2){\makebox(0,0)[c]{\scriptsize \mbox{$U$}}}
\end{picture}
}
 \, .  \label{trg_cdl_svd}
\\
\nonumber
\end{align} 
An important point is that the matrix rank of $W^*$ is automatically reduced to $\chi$ from $\chi^2$ due to the CDL property, since the CTMs at the NE and SW corners in $W^*$ are decoupled from each other.
This implies that the singular values are represented as doubling of the CTM spectra, $ \Sigma^* = \Omega \otimes \Omega$, which are illustrated as two dotted lines respectively carrying the bond dimension $\sqrt{\chi}$ in Eq. (\ref{trg_cdl_svd}),

Assuming Eq. (\ref{trg_cdl_svd}), we then rewrite the renormalized vertex of Eq. (\ref{trg_contraction}) as
\begin{align}
W^{\prime *}(\sigma_1\, \sigma_2\, \sigma_3\, \sigma_4) 
=\hspace{1mm}{\setlength\unitlength{2mm}
\begin{picture}(6,4)(0,-0.3)
\linethickness{1.0pt}
\put(2.3,1.7){\line(1,0){2.5}}
\put(2.3,-0.7){\line(1,0){2.5}}
\put(2.3,-0.7){\line(0,1){2.5}}
\put(4.7,-0.7){\line(0,1){2.5}}
\put(2.1,2.1){\line(1,0){2.8}}
\put(2.0,-1.1){\line(1,0){2.8}}
\put(1.9,-0.9){\line(0,1){2.8}}
\put(5.1,-0.9){\line(0,1){2.8}}
\multiput(0.4,-2.4)(0.25,0.25){7}{\circle*{0.1}}
\multiput(0.6,-2.6)(0.25,0.25){7}{\circle*{0.1}}
\multiput(5.1,1.9)(0.25,0.25){7}{\circle*{0.1}}
\multiput(4.9,2.1)(0.25,0.25){7}{\circle*{0.1}}
\multiput(5.1,-0.9)(0.25,-0.25){7}{\circle*{0.1}}
\multiput(4.9,-1.1)(0.25,-0.25){7}{\circle*{0.1}}
\multiput(2.1,2.1)(-0.25,0.25){7}{\circle*{0.1}}
\multiput(1.9,1.9)(-0.25,0.25){7}{\circle*{0.1}}
\put(1.5,0.5){\makebox(0,0)[c]{\scriptsize \mbox{$s_1$}}}
\put(3.5,-1.6){\makebox(0,0)[c]{\scriptsize \mbox{$s_{2}$}}}
\put(5.6,0.5){\makebox(0,0)[c]{\scriptsize \mbox{$s_3$}}}
\put(3.5,2.5){\makebox(0,0)[c]{\scriptsize \mbox{$s_{4}$}}}
\put(0.3,-1.7){\makebox(0,0)[c]{\scriptsize \mbox{$\sigma_1$}}}
\put(6.7,-1.7){\makebox(0,0)[c]{\scriptsize \mbox{$\sigma_2$}}}
\put(0.3,2.7){\makebox(0,0)[c]{\scriptsize \mbox{$\sigma_4$}}}
\put(6.7,2.7){\makebox(0,0)[c]{\scriptsize \mbox{$\sigma_3$}}}
\end{picture}
}  \quad  \Rightarrow 
\hspace{2mm}{\setlength\unitlength{2mm}
\begin{picture}(10,5)(0,-0.3)
\multiput(3.6,0.5)(0.25,0.25){8}{\circle*{0.1}}
\multiput(3.6,0.5)(0.25,-0.25){8}{\circle*{0.1}}
\multiput(3.0,0.5)(-0.25,0.25){8}{\circle*{0.1}}
\multiput(3.0,0.5)(-0.25,-0.25){8}{\circle*{0.1}}
\multiput(3.3,0.8)(0.25,0.25){8}{\circle*{0.1}}
\multiput(3.3,0.2)(0.25,-0.25){8}{\circle*{0.1}}
\multiput(3.3,0.8)(-0.25,0.25){8}{\circle*{0.1}}
\multiput(3.3,0.2)(-0.25,-0.25){8}{\circle*{0.1}}
\put(0.5,-2){\makebox(0,0)[c]{\scriptsize \mbox{$\sigma_1$}}}
\put(6,-2){\makebox(0,0)[c]{\scriptsize \mbox{$\sigma_2$}}}
\put(0.5,3){\makebox(0,0)[c]{\scriptsize \mbox{$\sigma_4$}}}
\put(6,3){\makebox(0,0)[c]{\scriptsize \mbox{$\sigma_3$}}} 
\put(7.0,0.5){\makebox(0,0)[c]{\mbox{$\times$}}}
\linethickness{1pt}
\put(9,-0.5){\line(1,0){2}}
\put(9,1.5){\line(1,0){2}}
\put(9,-0.5){\line(0,1){2}}
\put(11,-0.5){\line(0,1){2}}
\end{picture}
}
\, \hspace{4mm} ,\label{trg_cdl_contraction} \\
\nonumber
\end{align} 
where  the closed internal loop corresponds to a contraction of the four inner CTMs.
Since the dotted lines of the four outer CTMs are decoupled from the internal loop,  $W^{\prime *}$ becomes a CDL constituting of the four CTMs again, with the overall scalar coefficient of ${\bf \square} \equiv {\rm Tr} C^4 $.
Of course, the same CDL decoupling also occurs in the next step of $W^{\prime *} \to \tilde{W}^*$. [Fig.\ref{trg_iteration}(c) and (d)]
In this sense, the fixed point of the TRG is basically explained by the CTM in the CTMRG with the number of kept basis $\sqrt{\chi}$.
In the TRG iteration, however, the number of spins contained in the renormalized vertex exponentially increases by the scale factor $2$, and thus the convergence to the bulk limit is  much faster, compared with the linear increase of the system size in the CTMRG.
Instead, the CDL tensors in the TRG may not satisfy the variational equations (\ref{ctmrg_fix1}) and (\ref{ctmrg_fix2}).
This is because any CDL tensor satisfies the fixed-point condition of the TRG iteration, even if it deviates from the CTMRG fixed point.
In other words, once $W$ causes the CDL decoupling, it can never be improved with further TRG iterations. 
Thus, we should always pay attention to the above-noted features, if using the TRG algorithm. 

The TRG is a one-way algorithm starting from a small system size and has no mechanism of refining the fixed point tensor after the CDL decoupling.
The second renormalization (SRG) is an improved algorithm allowing further optimization of the renormalized vertex tensors with use of backward iteration of TRG.\cite{SRG2009,Zhao2010}
Constructing the environment tensor for the vertex tensor of $2^n \times 2^n$ from the renormalized vertex of the size $2^{n+1} \times 2^{n+1}$, and then update the building block isometry, i.e. the singular vectors $V$ connecting $n+1$ and $n$, toward the smaller size vertex.
After repeating the backward iterations down to $n=1$,  we turn to the forward updating of the isometries again.
And finally, we obtain the improved renormalized vertex, which certainly provides a more accurate result than the TRG.  
Note that, in the context of the TTN,  the SRG approach is to improve the quality of the isometry tensors, but does not change the tree network structure of the tensors.

Finally, we would like to comment on the critical behavior of the TRG-based algorithm.
At the critical point, the intrinsic correlation length of the system diverges, while the effective length described by the truncated tensors in the TRG is always finite. 
This implies that the effective tensors in the TRG undergo the CDL decoupling after TRG iterations exceeding the effective correlation length governed by the bond dimension $\chi$.  
Thus, the CDL tensors embedded in $W^*$ always give the mean-field nature, which may mask the true critical nature of the system.
In order to eliminate such mean-field behavior, we need to introduce a disentangler, as will be discussed for the TNR.\cite{TNR2015}

\subsection{HOTRG}

The TRG is a simple algorithm to deal with 2D statistical lattice models, but its generalization to higher dimensional systems is not so easy, because the shape of the lattice is changed under a renormalization process, where the number of tensor legs may rapidly increase.
Meanwhile, the HOTRG, which is an abbreviation of tensor renormalization group assisted with higher-order SVD\cite{HOSVD1,HOSVD2}, was introduced as a TRG based approach  easy to generalize to higher dimensions\cite{HOTRG2012}.
This could be a reason why it attracts much interest particularly in the community of lattice gauge theory\cite{Liu_gauge2013,Yoshimura2018,Akiyama2019,Carmen_Banuls_LG2020}, although its fixed point is also described by the CDL.

In the HOTRG, we recursively update the renormalized vertex tensors, starting from the single site vertex tensor.
The differences from the TRG algorithm in the previous subsection are summarized as the following two points.
\begin{quote}
(I) The vertex tensor is sequentially renormalized respectively in the $x,y,\cdots $ directions.
This makes a higher dimensional generalization straightforward.

(II) In order to extract important degrees of freedom in the vertex tensor, the HOSVD for the vertex is employed, instead of the SVD.
\end{quote}
In practical situations, the isometry tensor (RG transformation matrix) can be calculated with the usual matrix diagonalization for a reduced density matrix constructed from a certain contraction of the vertex tensors [See Eq. (\ref{hotrg_rd})], which is basically equivalent to a direct application of the HOSVD to the vertex tensor.

Let us  briefly describe the above process (I).
For the case of square-lattice models, two adjacent vertices of $W$ in the $y$ (vertical) direction are contracted, and then the legs in the $x$ (horizontal) direction are renormalized by the isometry $P$ generated with the HOSVD. 
The graphical representation of this process is given by 
\begin{align}
\hspace{3mm}{\setlength\unitlength{2mm}
\begin{picture}(11,5)(0,-0.3)
\put(4,-4){\line(0,1){8}}
\put(3.4,2){\line(1,0){1.4}}
\put(3.4,-2){\line(1,0){1.4}}
\qbezier(1.2,0)(2,2)(3.4,2)
\qbezier(1.2,0)(2,-2)(3.4,-2)
\qbezier(6.9,0)(5.9,2)(4.8,2)
\qbezier(6.9,0)(5.9,-2)(4.8,-2)
\multiput(6.9,0)(0.3,0){8}{\circle*{0.1}}
\multiput(1.1,0)(-0.3,0){8}{\circle*{0.1}}
\put(-2,0){\makebox(0,0)[c]{\scriptsize \mbox{$\tilde{s}_1$}}}
\put(10,0){\makebox(0,0)[c]{\scriptsize \mbox{$\tilde{s}_{3}$}}}
\put(4,-4.8){\makebox(0,0)[c]{\scriptsize \mbox{$s_2$}}}
\put(4,4.8){\makebox(0,0)[c]{\scriptsize \mbox{$s_{4}$}}}
\put(0.7,-0.8){\makebox(0,0)[c]{\scriptsize \mbox{$P$}}}
\put(7.4,-0.8){\makebox(0,0)[c]{\scriptsize \mbox{$P$}}}
\put(5,3){\makebox(0,0)[c]{\scriptsize \mbox{$W$}}}
\put(5,-3){\makebox(0,0)[c]{\scriptsize \mbox{$W$}}}
\put(7,0){\makebox(0,0)[c]{\scriptsize \mbox{$\blacktriangleright$}}}
\put(1,0){\makebox(0,0)[c]{\scriptsize \mbox{$\blacktriangleleft$}}}
\end{picture}
} \Rightarrow 
\hspace{3mm}{\setlength\unitlength{2mm}
\begin{picture}(5,5)(0,-0.3)
\put(3,-2.0){\line(0,1){1.7}}
\put(3,2.3){\line(0,-1){2.1}}
\multiput(1,0)(0.3,0){6}{\circle*{0.1}}
\multiput(5,0)(-0.3,0){6}{\circle*{0.1}}
\put(3.0,-0.1){\makebox(0,0)[c]{\scriptsize $\square$}}
\put(0,0){\makebox(0,0)[c]{\scriptsize \mbox{$\tilde{s}_1$}}}
\put(6,0){\makebox(0,0)[c]{\scriptsize \mbox{$\tilde{s}_{3}$}}}
\put(3,-3){\makebox(0,0)[c]{\scriptsize \mbox{$s_2$}}}
\put(3,3.4){\makebox(0,0)[c]{\scriptsize \mbox{$s_{4}$}}}
\put(4,1){\makebox(0,0)[c]{\scriptsize \mbox{$W'$}}}
\end{picture}
} \qquad ,
 \label{hotrg_rg1} \\
\nonumber
\end{align}
where $W'$ is a new vertex coarse grained in the $y$ direction and $P$ denotes the RG transformation bunching the two horizontal legs of $W$.
The next step is to do the similar process  for two $W'$ connected in the $x$ direction, 
\begin{align}
\hspace{3mm}{\setlength\unitlength{2mm}
\begin{picture}(11,5)(0,-0.3)
\put(2.2,-0.8){\line(0,1){0.4}}
\put(6,-0.8){\line(0,1){0.4}}
\put(2.2,0.8){\line(0,-1){0.6}}
\put(6,0.8){\line(0,-1){0.6}}
\qbezier(2.2,0.6)(2.2,2)(4.1,2.6)
\qbezier(2.2,-0.8)(2.2,-2)(4.1,-2.5)
\qbezier(6,0.6)(6,2)(4.1,2.6)
\qbezier(6,-0.8)(6,-2)(4.1,-2.5)
\multiput(-0.6,0)(0.3,0){9}{\circle*{0.1}}
\multiput(2.8,0)(0.3,0){10}{\circle*{0.1}}
\multiput(6.5,0)(0.3,0){8}{\circle*{0.1}}
\multiput(4.1,2.6)(0,0.3){5}{\line(0,1){0.2}}
\multiput(4.1,-2.6)(0,-0.3){5}{\line(0,-1){0.2}}
\put(2.2,-0.1){\makebox(0,0)[c]{\scriptsize $\square$}}
\put(6.0,-0.1){\makebox(0,0)[c]{\scriptsize $\square$}}
\put(-2,0){\makebox(0,0)[c]{\scriptsize \mbox{$\tilde{s}_1$}}}
\put(10,0){\makebox(0,0)[c]{\scriptsize \mbox{$\tilde{s}_{3}$}}}
\put(4,-4.8){\makebox(0,0)[c]{\scriptsize \mbox{$\tilde{s}_2$}}}
\put(4,4.8){\makebox(0,0)[c]{\scriptsize \mbox{$\tilde{s}_{4}$}}}
\put(3.3,-3.2){\makebox(0,0)[c]{\scriptsize \mbox{$Q$}}}
\put(3.3,3){\makebox(0,0)[c]{\scriptsize \mbox{$Q$}}}
\put(4.1,2.7){\makebox(0,0)[c]{\scriptsize \mbox{$\blacktriangle$}}}
\put(4.1,-2.7){\makebox(0,0)[c]{\scriptsize \mbox{$\blacktriangledown$}}}
\put(1.0,-1){\makebox(0,0)[c]{\scriptsize \mbox{$W'$}}}
\put(7.2,-1){\makebox(0,0)[c]{\scriptsize \mbox{$W'$}}}
\end{picture}
} \Rightarrow 
\hspace{3mm}{\setlength\unitlength{2mm}
\begin{picture}(5,5)(0,-0.3)
\multiput(3,0.5)(0,0.3){7}{\line(0,-1){0.2}}
\multiput(3,-2.1)(0,0.3){6}{\line(0,1){0.2}}
\multiput(0.9,0)(0.3,0){7}{\circle*{0.1}}
\multiput(3.3,0)(0.3,0){7}{\circle*{0.1}}
\put(3.0,-0.1){\makebox(0,0)[c]{\scriptsize $\diamondsuit$}}
\put(0,0){\makebox(0,0)[c]{\scriptsize \mbox{$\tilde{s}_1$}}}
\put(6,0){\makebox(0,0)[c]{\scriptsize \mbox{$\tilde{s}_{3}$}}}
\put(3,-3){\makebox(0,0)[c]{\scriptsize \mbox{$\tilde{s}_2$}}}
\put(3,3.5){\makebox(0,0)[c]{\scriptsize \mbox{$\tilde{s}_{4}$}}}
\put(4,1){\makebox(0,0)[c]{\scriptsize \mbox{$\tilde{W}$}}}
\end{picture}
} \qquad ,
 \label{hotrg_rg2} \\
\nonumber
\end{align}
where $\tilde{W}$ is the coarse-grained tensor in both of the $x$ and $y$ directions, and $Q$ merges the two vertical legs of $W'$.
The resulting $\tilde{W}$ contains $2\times 2$ original vertices of $W$, implying that its lattice space is doubled.
The isometry $P$ ($Q$) is constructed from the vertex $W$ ($W'$) with the use of the HOSVD.
However, the HOSVD is not so popular algorithm. 
Instead, we can equivalently calculate  $P$ ($Q$) from usual matrix diagonalization of an effective reduced density matrix for $W$ ($W'$).
We will explain the process (II) below.

Assume that the bond dimension of a vertex $W$ is $\chi$.
We then align two adjacent $W$ in the $y$ direction and then take a sum with respect to the vertically connected link, 
\begin{align}
{ M}^x_{s_1s_2, s'_1s'_2}(t_1, t_2) \equiv  \hspace{5mm}{\setlength\unitlength{2mm}
\begin{picture}(5,5)(0,-0.3)
\put(2,-3){\line(0,1){7}}
\put(0,-1.2){\line(1,0){4}}
\put(0,2.2){\line(1,0){4}}
\put(-1,-1){\makebox(0,0)[c]{\scriptsize \mbox{$s_1$}}}
\put(5,-1){\makebox(0,0)[c]{\scriptsize \mbox{$s'_{1}$}}}
\put(-1,2){\makebox(0,0)[c]{\scriptsize \mbox{$s_2$}}}
\put(5,2){\makebox(0,0)[c]{\scriptsize \mbox{$s'_{2}$}}}
\put(2,-4){\makebox(0,0)[c]{\scriptsize \mbox{$t_1$}}}
\put(2,5){\makebox(0,0)[c]{\scriptsize \mbox{$t_{2}$}}}
\put(3,3){\makebox(0,0)[c]{\scriptsize \mbox{$W$}}}
\put(3,-2){\makebox(0,0)[c]{\scriptsize \mbox{$W$}}}
\end{picture}
} \qquad ,
\label{hotrg_ww}
\\
\nonumber
\end{align}
which may be viewed as a minimal unit of column to column transfer matrix.
We then construct an effective reduced density matrix with ${M }^x $, 
\begin{align}
 { \rho}^x \equiv \sum_{t_1,t_2} { { M}^x}^\dagger(t_1,t_2) { M}^x(t_1, t_2) 
 =  \hspace{5mm}{\setlength\unitlength{2mm}
\begin{picture}(6,5)(0,-0.3)
\put(1,-2.8){\line(0,1){6.6}}
\put(-0.7,-3.0){\line(1,1){2.5}}
\put(-0.7,0.4){\line(1,1){2.5}}
\put(4,-2.8){\line(0,1){6.6}}
\put(2.7,-3.0){\line(1,1){2.5}}
\put(2.7,0.4){\line(1,1){2.5}}
\put(2.5,3.4){\oval(3,2)[t]}
\put(2.5,-2.5){\oval(3,2)[b]}
\qbezier(1.8,-0.5)(3,0.4)(3.7,0.4)
\qbezier(4.2,0.4)(5.5,0.4)(5.2,-0.5)
\qbezier(1.8,2.9)(3,3.8)(3.7,3.8)
\qbezier(4.2,3.8)(5.5,3.8)(5.2,2.9)
\put(-1.2,-2.3){\makebox(0,0)[c]{\scriptsize \mbox{$s_1$}}}
\put(2.1,-2.3){\makebox(0,0)[c]{\scriptsize \mbox{$s'_{1}$}}}
\put(-1.2,0.9){\makebox(0,0)[c]{\scriptsize \mbox{$s_2$}}}
\put(2.1,0.9){\makebox(0,0)[c]{\scriptsize \mbox{$s'_{2}$}}}
\put(2.5,-4.2){\makebox(0,0)[c]{\scriptsize \mbox{$t_1$}}}
\put(2.5,5){\makebox(0,0)[c]{\scriptsize \mbox{$t_{2}$}}}
\put(0,2){\makebox(0,0)[c]{\scriptsize \mbox{$W$}}}
\put(0,-1){\makebox(0,0)[c]{\scriptsize \mbox{$W$}}}
\put(5,1.5){\makebox(0,0)[c]{\scriptsize \mbox{$W$}}}
\put(5,-2){\makebox(0,0)[c]{\scriptsize \mbox{$W$}}}
\end{picture}
} \quad ,
\nonumber \\
\label{hotrg_rd}
\end{align}
and calculate eigenvalues of $\rho^x$ and the corresponding eigenvectors with the conventional matrix diagonalization (or SVD),
\begin{align}
\rho^x = P \Sigma  P^\dagger \, ,
\label{hotrg_svd}
\end{align}
where $\Sigma$ is the eigenvalue matrix of dimension $\chi^2$ and $P$ is the $\chi^2 \times \chi^2$-dimensional orthonormal matrix.
Note that $\Sigma$ is positive semidefinite by construction.
The low-rank approximation based on the larger $\chi$ eigenvalues in $\Sigma$ yields the transformation matrix $P$ for Eq. (\ref{hotrg_rg1}), where its matrix size is basically $\chi^2 \times \chi$.
In the similar way, we construct ${M}^y(s_1,s_2) \equiv \sum_{x~\rm legs} W' W'$ with tracing out the leg indices in the $x$ direction, and then the low-rank approximation for the reduced density matrix $\rho^y \equiv \sum_{s_1,s_2}{ M}^{y \dagger}(s_1,s_2) { M}^y(s_1,s_2)$ also gives the isometry $Q$ of $\chi^2 \times \chi$ for Eq. (\ref{hotrg_rg2}).
 Note that the above sequential RG transformations in the $y$ and $x$ directions may break the 90$^\circ$ rotational symmetry of the square lattice.
However, the numerical accuracy based on the low-rank approximation is sufficiently good except at the critical point.
The practical algorithm is summarized in Fig. \ref{HOTRG_loop}.

\begin{figure}[t]
\begin{center} \includegraphics[width = 8.0 cm]{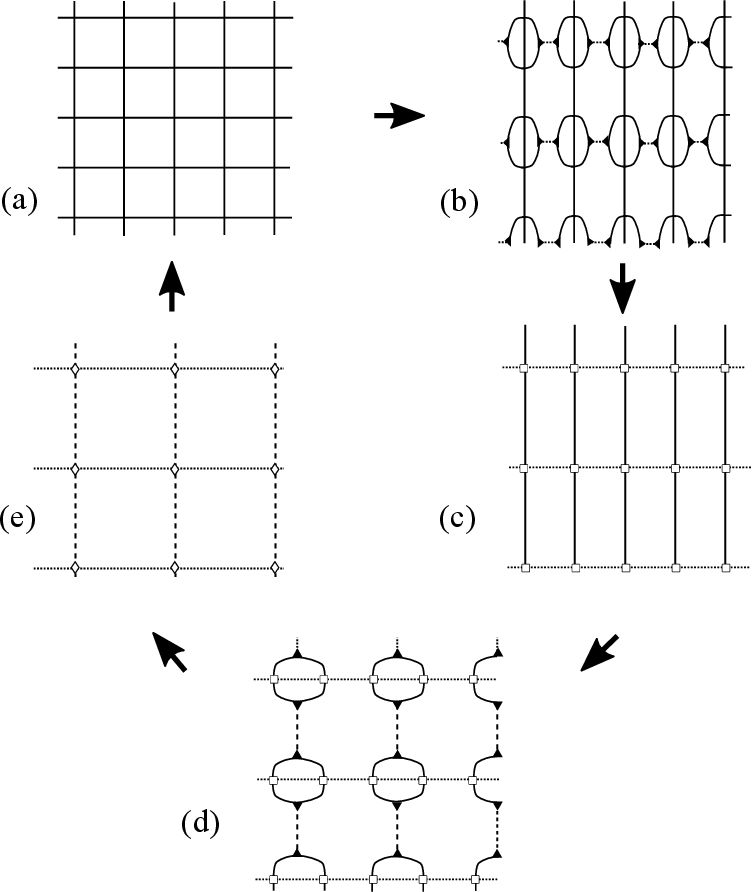} \end{center}
\caption{Recursive algorithm of the HOTRG.
(a) The square-lattice vertex model in a certain length scale. 
(b) Every two horizontal links are renormalized with Eq. (\ref{hotrg_rg1}), where the transformation matrix (isometry) $P$ is generated by Eqs. (\ref{hotrg_rd}) and (\ref{hotrg_svd}).
(c) The resulting vertex model with the renormalized horizontal links. 
(d) Every two vertical links are renormalized with Eq. (\ref{hotrg_rg2}).
(e) The coarse-grained square-lattice vertex model. 
By the scale transformation, the vertex model in the panel (e) goes back to the vertex model in the panel (a) with the renormalized vertex tensor. }
\label{HOTRG_loop}
\end{figure}

After $N$ times of HOTRG iterations, the number of vertex weight included in the effective renormalized vertex is $2^N \times 2^N$, implying that the situation of the renormalized vertex is basically the same as the TRG case.
Thus, the fixed point of the HOTRG can be also described by the CDL tensor of Eq. (\ref{trg_cdl}),  as precisely discussed in Ref. [\citen{Ueda2014}].
For the CDL, the reduced density matrix of Eq. (\ref{hotrg_rd}) is decoupled into 
\begin{align}
\rho^x = \alpha  C^2 \otimes C^2 \otimes ({\rm  internal~lines})\, ,
\label{hotrg_cdl_dm}
\end{align}
with $\alpha = (\tr C^4 )(\tr C^2)^2$, where $C$ denotes a CTM of the dimension $\sqrt{\chi}\times \sqrt{\chi}$ defined in Eq. (\ref{trg_cdl_vertex}).  
The decoupled internal lines consist of the leg-indices of the dimension $\chi\times \chi$ that do not contribute to the eigenvalue problem of $\rho^x$.
Thus,  the effective matrix rank of $\rho^x$ is reduced to be $\chi \times \chi$ from $\chi^2 \times \chi^2$, as in the case of TRG.  
The eigenvalues and the corresponding eigenvectors of Eq. (\ref{hotrg_cdl_dm}) are described by 
\begin{align}
& \Sigma = \alpha \Omega^2 \otimes \Omega^2 \oplus ({\rm zero~eigenvalues})\, , \\
&  P = U\otimes U\otimes({\rm internal~lines}) \,,
\label{hotrg_cdl_internal}
\end{align} 
where $\Omega$ and $U$ respectively denote the singular values and the corresponding singular vectors of the CTM defined by Eq. (\ref{trg_ctm_ev}), and $\alpha$ is the overall constant which can be absorbed into the normalization. 
The decoupled internal lines originate from the zero eigenvalues of $\rho^x$, implying that the matrix-rank reduction automatically occurs for Eq. (\ref{hotrg_cdl_dm}), and thus the dimension of $P$ becomes $\chi^2 \times \chi$.
Of course, the CDL decoupling also occurs in $W'$ and $\rho^y$, implying that the corresponding dimension of $Q$ is $\chi^2 \times \chi$.
In Ref. [\citen{Ueda2014}], it was actually demonstrated that the fixed point spectrum of the HOTRG is described by the doubling of the CTM spectrum.

As seen above, the fixed point of the HOTRG is described by the CDL tensor, which is basically the same as that of the TRG. 
The dominant computational cost of the HOTRG, which is attributed to Eqs. (\ref{hotrg_rg1}), (\ref{hotrg_rg2}) and (\ref{hotrg_ww}), is estimated to be ${\cal O}( m^{4d-1})$ with $d$ begin the spatial dimension.
For the 2D system, thus, the HOTRG is more expensive than the TRG approach.
In the HOTRG, however,   sequential updating of the renormalized vertex with respect to the spatial directions makes an application to higher dimensions straightforward.\cite{HOTRG2012}
Thus, the HOTRG algorithm is often applied to higher dimensional systems toward 4D lattice gauge models.\cite{Yoshimura2018,Akiyama2019}. 
Then,  a significant problem from a practical viewpoint is how to reduce the computational cost of tensor contraction.
Recently,  anisotropic tensor renormalization group (ATRG) provides an efficient algorithm whose computational cost to realize the same accuracy as the HOTRG is scaled with ${\cal O}(m^{2d+1})$}.\cite{ATRG}

\subsection{TNR}
\label{Sec_TNR}

The TRG-type algorithms can be considered as real-space RGs consisting of contraction and truncation of local tensors.
In particular, both of coarse graining and scale transformation are eventually governed by the SVD for the vertex weight. 
As shown in the previous subsections, however, the TRG-type algorithms with a finite $\chi$ generally cause the CDL decoupling in renormalized tensors, where the finite $\chi$ effect always introduces an effective length scale into the tensors even at criticality.
In the context of the entanglement, this CDL decoupling is attributed to the fact that the maximum EE that can be involved in the TTN is basically bounded by $\log \chi$.
Thus, the TRG approaches are not able to handle the log correction to the area law of EE at the criticality.

\begin{figure}[bt]
\begin{center} \includegraphics[width = 8.5 cm]{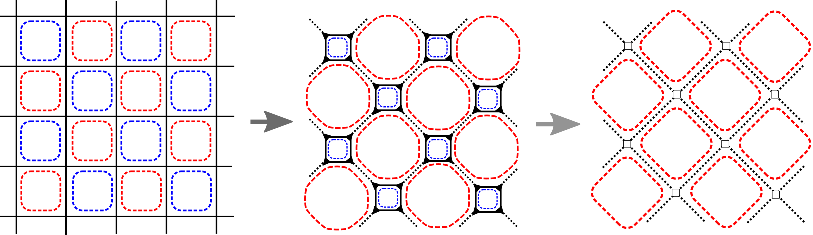} \end{center}
\caption{(Color online) 
Short range entanglements in the TRG algorithm [Fig. \ref{trg_iteration}].
Left panel: 
Short-range entanglements embedded in the vertex tensor are schematically represented as blue and red dashed lines, which correspond to the CDL tensor.
Center panel: 
The short-range entanglements of the blue dashed lines can be traced out by contraction of the isometry tensors generated by SVD of the vertex tensors [Eq. (\ref{trg_contraction})]. 
Right panel:
The short-range entanglements of the red dashed lines always survive in the renormalized vertex model in the larger scale,  consistentwith the CDL fixed point of the TRG algorithm. 
In order to maintain these short range entanglements,  the critical long-range entanglement is excluded in an early stage of the TRG iteration. 
}
\label{TRG_eloop}
\end{figure}

More precisely, the above difficulty in the TRG can be illustrated as Fig. \ref{TRG_eloop}.
Short-range entanglements associated with the CDL factorization are represented as blue and red dashed lines on plaquettes in the vertex model.
The TRG iteration can trace out the short-range entanglements of the blue dashed line by the contraction of Eq. (\ref{trg_contraction}).
However,  the red dashed lines are never eliminated by such a renormalization process and thus the corresponding short-range entanglements survive in the larger length scale, resulting in the trivial CDL fixed point of the TRG.
In other words, the TRG has to always  spend its expression capacity of entanglements for maintaining the short-range entanglements, so that the long-range critical entanglement has been missed in an early stage of the TRG iteration.

\begin{figure}[tbh]
\begin{center} \includegraphics[width = 8.2 cm]{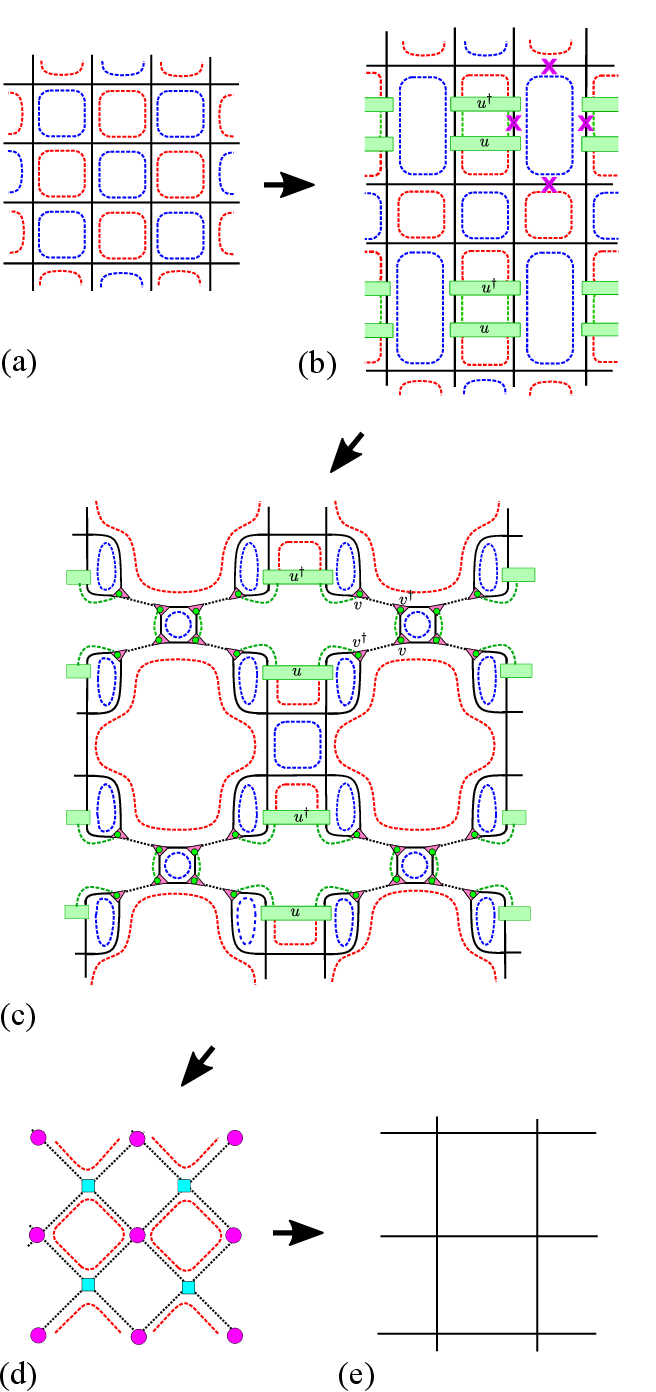} \end{center}
\caption{(Color online) 
The TNR algorithm and reduction of short-range entanglements.
(a) Short-range entanglements in the 2D vertex model are illustrated as red and blue dashed lines (See also Fig. \ref{TRG_eloop}).
(b) Insertion of a pair of disentaglers $u^\dagger u= 1$, which converts the red dashed lines into the green dashed lines.  
(c) A pair of isometries $v v^\dagger \approx 1 $ are inserted at {\sf x} symbols and their equivalent positions on the lattice in the panel (b).
Note that $u$ and $v $ are obtained by minimizing $\delta$ defined by Fig. \ref{role_disentangler}(b).
Then, the blue dashed loops are traced out by the isometries. 
Also the greed dashed lines are terminated at the isometries. 
(d) The four-leg composite tensors of the blue square and of the pink circle are respectively defined in Fig. \ref{TNR_SVD}(a) and (b).
(e) The remaining red-dashed loops can be traced out by the SVD and contraction depicted in Fig. \ref{TNR_SVD}(c).
}
\label{TNR_disentangler}
\end{figure}

In order to properly deal with the critical entanglement, a scale-dependent controlling mechanism of correlations/entanglements is needed. 
For this purpose, an important concept is the disentangler, which is a unitary operator firstly introduced in the MERA for 1D quantum systems \cite{MERA2007} and later reformulated in the TNR for 2D classical systems.\cite{Evenbly2016}
An essential point of the disentangler is that it modifies the connectivity of tensors from the tree type into a scale-dependent loop network, which allows us to represent the EE up to the log correction to the area law. 
 In contrast to the TRG, the disentangler in the TNR  systematically filters out the short-range entanglements that mask the long-range entanglement intrinsic to the bulk critical behavior. 
As a result, the long-range entanglement can be properly maintained in the bulk fixed point.

\begin{figure}[bt]
\begin{center} \includegraphics[width = 7.8 cm]{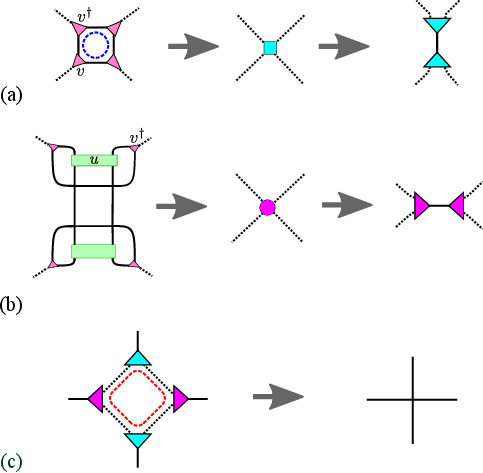} \end{center}
\caption{(Color online) 
The renormalization of the composite tensors included in Figs. \ref{TNR_disentangler}(c), (d) and (e).
(a) The cluster of four isometries is regarded as a four-leg tensor with the blue square symbol.
The SVD in the vertical direction yield  two isometries of the blue triangles.
(b) The composite tensor containing two disentanglers is regarded as a four-leg tensor with the pink circle.  
The SVD in the horizontal direction yields the corresponding isometries with the pink triangle.
(c) The red dotted loops in Fig. \ref{TNR_disentangler}(d) can be traced out by a contraction of the isometries of (a) and (b), which provides a new renormalized vertex in Fig. \ref{TNR_disentangler}(e). }
\label{TNR_SVD}
\end{figure}

The disentangler is implemented so as to trace out loops of the red-dashed lines in Fig. \ref{TNR_disentangler} (a). 
As depicted as green square box symbols in Fig. \ref{TNR_disentangler} (b), a pair of  disentanglers $u^\dagger u =1$ is inserted in every two plaquettes corresponding to the red-dashed loops.
The disentangler is a unitary operator, which transforms the red lines into green lines such that the entanglement carried by the red dashed line is disentangled.
We next insert pairs of isometries $v$ and $v^\dagger$  at pink ``{\sf x}" symbols and their equivalent points around the disentanglers in Fig. \ref{TNR_disentangler} (b). 
Here, we note that the position of ``{\sf x}" is basically the same as that in the TRG, if there is no disentangler.

In Fig. \ref{TNR_disentangler} (c), then, small plaquettes consisting of four isometries illustrated as pink triangles trace out the blue dashed loop, like in the case of the TRG.
On the other hand, the green dashed lines are terminated at the isometry, implying that the entanglements originating from the red dashed lines can be eliminated by the combination of the disentangler and isometry.
As shown in Fig. \ref{TNR_SVD}(a), we regard the plaquette, which contains the blue dashed loop,  as a four-leg renormalized vertex with a blue square symbol. 
Also, we construct the four-leg composite tensor with the pink circle as in Fig. \ref{TNR_SVD}(b) to  obtain the vertex model on the 45$^\circ$-rotated lattice as shown in Fig  \ref{TNR_disentangler} (d),  where the entanglements of the red dashed loops are partially traced out. 
As in Fig. \ref{TNR_SVD}(a) and (b), we further perform SVD of the vertices and then combine the corresponding isometries to trace out the remaining loop entanglements [Fig. \ref{TNR_SVD}(c)].
The renormalized vertex model with the doubled lattice space is finally obtained in Fig  \ref{TNR_disentangler} (e), where the short-range entanglements in Fig. \ref{TNR_disentangler} (a) are totally eliminated.

 Let us illustrate roles of the disentangler and the isometry in Fig. \ref{role_disentangler}(a), which shows a magnification of the corresponding part in Fig. \ref{TNR_disentangler} (c).
A main issue is how to trace out the red dashed line connecting two vertices labeled by {\sf A} and {\sf B}.
Since this red dashed line is independent of the other loops, its entanglement can be controlled by a unitary acting on the two legs coming out from {\sf A} and {\sf B}; 
The disentangler $u$ in Fig. \ref{role_disentangler}(a) converts the red dashed line of the entanglement into the left and right green lines so as to be classically separable from each other.
However, the unitary never changes the size of the (local) Hilbert space of physical degrees of freedom.
We further need to implement the isometry $v$ to terminate the green lines.

\begin{figure}[tb]
\begin{center} \includegraphics[width = 7 cm]{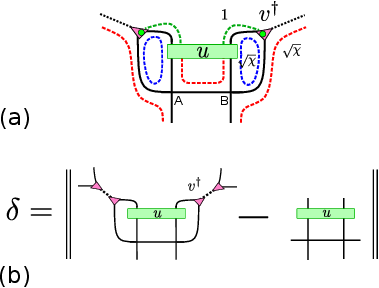} \end{center}
\caption{(Color online) 
(a) Roles of the disentangler and isometry in the TNR.
The disentangler converts the red dashed line representing the short range entanglement into the green dashed lines such that their left and right branches are separable from each other, and can be terminated at the isometries.
Accordingly, the uplinks of the dashed lines from the isometry have capacity of encoding nontrivial long-range entanglement by $\sqrt{\chi}$.
(b) The definition of the residual norm $\delta$. 
The optimal disentangler and isometry in the TNR can be constructed by minimizing $\delta$. 
} 
\label{role_disentangler}
\end{figure}

For a more precise discussion, we assume that the effective bond dimension for the red and blue dashed lines is $\sqrt{\chi}$,
and thus the total bond dimension of the tensor leg is $\chi$.
The closed loop of the blue dashed lines is traced out by the isometry, as in the case of TRG, whereas the red dashed lines in the outer side of {\sf A} or {\sf B}, which are supported by the bond dimension $\sqrt{\chi}$,  go to the upper layer passing through the isometry $v^\dagger$.
As in the TRG case, these red dashed lines will be eliminated in the next step of the TNR [Fig.\ref{TNR_disentangler}(d) and (e)].
On the other hand,  the bond dimension required for describing the entanglement of the green lines is in principle reduced to one by the disentangler, implying that the Hilbert space corresponding  to the green dashed lines can be projected out by appropriate isometries.
As a result,  a buffer of $\sqrt{\chi}$ of the effective bond dimension is generated in the up connecting link from the isometry, in which one can encode intrinsic long-range entanglements instead of the short-range entanglement eliminated.

In a practical situation, balance of the short-range and long-range entanglements encoded in the up connecting link depends on the length scale of the tensors(or equivalently renormalization steps).
In order to realize the optimal disentangler and isometry,  we consider the residual norm $\delta$ between approximated (truncated) tensors and the full tensors, which is depicted in Fig. \ref{role_disentangler}(b).
An essential point is that the disentangler and isometry are simultaneously optimized so as to minimize the cost function of  $\delta$.
In practice, this optimization can be formulated through SVD of quasi-local environment tensors including both of the disentanglers and the isometries.
For the detailed algorithm of the optimization scheme, we would like to refer readers to Refs. [\citen{Evenbly2016}] and [\citen{Evenbly2017}].
Here, we particularly suggest a stable version of the TNR algorithm in Ref. [\citen{Evenbly2017}].

The above TNR algorithm actually reproduces the numerically exact scaling dimensions for the 2D Ising model with the bond dimension $\chi \sim 24$.
Then, an interesting aspect of the TNR is that the optimization based on the quasi-local environment tensors associated with $\delta$ in Fig. \ref{role_disentangler}(b) does not refer to the global TN structure corresponding to  the total free energy.
In other words, the isometry and disentangler can be determined with the tensors of a particular length scale, allowing us to eliminate the short-range entanglement without taking variation of the global free energy. 
As will be discussed in the next subsection, the TNR for the 2D vertex model is basically equivalent to the MERA for the 1D quantum system, which was originally formulated as a variational method for the global ground-state energy.
By contrast, the quasi-locality in the TNR algorithm is more important to discuss the connection to the Wilsonian renormalization group from the field theoretical viewpoint. 
To our best knowledge, the TNR is the first numerical real-space RG approach that overcame the difficulty in conventional real-space RGs since the Kadanoff and Wilson, and  succeeds in extracting the (numerically) exact scaling dimensions.     
We therefore think that the TNR(and MERA) can be a milestone of theoretical physics.

Recently, there are  a couple of interesting and practical entanglement filtering approaches similar to the TNR, e.g. Loop TNR [\citen{LoopTNR2017}], entanglement branching operator [\citen{Harada2018}] and graph-independent-local-truncation (GILT) TNR [\citen{GILT2018}], etc.
These approaches are also based on quasi-local environment tensors to filter out short-range entanglements associated with the CDL tensor and successfully extract correct scaling dimensions for critical 2D classical systems with the computational cost of ${\cal O}(\chi^6)$(Loop TNR, GILT TNR) $\sim {\cal O}(\chi^7)$(HOTRG with branching operator).
However, an efficient controlling of short-range entanglements in higher dimensional systems has not been established yet.
It may be also necessary to clarify the nature of the 3D version of a CDL tensor, which may be attributed to the failure of CTTRG as discussed in Sec. \ref{sec_TPS}.

\subsection{MERA}

\label{Sec_MERA}

\begin{figure}[tb]
\begin{center} \includegraphics[width = 7.2 cm]{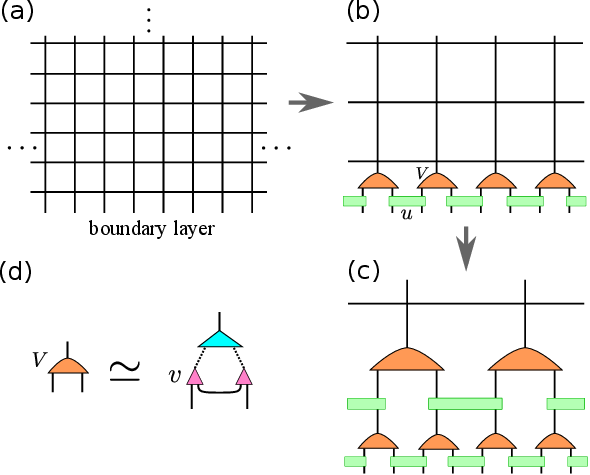} \end{center}
\caption{(Color online) 
(a) A half-infinite lattice system with the boundary at the bottom layer.
(b) The TNR iteration explicitly inserts a set of disentanglers and isometries in the bottom layer, where the four-leg tensor with the green box denotes the disentangler $u$ and the three leg tensor with the orange symbol indicate the isometry $V$ in the MERA network.
(c) The next step of the TNR iteration also generates a set of disentanglers and isometries in the next layer.
(d) The isometry $V$ in the MERA network basically corresponds to a composite of the isometries in Fig. \ref{TNR_SVD}(a).
}
\label{TNRtoMERA}
\end{figure}

The MERA is a variational method with respect to the global ground-state energy of quantum many-body systems,  based on the multilayered TN wavefunction.\cite{MERA2007} 
To be specific, we discuss the MERA for the ground state of 1D critical quantum systems in the following.
The most important feature of the MERA is that the disentangler is inserted into the TTN of the conventional real-space RG  and the resulting MERA network turns out to be capable of representing up to the logarithmic correction to the area law of EE.
Also, its connection to the holographic EE\cite{RT_PRL2006,RT_JHEP2006} attracts much attention to physics of the TN from the viewpoints of quantum information and quantum gravity.\cite{Swingle2012,Matsueda2013}

Historically, the concept of disentangler was first introduced in the MERA network by G. Vidal prior to the TNR.
Thus the TNR could not be invented without the establishment of MERA, possibly. 
In this subsection, we rather discuss the MERA from the standpoint of the TNR.\cite{TNRtoMERA}
In particular, we illustrate the variational optimization algorithm of the MERA as a finite-layer version of TNR for 1D quantum systems, which provides a unified view for the TNR/MERA in the context of the real-space RG.

As in Fig. \ref{TNRtoMERA}(a), let us {begin with a half-infinite vertex model representing the world sheet of the ground-state wavefunction for a 1D quantum spin chain, where tensor legs at the bottom boundary layer correspond to  physical spin degrees of freedom. 
Following the prescription in Fig. \ref{TNR_disentangler},  we perform the TNR iteration with inserting disentanglers $u^\dagger u =1$ and isometries $v v^\dagger \sim1$ into the lattice of Fig. \ref{TNRtoMERA}(a). 
However, the disentangler and the isometry that emerged at the boundary layer cannot find the proper partners to form the bulk renormalized tensors.  
In Fig. \ref{TNRtoMERA}(b), thus, such disentangler and the isometry at the boundary layer explicitly connect the leg degrees of freedom in the adjacent length scales.
Note that the isometry $V$ with the orange symbol in the MERA is a composite tensor consisting of those in the TNR as illustrated in Fig. \ref{TNRtoMERA}(d).
Applying the TNR iteration recursively to the bulk region of tensors, we obtain the disentanglers and isometries bridging to the next layer, as depicted in Fig. \ref{TNRtoMERA}(c).
With repeating the TNR iteration, thus, we can systematically generate the MERA network.
In comparison with the TNR, an important feature of the MERA network is that the connectivity of the tensors in the different length scales is  manifest.
As the depth of layers increases, in particular, the number of tensors decreases exponentially,  consistently with the framework of the real-space RG.

\begin{figure}[tb]
\begin{center} \includegraphics[width = 6.4 cm]{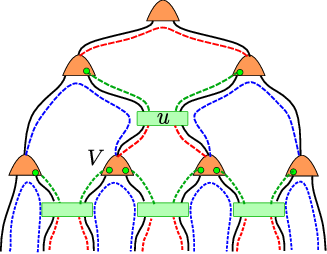} \end{center}
\caption{(Color online) 
Flow of the scale-dependent entanglements associated with the CDL tensor in the TNR.
The red-dashed lines of local entanglements are converted into the green-dashed lines representing classically separable states by the disentangler $u$, which can be terminated at the isometry $V$ of the orange symbol. 
The multi-layered structure of the MERA network provides the scale-dependent entanglement-filtering mechanism.
}
\label{MERA_CDL}
\end{figure}

The TNR is basically a one-way algorithm toward the bulk fixed point of the system, while the depth of layers in the practical variational MERA algorithm is usually terminated at a finite number.
This implies that the MERA algorithm can be interpreted as a finite layer version of the TNR. 
More precisely, in the MERA algorithm,  each tensor can be updated with referring to the global energy expectation value, through finite-size sweeps in the up and down, and left and right directions in  the MERA network. 
This relation between the TNR and the MERA is reminiscent of that of the infinite-size} DMRG and the finite-size DMRG. 
Of course, the MERA network structure is much more complicated than the MPS in the DMRG.
However, the explicit connectivity of tensors between adjacent scale layers in the MERA network visualizes the causal structure of tensors in the network.
From a practical viewpoint, the unitarity of the disentangler $u^\dagger u = 1 $ and the orthonormal condition of the isometry $V^\dagger V=1 $ ensure significant simplification of contraction of tensors in the MERA wavefunction.
Then one can update a set of tensors with finite-size sweeps combined with SVD of environment tensors, where the dominant computational cost is ${\cal O}(\chi^7)$, which is of the same order as the TNR.
The computational details can be found in Ref. [\citen{Evenbly2009}].
Here, it should be remarked that the scale-invariant MERA that sets up a recursive relation of tensors in the scale-invariant layers would contain a transitional aspect between the variational MERA algorithm and the TNR algorithm.\cite{Evenbly2013quantum}

Figure \ref{MERA_CDL} demonstrates how the scaled-dependent entanglements associated with the CDL tensor can be eliminated in the MERA network.
The relation between the TNR and the MERA in Fig. \ref{TNRtoMERA} also provides an illustration of the entanglement flow in the MERA network.
The disentangler converts the red dashed line coming from the lower layer into the two green dashed branches disentangled from each other.
Then these green dashed lines can be terminated at the isometry of the orange symbol consisting of three pieces of isometries in the TNR, which was defined in Fig. \ref{TNRtoMERA}(d). 
Although the pink small triangle of $v$ in the TNR can eliminate only a single green broken line,  the two triangles of $v$ involved in $V$ can filter out the two green broken lines coming up from both sides of $V$.
Thus, the isometry in the MERA network also filters out the short-range entanglements, so that the longer-scale entanglements, which are illustrated as red and blue dashed lines coming out from the top of the orange symbol of $V$, can be encoded into the links to the upper layer.

The situation where the power of MERA can be strikingly demonstrated is the ground state of 1D quantum critical systems. 
Actually, the MERA algorithm has succeeded in extracting correct critical exponents for various models associated with 2D CFTs in the framework of the real-space RG\cite{Giovannetti2008,Pfeifer2009}.
A more direct connection to 2D CFTs is also studied in Refs. [\citen{Evenbly2016,Milsted2018}].
Moreover, the multi-scaled-layer structure in the MERA network provides an intuitive view for the connectivity of tensors capable of representing the log-correction to the area law of EE.
This nature of the MERA network revealed an interesting connection of TN physics to quantum gravity; 
the entanglement between a finite-length region (system part) and its complement in the bottom layer is supported by the tensor legs along the minimal surface for the system part,\cite{Swingle2012} as in the case of the  holographic EE in the AdS space-time \cite{RT_PRL2006}.
This graphical correspondence between the MERA and AdS/CFT is recently termed as AdS/MERA.
A continuous field theory version of the MERA\cite{cMERA2013} was also designed for a Gaussian model\cite{Nozaki2012}, which stimulates further development of the entanglement renormalization in quantum field theories.

As discussed so far, the MERA/TNR algorithm has established a modern formulation of the quantum renormalization group in the real space.
However, it is still difficult to deduce theoretical structure directly from very complex numeric of tensor elements.
In this sense, a simple model which allows us to construct the analytic representation of a MERA wavefunction is highly desired.
Recently, a real-space representation of the MERA wavefunction is constructed with the wavelet basis for free fermion systems.
A hint to the wavelet representation can be seen in the fast Fourier transformation (FFT) of Danielson and Lanczos algorithm\cite{FFT1,FFT2,NumRecipes}.
To be specific, we consider single-particle wavefunctions of a $N$ site system, for which we need to calculate $N^2$ number of Fourier coefficients. 
In the FFT,  it is well established that the computational cost is down to $N\log N$ with the use of the tree network structure based on recursive bipartitioning with respect to even or odd sites.
Moreover,  a TN representation of the FFT is interestingly realized as a quantum circuit by using a series of unitaries.\cite{Ferris2014}

A similar reduction of computational cost is also possible for the wavelet basis, which results in the MERA-type network of local unitary gates with $\log N$ layers.\cite{Fishman2015} 
Moreover, the many-body ground state of critical free fermion systems can be properly reproduced by the modified wavelet basis taking the Fermi point effect into account,\cite{Evenbly_White} which also provides a quantum circuit interpretation of the scale-dependent filtering of entanglements.
Although the above approach is specific to the free fermion case, it gives rise to an analytic construction of the MERA, which will be of significant importance in discussing the hierarchical structure of general critical wavefunctions.



\section{Other trends and prospects of tensor networks}

In this review, we explained essential concepts included in the formulation of various TNs, basically assuming bulk uniform many-body systems with short-range interactions such as Ising model and $S=1/2$ Heisenberg model. 
Recently, however, the application range of TNs has been rapidly expanding beyond such bulk uniform systems. 
In this section, we would like to discuss some prospects of TN physics. 

\noindent{\bf Nonuniform and complex systems and TNs}: 
One of interesting topics of the TN for nonuniform systems is random quantum spin systems.
In the analytic level, the perturbative strong-disorder RG (SDRG) was established as an asymptotically exact real-space RG for the random singlet fixed point.\cite{SDRG1, DSFisher1994}
Recently, the SDRG was reformulated as tree-type TN algorithms \cite{Hikihara1999, Goldsborough2014, Seki2020}, which actually improved the quantitative reliability of the perturbative SDRG.
To what extent the tree TN state is capable of representing possible fixed points induced by randomness other than the random singlet phase may be an interesting problem. 

Another fascinating trend of the TN is quantum chemistry problems, which also have complicated interactions among electrons on atomic sites in the real-space representation.
For example, the finite-size algorithm of DMRG was examined for small molecules by White and Martin in 1999.~\cite{White1999}
After that, several DMRG based algorithms including tree TNs have been applied to quantum chemistry problems such as organic molecules.\cite{Daul2000, Chan2002, LegezaChem2003, Chan2011, Barcza2011, Nakatani2013, ChemRev2015, Chan2016}
In addition, quantum chemistry is one of the most promising targets of near-term quantum computers, which also attract much attention from the viewpoint of quantum circuit representation of TNs.

From the statistical mechanical viewpoint,  neural networks and  machine learning problems are  promising application targets of the TN.
Actually, an MPS can be used for approximating the probability distribution for hand writing images of numbers.~\cite{Stoudenmire2016}
Appearance frequency of words and their alignments in English texts is examined by means of MPS~\cite{Gallego2019}, where exceptions in grammar can be naturally categorized.
Equivalence between a restricted Boltzmann machine and an MPS is also  discussed by Chen et al.~\cite{Chen2018} 
Moreover, it is interesting that an RG structure can be seen in the neural machine learning assisted by TNs.\cite{Li2018}

\noindent{\bf Quantum circuits and TNs}: 
In accordance with the hardware development of noisy intermediate scale quantum (NISQ) computers\cite{Google2019}, the connection between TNs and quantum circuit models attracts much attention from both of theoretical and practical viewpoints.
Then, we can consider two directions of researches.
The first one is TN simulations of quantum circuits on a classical computer.
For example, well-known algorithms such as Grover's algorithm and Shor's algorithm are actually implemented in the MPS \cite{Kawaguchi2004, Dang2019} or TTN frameworks\cite{Dumitrescu2017}.
Recently, the efficiency of TN simulations of quantum circuits has come under the spotlight again,\cite{Zhou2020, Huang2020} triggered by the recent development of NISQ devices.
Another direction of the research is direct optimizations of TN states with the use of quantum computers.
For example, optimization of TN states with the use of a quantum circuit was proposed in Refs. [\citen{Liu2019,RanMPS2020}].
Also, a combination of the variational quantum eigensolvers\cite{Peruzzo2014,VQE2021review} and the TN formalism is expected to become important in connection with quantum chemistry problems.

Inspired by the development of quantum computers, nontrivial entanglement physics associated with quantum circuit models has been intensively studied from the viewpoint of theoretical physics.
For example, a measurement-induced entanglement transition in random quantum circuit models has attracted much attention.\cite{LiChenFisher2019, Goto2020,XianLiFisher2021}
Of course, TN methods are efficient numerical tools for simulating such quantum circuit models. 
On the other hand, such a TN algorithm for thermal equilibrium states as minimally entangled typical quantum states\cite{METTS2009, Stoudenmire2010,Iitaka2020,Goto2021,Iwaki2021} may be viewed as  a measurement-based sampling algorithm for the MPS, which involves a certain similarity to random quantum circuit models.
Further development of TNs for dynamical behaviors of quantum circuits associated with measurements is also an interesting problem.



\noindent{\bf Holography and TNs}: 
Finally, we would like to mention the close connection between the TN in Sec \ref{Sec_TNR} and \ref{Sec_MERA} and the holographic descriptions of quantum many-body systems.
According to the connection between MERA/TNR and Ryu-Takayanagi formula for the holographic EE, the TN has become an essential tool for understanding the interdisciplinary physics among quantum many-body systems, quantum gravity, and quantum information. 
In addition to the continuous MERA approach\cite{cMERA2013,Nozaki2012},  recently, such a concept as quantum circuit complexity in the path integral of quantum field theories has been developed.\cite{Caputa2017, Jefferson2017,Molina-Vilaplana2018} 
So far, the network structure of a TN was designed by hands on the basis of qualitative consideration about the physical situation of a target system at the UV boundary layer.
The path integral approach incorporating the quantum information geometry seems very interesting in the context of the TN, since it may provide a new quantitative criterion for designing network structures of the TN state.

We can find another interesting direction of researches on the TN and the holography;
The MERA explained in this review were basically introduced as a variational state for a Hamiltonian (or transfer matrix) living on the UV boundary layer.
In contrast,  holographic TN models are often constructed by covering a hyperbolic plane with tensors having a particular property, away from the variational optimization.
In HaPPY code model \cite{HaPPY2015}, for example, tiling of perfect tensors of the pentagon shape on a hyperbolic plane generates a holographic quantum error collection code on the basis of the bulk boundary correspondence.
Also, physics of random TNs attracts much attention in connection with the holography\cite{Hayden2016,Vassur2019}, where a nontrivial phase transition can be driven by tensors arranged in the bulk regime rather than a Hamiltonian at the UV boundary.
How these two competing directions of the TN, i.e. from the boundary to bulk or from the bulk to boundary, can be consistent in the holographic description of the TN state may be an interesting problem.
It may be intriguing that the holographic geometry was numerically reproduced by a machine learning approach assisted by the MERA network.~\cite{Hu2020}


To summarize,  the TN has been established as a practical and useful numerical method for investigating quantum many-body systems and its application range has been currently expanding.
In accordance with revealing rich physics behind the TN , moreover,  it has become an essential theoretical concept beyond the conventional classification of research fields in physics. 
We hope that this review can play a role of comps in navigating foundries of TN physics.

 \begin{acknowledgment}
%


We would like to thank Y. Akutsu, Y. Hieida, N. Maeshima, M. Kikuchi, T. Hikihara, A. Gendier, H. Katsura, T. Okubo, and K. Harada for various valuable discussions.
K. O. is grateful to Y. Akutsu for guiding him to the research field of tensor network, although the term ``tensor network" did not exist in the 20th century.

This work is partially supported by KAKENHI Grant Nos. JP17H02931, JP17K05578, JP17K14359, JP21K03403, JP21H04446 and  a Grant-in-Aid for Transformative Research Areas "The Natural Laws of Extreme Universe---A New Paradigm for Spacetime and Matter from Quantum Information" (KAKENHI Grant Nos. JP21H05182, JP21H05191) from JSPS of Japan.
It is also supported by MEXT Q-LEAP Grant No. JPMXS0120319794,  JST PRESTO No. JPMJPR1911 and the COE research grant in computational science from Hyogo Prefecture and Kobe City through Foundation for Computational Science.

\end{acknowledgment}

\bibliography{90050}

\begin{thebibliography}{100}

\bibitem{Peschel1999}
{\em Density-Matrix Renormalization A New Numerical Method in Physics}, ed.
  I.~Peschel, M.~Kaulke, X.~Wang, and K.~Hallberg (Springer, 1998).

\bibitem{SchollwoeckRMP}
U.~Schollw{\"o}ck: Rev. Mod. Phys. {\bfseries 77} (2005) 259.

\bibitem{Schollwoeck}
U.~Schollw{\"o}ck: Ann. Phys. (N.Y.) {\bfseries 326} (2011) 96 .

\bibitem{Verstraete2008}
F.~Verstraete, V.~Murg, and J.~Cirac: Adv. Phys. {\bfseries 57} (2008) 143.

\bibitem{Orus2014review}
R.~Or\'{u}s: Ann. Phys. (N.Y.) {\bfseries 349} (2014) 117 .

\bibitem{Montangero2018}
S.~Montangero: {\em Introduction to Tensor Network Methods} (Springer).

\bibitem{Zeng_book}
B.~{Zeng}, X.~{Chen}, D.-L. {Zhou}, and X.-G. {Wen}: arXiv e-prints  (2015)
  arXiv:1508.02595.

\bibitem{Biamonte2019lectures}
J.~Biamonte: arXiv  (2019) 1912.10049.

\bibitem{Ran2020}
S.-J. Ran, E.~Tirrito, C.~Peng, X.~Chen, L.~Tagliacozzo, G.~Su, and
  M.~Lewenstein: {\em Tensor Network Contractions} (Springer, 2020).

\bibitem{Baxter1968}
R.~J. Baxter: J. Math. Phys. {\bfseries 9} (1968) 650.

\bibitem{Baxter1978}
R.~J. Baxter: J. Stat. Phys. {\bfseries 19} (1978) 461.

\bibitem{White1992}
S.~R. White: Phys. Rev. Lett. {\bfseries 69} (1992) 2863.

\bibitem{White1993}
S.~R. White: Phys. Rev. B {\bfseries 48} (1993) 10345.

\bibitem{Schmidt1907}
E.~Schmidt: Math. Ann {\bfseries 63} (1907) 433.

\bibitem{Ekert1995}
A.~Ekert and P.~L. Knight: Am. J. Phys. {\bfseries 63} (1995) 415.

\bibitem{MERA2007}
G.~Vidal: Phys. Rev. Lett. {\bfseries 99} (2007) 220405.

\bibitem{TNR2015}
G.~Evenbly and G.~Vidal: Phys. Rev. Lett. {\bfseries 115} (2015) 180405.

\bibitem{Kadanoff1966}
L.~P. Kadanoff: Physics Physique Fizika {\bfseries 2} (1966) 263.

\bibitem{Efrati2014}
E.~Efrati, Z.~Wang, A.~Kolan, and L.~P. Kadanoff: Rev. Mod. Phys. {\bfseries
  86} (2014) 647.

\bibitem{KW}
H.~A. Kramers and G.~H. Wannier: Phys. Rev. {\bfseries 60} (1941) 263.

\bibitem{KW_MPS}
Exactly speaking, the variational state in the Kramer-Wannier approximation is
  based on the representation of the interaction round a face model, where spin
  degrees of freedom are sitting on lattice sites. For quantum spin systems, on
  the other hand, the MPS is usually constructed with the vertex-model
  representation, where spin degrees of freedom are located on links of lattice
  points.

\bibitem{BaxterBook}
R.~J. Baxter: {\em Exactly solved models in statistical mechanics} (Academic
  Press, 1982).

\bibitem{AKLT1}
I.~Affleck, T.~Kennedy, E.~H. Lieb, and H.~Tasaki: Phys. Rev. Lett. {\bfseries
  59} (1987) 799.

\bibitem{AKLT2}
I.~Affleck, T.~Kennedy, E.~H. Lieb, and H.~Tasaki: Comm. Math. Phys. {\bfseries
  115} (1988) 477.

\bibitem{Fannes1989}
M.~Fannes, B.~Nachtergaele, and R.~F. Werner: Europhys. Lett. {\bfseries 10}
  (1989) 633.

\bibitem{Fannes1992}
M.~Fannes, B.~Nachtergaele, and R.~F. Werner: Comm. Math. Phys. {\bfseries 144}
  (1992) 443.

\bibitem{Fannes1996}
M.~Fannes, B.~Nachtergaele, and R.~F. Werner: Comm. Math. Phys. {\bfseries 174}
  (1996) 477.

\bibitem{Klumper1991}
A.~Kl\"umper, A.~Schadschneider, and J.~Zittartz: J. Phys. A: Math. Gen.
  {\bfseries 24} (1991) L955.

\bibitem{Klumper1992}
A.~Kl{\"u}mper, A.~Schadschneider, and J.~Zittartz: Z. Phys., B Condens. matter
  {\bfseries 87} (1992) 281.

\bibitem{Klumper1993}
A.~Kl\"umper, A.~Schadschneider, and J.~Zittartz: Europhys. Lett. {\bfseries
  24} (1993) 293.

\bibitem{Gu_Filtering2009}
Z.-C. Gu and X.-G. Wen: Phys. Rev. B {\bfseries 80} (2009) 155131.

\bibitem{Pollmann2010}
F.~Pollmann, A.~M. Turner, E.~Berg, and M.~Oshikawa: Phys. Rev. B {\bfseries
  81} (2010) 064439.

\bibitem{denNijs1}
K.~Rommelse and M.~den Nijs: Phys. Rev. Lett. {\bfseries 59} (1987) 2578.

\bibitem{denNijs2}
M.~den Nijs and K.~Rommelse: Phys. Rev. B {\bfseries 40} (1989) 4709.

\bibitem{Nightingale}
M.~P. Nightingale and H.~W.~J. Bl\"ote: Phys. Rev. B {\bfseries 33} (1986) 659.

\bibitem{Haldane}
F.~D.~M. Haldane: Phys. Rev. Lett. {\bfseries 50} (1983) 1153.

\bibitem{Baxter1972}
R.~J. Baxter: Ann. Phys. (N.Y.) {\bfseries 70} (1972) 193 .

\bibitem{Davies1993}
B.~Davies, O.~Foda, M.~Jimbo, T.~Miwa, and A.~Nakayashiki: Commun. Math. Phys.
  {\bfseries 151} (1993) 89.

\bibitem{Faddeev1995}
L.~Faddeev: Int. J. Mod. Phys. A {\bfseries 10} (1995) 1845.

\bibitem{Korepin1993}
V.~E. Korepin, N.~M. Bogoliubov, and A.~G. Izergin: {\em Quantum inverse
  scattering method and correlation functions} (Cambridge University Press,
  1993), Vol.~3.

\bibitem{Alcaraz2006}
F.~C. Alcaraz and M.~J. Lazo: J. Phys. A: Math. Gen. {\bfseries 39} (2006)
  11335.

\bibitem{Katsura2010}
H.~Katsura and I.~Maruyama: J. Phys. A: Math. Theor. {\bfseries 43} (2010)
  175003.

\bibitem{Murg2012}
V.~Murg, V.~E. Korepin, and F.~Verstraete: Phys. Rev. B {\bfseries 86} (2012)
  045125.

\bibitem{preDMRG}
At that time, such a conventional real-space RG as block spin transformation
  was thought of as a less reliable method for low-dimensional quantum
  many-body systems, since it often yielded even qualitatively wrong results.
  In modern terminology, this is because the block-spin transformation usually
  generates a tree-type TN, which may not be appropriate to describe strong
  quantum fluctuations in one dimension. We can guess that a lot of condensed
  matter theorists might have a cautious attitude to DMRG. What happened on
  Whites' DMRG paper was stated in the preface of Ref.[\citen{Peschel1999}].

\bibitem{Ostlund1995}
S.~\"Ostlund and S.~Rommer: Phys. Rev. Lett. {\bfseries 75} (1995) 3537.

\bibitem{Rommer1997}
S.~Rommer and S.~\"Ostlund: Phys. Rev. B {\bfseries 55} (1997) 2164.

\bibitem{PWFRG}
T.~Nishino and K.~Okunishi: J. Phys. Soc. Jpn. {\bfseries 64} (1995) 4084.

\bibitem{White1996}
S.~R. White: Phys. Rev. Lett. {\bfseries 77} (1996) 3633.

\bibitem{Hieida1997}
Y.~Hieida, K.~Okunishi, and Y.~Akutsu: Phys. Lett. A {\bfseries 233} (1997) 464
  .

\bibitem{NishinoDMRG1995}
T.~Nishino: J. Phys. Soc. Jpn. {\bfseries 64} (1995) 3598.

\bibitem{NishinoHOH1999}
T.~Nishino, T.~Hikihara, K.~Okunishi, and Y.~Hieida: Int. J. Mod. Phys. B
  {\bfseries 13} (1999) 1.

\bibitem{Bursill1996}
R.~J. Bursill, T.~Xiang, and G.~A. Gehring: J. Phys. Condens. Matter {\bfseries
  8} (1996) L583.

\bibitem{WangXiang1997}
X.~Wang and T.~Xiang: Phys. Rev. B {\bfseries 56} (1997) 5061.

\bibitem{Shibata1997}
N.~Shibata: J. Phys. Soc. Jpn. {\bfseries 66} (1997) 2221.

\bibitem{Shibata2003}
N.~Shibata: J. Phys. A: Math. Gen. {\bfseries 36} (2003) R381.

\bibitem{Okunishi1999}
K.~Okunishi: Phys. Rev. B {\bfseries 60} (1999) 4043.

\bibitem{Suzuki1987}
M.~Suzuki and M.~Inoue: Prog. Theor. Phys. {\bfseries 78} (1987) 787.

\bibitem{Trotter1959}
H.~F. Trotter: Proc. Am. Math. Soc. {\bfseries 10} (1959) 545.

\bibitem{Suzuki1976}
M.~Suzuki: Prog. Theor. Phys. {\bfseries 56} (1976) 1454.

\bibitem{Okunishi1996}
K.~Okunishi: Master hesis, Osaka University (in Japanese)  (1996).

\bibitem{CTMRG1}
T.~Nishino and K.~Okunishi: J. Phys. Soc. Jpn. {\bfseries 65} (1996) 891.

\bibitem{CTMRG2}
T.~Nishino and K.~Okunishi: J. Phys. Soc. Jpn. {\bfseries 66} (1997) 3040.

\bibitem{Peschel_Kaulke_Legeza1999}
I.~Peschel, M.~Kaulke, and {\"O}.~Legeza: Ann. Phys. (Berl.) {\bfseries 8}
  (1999) 153.

\bibitem{OHA}
K.~Okunishi, Y.~Hieida, and Y.~Akutsu: Phys. Rev. E {\bfseries 59} (1999)
  R6227.

\bibitem{Lefevre2008}
P.~Calabrese and A.~Lefevre: Phys. Rev. A {\bfseries 78} (2008) 032329.

\bibitem{Cho2017}
G.~Y. Cho, A.~W.~W. Ludwig, and S.~Ryu: Phys. Rev. B {\bfseries 95} (2017)
  115122.

\bibitem{Orus2009}
R.~Or\'us and G.~Vidal: Phys. Rev. B {\bfseries 80} (2009) 094403.

\bibitem{Fishman2018}
M.~T. Fishman, L.~Vanderstraeten, V.~Zauner-Stauber, J.~Haegeman, and
  F.~Verstraete: Phys. Rev. B {\bfseries 98} (2018) 235148.

\bibitem{Derrida1993}
B.~Derrida, M.~R. Evans, V.~Hakim, and V.~Pasquier: J. Phys. A: Math. Gen.
  {\bfseries 26} (1993) 1493.

\bibitem{Derrida1998}
B.~Derrida: Phys. Rep. {\bfseries 301} (1998) 65 .

\bibitem{ASEP2007}
R.~A. Blythe and M.~R. Evans: J. Phys. A: Math. Theor. {\bfseries 40} (2007)
  R333.

\bibitem{Hieida1998}
Y.~Hieida: J. Phys. Soc. Jpn. {\bfseries 67} (1998) 369.

\bibitem{Nagy2002}
Z.~Nagy, C.~Appert, and L.~Santen: J. Stat. Phys. {\bfseries 109} (2002) 623.

\bibitem{Carlon1999}
E.~Carlon, M.~Henkel, and U.~Schollw{\"o}ck: Eur. Phys. J. B {\bfseries 12}
  (1999) 99.

\bibitem{pnorm}
For a vector $\vec{r}=(x_1, x_2,\cdots x_N)$, the $p$-norm is defined as
  $||\vec{r}||_p := \sqrt[p]{|x_1|^p+|x_2|^p+\cdots + |x_N|^p}$.

\bibitem{Enss2001}
T.~Enss and U.~Schollw\"{o}ck: J. Phys. A: Math. Gen. {\bfseries 34} (2001)
  7769.

\bibitem{Johnson2010}
T.~H. Johnson, S.~R. Clark, and D.~Jaksch: Phys. Rev. E {\bfseries 82} (2010)
  036702.

\bibitem{Harada2019}
K.~Harada and N.~Kawashima: Phys. Rev. Lett. {\bfseries 123} (2019) 090601.

\bibitem{Liang1994}
S.~Liang and H.~Pang: Phys. Rev. B {\bfseries 49} (1994) 9214.

\bibitem{Jeckelmann1998}
E.~Jeckelmann and S.~R. White: Phys. Rev. B {\bfseries 57} (1998) 6376.

\bibitem{Jeckelmann2002}
E.~Jeckelmann: Phys. Rev. B {\bfseries 66} (2002) 045114.

\bibitem{Hida1996}
K.~Hida: Journal of the Physical Society of Japan {\bfseries 65} (1996) 895.

\bibitem{Xiang1996}
T.~Xiang: Phys. Rev. B {\bfseries 53} (1996) R10445.

\bibitem{Shibata2001}
N.~Shibata and D.~Yoshioka: Phys. Rev. Lett. {\bfseries 86} (2001) 5755.

\bibitem{White1999}
S.~R. White and R.~L. Martin: J. Chem. Phys. {\bfseries 110} (1999) 4127.

\bibitem{Vidal_Latorre2003}
G.~Vidal, J.~I. Latorre, E.~Rico, and A.~Kitaev: Phys. Rev. Lett. {\bfseries
  90} (2003) 227902.

\bibitem{Calabrese2004}
P.~Calabrese and J.~Cardy: J. Stat. Mech. Theory Exp. {\bfseries 2004} (2004)
  P06002.

\bibitem{TEBD}
G.~Vidal: Phys. Rev. Lett. {\bfseries 93} (2004) 040502.

\bibitem{iTEBD}
G.~Vidal: Phys. Rev. Lett. {\bfseries 98} (2007) 070201.

\bibitem{Zauner-Stauber2018}
V.~Zauner-Stauber, L.~Vanderstraeten, M.~T. Fishman, F.~Verstraete, and
  J.~Haegeman: Phys. Rev. B {\bfseries 97} (2018) 045145.

\bibitem{Daley2004}
A.~J. Daley, C.~Kollath, U.~Schollw\"{o}ck, and G.~Vidal: J. Stat. Mech. Theory
  Exp. {\bfseries 2004} (2004) P04005.

\bibitem{WhiteFeiguin2004}
S.~R. White and A.~E. Feiguin: Phys. Rev. Lett. {\bfseries 93} (2004) 076401.

\bibitem{Xiang2001}
T.~Xiang, J.~Lou, and Z.~Su: Phys. Rev. B {\bfseries 64} (2001) 104414.

\bibitem{Legeza2003}
O.~Legeza and J.~S\'olyom: Phys. Rev. B {\bfseries 68} (2003) 195116.

\bibitem{Legeza2004}
O.~Legeza and J.~S\'olyom: Phys. Rev. B {\bfseries 70} (2004) 205118.

\bibitem{McCulloch2007}
I.~P. McCulloch: J. Stat. Mech. Theory Exp. {\bfseries 2007} (2007) P10014.

\bibitem{Eisert_RMP2010}
J.~Eisert, M.~Cramer, and M.~B. Plenio: Rev. Mod. Phys. {\bfseries 82} (2010)
  277.

\bibitem{MERA2008}
G.~Vidal: Phys. Rev. Lett. {\bfseries 101} (2008) 110501.

\bibitem{RT_PRL2006}
S.~Ryu and T.~Takayanagi: Phys. Rev. Lett. {\bfseries 96} (2006) 181602.

\bibitem{RT_JHEP2006}
S.~Ryu and T.~Takayanagi: J. High Energy Phys. {\bfseries 2006} (2006) 045.

\bibitem{CTTRG}
T.~Nishino and K.~Okunishi: J. Phys. Soc. Jpn. {\bfseries 67} (1998) 3066.

\bibitem{KW3D}
K.~Okunishi and T.~Nishino: Prog. Theor. Phys. {\bfseries 103} (2000) 541.

\bibitem{TPVA1}
T.~Nishino, K.~Okunishi, Y.~Hieida, N.~Maeshima, and Y.~Akutsu: Nucl. Phys. B
  {\bfseries 575} (2000) 504 .

\bibitem{TPVA2}
A.~Gendiar and T.~Nishino: Phys. Rev. E {\bfseries 65} (2002) 046702.

\bibitem{TPS1}
T.~Nishino, Y.~Hieida, K.~Okunishi, N.~Maeshima, Y.~Akutsu, and A.~Gendiar:
  Prog. Theor. Phys. {\bfseries 105} (2001) 409.

\bibitem{TPS2}
A.~Gendiar, N.~Maeshima, and T.~Nishino: Prog. Theor. Phys. {\bfseries 110}
  (2003) 691.

\bibitem{Corboz2016}
P.~Corboz: Phys. Rev. B {\bfseries 94} (2016) 035133.

\bibitem{iPEPS}
J.~Jordan, R.~Or\'us, G.~Vidal, F.~Verstraete, and J.~I. Cirac: Phys. Rev.
  Lett. {\bfseries 101} (2008) 250602.

\bibitem{Niggemann1997}
H.~Niggemann, A.~Kl\"{u}mper, and J.~Zittartz: Z. Phys., B Condens. matter
  {\bfseries 104} (1997) 103.

\bibitem{Hieida1999}
Y.~Hieida, K.~Okunishi, and Y.~Akutsu: New J. Phys. {\bfseries 1} (1999) 7.

\bibitem{qTPVA}
Y.~Nishio, N.~Maeshima, A.~Gendiar, and T.~Nishino: arXiv cond-mat/0401115
  (2004).

\bibitem{VDMA}
N.~Maeshima, Y.~Hieida, Y.~Akutsu, T.~Nishino, and K.~Okunishi: Phys. Rev. E
  {\bfseries 64} (2001) 016705.

\bibitem{Maeshima2004}
N.~Maeshima: J. Phys. Soc. Jpn. {\bfseries 73} (2004) 60.

\bibitem{PEPS}
F.~Verstraete and J.~Cirac: arXiv:cond-mat/0407066  (2004).

\bibitem{Murg2007}
V.~Murg, F.~Verstraete, and J.~I. Cirac: Phys. Rev. A {\bfseries 75} (2007)
  033605.

\bibitem{Xiang_simple2008}
H.~C. Jiang, Z.~Y. Weng, and T.~Xiang: Phys. Rev. Lett. {\bfseries 101} (2008)
  090603.

\bibitem{Gracia2008}
D.~P\'erez-Garc\'{\i}a, M.~M. Wolf, M.~Sanz, F.~Verstraete, and J.~I. Cirac:
  Phys. Rev. Lett. {\bfseries 100} (2008) 167202.

\bibitem{Gu_String2009}
Z.-C. Gu, M.~Levin, B.~Swingle, and X.-G. Wen: Phys. Rev. B {\bfseries 79}
  (2009) 085118.

\bibitem{Schuch2010}
N.~Schuch, I.~Cirac, and D.~P\'{e}rez-Garc\'{i}a: Ann. Phys. (N.Y.) {\bfseries
  325} (2010) 2153 .

\bibitem{Schuch2011}
N.~Schuch, D.~P\'erez-Garc\'{\i}a, and I.~Cirac: Phys. Rev. B {\bfseries 84}
  (2011) 165139.

\bibitem{Cirac2020}
I.~Cirac, D.~Perez-Garcia, N.~Schuch, and F.~Verstraete: arXiv:2011.12127 .

\bibitem{Wei2016}
C.-Y. Huang and T.-C. Wei: Phys. Rev. B {\bfseries 93} (2016) 155163.

\bibitem{Wei2020}
N.~Pomata and T.-C. Wei: Phys. Rev. Lett. {\bfseries 124} (2020) 177203.

\bibitem{MBQC2001}
R.~Raussendorf and H.~J. Briegel: Phys. Rev. Lett. {\bfseries 86} (2001) 5188.

\bibitem{GrossEisert2007}
D.~Gross and J.~Eisert: Phys. Rev. Lett. {\bfseries 98} (2007) 220503.

\bibitem{VerstraeteCirac2004}
F.~Verstraete and J.~I. Cirac: Phys. Rev. A {\bfseries 70} (2004) 060302.

\bibitem{FujiiMorimae2012}
K.~Fujii and T.~Morimae: Phys. Rev. A {\bfseries 85} (2012) 032338.

\bibitem{Burkhardt1982}
T.~W. Burkhardt and J.~M.~J. van Leeuwen: {\em Real-Space Renormalization}
  (Topics in Current Physics 30. Springer, 1982), Topics in Current Physics 30.

\bibitem{TRG2007}
M.~Levin and C.~P. Nave: Phys. Rev. Lett. {\bfseries 99} (2007) 120601.

\bibitem{HOTRG2012}
Z.~Y. Xie, J.~Chen, M.~P. Qin, J.~W. Zhu, L.~P. Yang, and T.~Xiang: Phys. Rev.
  B {\bfseries 86} (2012) 045139.

\bibitem{Yoshimura2018}
Y.~Yoshimura, Y.~Kuramashi, Y.~Nakamura, S.~Takeda, and R.~Sakai: Phys. Rev. D
  {\bfseries 97} (2018) 054511.

\bibitem{Akiyama2019}
S.~Akiyama, Y.~Kuramashi, T.~Yamashita, and Y.~Yoshimura: Phys. Rev. D
  {\bfseries 100} (2019) 054510.

\bibitem{Wang2014}
S.~Wang, Z.-Y. Xie, J.~Chen, B.~Normand, and T.~Xiang: Chin. Phys. Lett.
  {\bfseries 31} (2014) 070503.

\bibitem{Gu_TRG2008}
Z.-C. Gu, M.~Levin, and X.-G. Wen: Phys. Rev. B {\bfseries 78} (2008) 205116.

\bibitem{Ueda2014}
H.~Ueda, K.~Okunishi, and T.~Nishino: Phys. Rev. B {\bfseries 89} (2014)
  075116.

\bibitem{Evenbly2009}
G.~Evenbly and G.~Vidal: Phys. Rev. B {\bfseries 79} (2009) 144108.

\bibitem{TNRtoMERA}
G.~Evenbly and G.~Vidal: Phys. Rev. Lett. {\bfseries 115} (2015) 200401.

\bibitem{LoopTNR2017}
S.~Yang, Z.-C. Gu, and X.-G. Wen: Phys. Rev. Lett. {\bfseries 118} (2017)
  110504.

\bibitem{Harada2018}
K.~Harada: Phys. Rev. B {\bfseries 97} (2018) 045124.

\bibitem{GILT2018}
M.~Hauru, C.~Delcamp, and S.~Mizera: Phys. Rev. B {\bfseries 97} (2018) 045111.

\bibitem{Swingle2012}
B.~Swingle: Phys. Rev. D {\bfseries 86} (2012) 065007.

\bibitem{cMERA2013}
J.~Haegeman, T.~J. Osborne, H.~Verschelde, and F.~Verstraete: Phys. Rev. Lett.
  {\bfseries 110} (2013) 100402.

\bibitem{Nozaki2012}
M.~Nozaki, S.~Ryu, and T.~Takayanagi: J. High Energy Phys. {\bfseries 2012}
  (2012) 193.

\bibitem{Pfeifer2009}
R.~N.~C. Pfeifer, G.~Evenbly, and G.~Vidal: Phys. Rev. A {\bfseries 79} (2009)
  040301.

\bibitem{Evenbly2016}
G.~Evenbly and G.~Vidal: Phys. Rev. Lett. {\bfseries 116} (2016) 040401.

\bibitem{ketspace}
The matrix size of the row-to-row transfer matrix $T$ is $2^N \times 2^N$.
  Unless otherwise noted, we assume that the ket space of $|\Psi\rangle $ is
  represented by $2^N$ configurations of the spins in the row direction.

\bibitem{mhalf}
We consider $M/2 -1 $ for the power of $T$ in $|\Psi^{(M)}\rangle $ with $M(\ge
  2)$ beeing an even number, where ``$-1$'' comes from the contribution of the
  initial boundary row. Accordingly, the norm of $\langle \Psi^{(M)}|
  \Psi^{(M)}\rangle $ corresponds to the partition function of the system
  having $M$ rows.

\bibitem{ket_phi}
We assume that the ket space of $|\Phi_{X(Y)}\rangle $ is represented by $2
  m^2(m^2)$ configurations of the renormalized spins in the column direction,
  corresponding to the $X(Y)$ tensors.

\bibitem{C_leg}
The order of renormalized index $\mu$ and $\nu$ in $C(\mu|\nu)$ is inverted in
  comparison with Eq. (\ref{eq_ctm_step2}) where $s'_n$ is yet a bare leg spin
  index in $C_N(s_N|\mu_N)$.

\bibitem{imaginary_d}
Here, the vertical direction of the 2D lattice is assumed to be the imaginary
  time direction in the corresponding quantum system.

\bibitem{site_matrix}
In the present convention of the matrix representation, the site index $n$ is
  arranged in the descending order from left to right, which is the opposite
  order in the diagrammatic representation.

\bibitem{Ueda2010}
H.~Ueda, A.~Gendiar, and T.~Nishino: J. Phys. Soc. Jpn. {\bfseries 79} (2010)
  044001.

\bibitem{cutoffm}
In recent TN literature, $\chi$ is often assigned for the cutoff dimension of
  tensors, instead of $m$ in the DMRG/CTMRG cases.

\bibitem{Baxter1976}
R.~J. Baxter: J. Stat. Phys. {\bfseries 15} (1976) 485.

\bibitem{Baxter1977}
R.~J. Baxter: J. Stat. Phys. {\bfseries 17} (1977) 1.

\bibitem{GarnetChan2002}
G.~K.-L. Chan, P.~W. Ayers, and E.~S. Croot: J. Stat. Phys {\bfseries 109}
  (2002) 289.

\bibitem{Tsang1979}
S.~Tsang: J. Stat. Phys. {\bfseries 20} (1979) 95.

\bibitem{NOK1996}
T.~Nishino, K.~Okunishi, and M.~Kikuchi: Phys. Lett. A {\bfseries 213} (1996)
  69.

\bibitem{Tagliacozzo2008}
L.~Tagliacozzo, T.~R. de~Oliveira, S.~Iblisdir, and J.~I. Latorre: Phys. Rev. B
  {\bfseries 78} (2008) 024410.

\bibitem{Pollmann2009}
F.~Pollmann, S.~Mukerjee, A.~M. Turner, and J.~E. Moore: Phys. Rev. Lett.
  {\bfseries 102} (2009) 255701.

\bibitem{UedaIcosa}
H.~Ueda, K.~Okunishi, R.~Kr\ifmmode~\check{c}\else \v{c}\fi{}m\'ar, A.~Gendiar,
  S.~Yunoki, and T.~Nishino: Phys. Rev. E {\bfseries 96} (2017) 062112.

\bibitem{UedaBKT}
H.~Ueda, K.~Okunishi, K.~Harada, R.~Kr\ifmmode~\check{c}\else \v{c}\fi{}m\'ar,
  A.~Gendiar, S.~Yunoki, and T.~Nishino: Phys. Rev. E {\bfseries 101} (2020)
  062111.

\bibitem{UedaDodeca}
H.~Ueda, K.~Okunishi, S.~Yunoki, and T.~Nishino: Phys. Rev. E {\bfseries 102}
  (2020) 032130.

\bibitem{Pirvu2012}
B.~Pirvu, G.~Vidal, F.~Verstraete, and L.~Tagliacozzo: Phys. Rev. B {\bfseries
  86} (2012) 075117.

\bibitem{Schmoll2019}
P.~Schmoll, A.~Haller, M.~Rizzi, and R.~Or\'us: Phys. Rev. B {\bfseries 99}
  (2019) 205121.

\bibitem{Huang2012}
Y.-K. Huang, P.~Chen, and Y.-J. Kao: Phys. Rev. B {\bfseries 86} (2012) 235102.

\bibitem{comment_tilde}
In this subsection, we have omitted ``$\sim$" assigned for renormalized
  tensors/matrices.

\bibitem{Mcculloch2008}
I.~P. McCulloch: arXiv: 0804.2509  (2008).

\bibitem{Hofstetter2000}
W.~Hofstetter: Phys. Rev. Lett. {\bfseries 85} (2000) 1508.

\bibitem{Pirvu2010}
B.~Pirvu, V.~Murg, J.~I. Cirac, and F.~Verstraete: New J. Phys. {\bfseries 12}
  (2010) 025012.

\bibitem{Cazalilla2002}
M.~A. Cazalilla and J.~B. Marston: Phys. Rev. Lett. {\bfseries 88} (2002)
  256403.

\bibitem{Ferstraete_MPDO2004}
F.~Verstraete, J.~J. Garc\'{\i}a-Ripoll, and J.~I. Cirac: Phys. Rev. Lett.
  {\bfseries 93} (2004) 207204.

\bibitem{Zwolak2004}
M.~Zwolak and G.~Vidal: Phys. Rev. Lett. {\bfseries 93} (2004) 207205.

\bibitem{Ueda_condmat}
K.~{Ueda}, C.~{Jin}, N.~{Shibata}, Y.~{Hieida}, and T.~{Nishino}:
  arXiv:cond-mat/0612480  (2006).

\bibitem{TDVP2011}
J.~Haegeman, J.~I. Cirac, T.~J. Osborne, I.~Pi\ifmmode~\check{z}\else
  \v{z}\fi{}orn, H.~Verschelde, and F.~Verstraete: Phys. Rev. Lett. {\bfseries
  107} (2011) 070601.

\bibitem{TDVP2016}
J.~Haegeman, C.~Lubich, I.~Oseledets, B.~Vandereycken, and F.~Verstraete: Phys.
  Rev. B {\bfseries 94} (2016) 165116.

\bibitem{Dirac_1930}
P.~A.~M. Dirac: Mathematical Proceedings of the Cambridge Philosophical Society
  {\bfseries 26} (1930) 376–385.

\bibitem{Frenkel1934}
J.~Frenkel: {\em Wave Mechanics, Advanced General Theory} (Clarendon Press,
  Oxford, 1934).

\bibitem{Paeckel_TD2019}
S.~Paeckel, T.~K\"{o}hler, A.~Swoboda, S.~R. Manmana, U.~Schollw\"{o}ck, and
  C.~Hubig: Ann. Phys. (N.Y.) {\bfseries 411} (2019) 167998.

\bibitem{tensorproduct}
The ``tensor product" state here indicates a state represented as a contraction
  of local tensors, which is just an analogy of the matrix-product state. Thus
  it is a different object from the tensor product, i.e. the direct product of
  matrices in mathematics.

\bibitem{comment_TPS}
We concentrate on the layer-to-layer transfer matrix with the TPS. The
  correspondence to the PEPS is straightforward.

\bibitem{Corboz2010}
P.~Corboz, R.~Or\'us, B.~Bauer, and G.~Vidal: Phys. Rev. B {\bfseries 81}
  (2010) 165104.

\bibitem{Zaletel2020}
M.~P. Zaletel and F.~Pollmann: Phys. Rev. Lett. {\bfseries 124} (2020) 037201.

\bibitem{Haghshenas2019}
R.~Haghshenas, M.~J. O'Rourke, and G.~K.-L. Chan: Phys. Rev. B {\bfseries 100}
  (2019) 054404.

\bibitem{Gradient2016}
L.~Vanderstraeten, J.~Haegeman, P.~Corboz, and F.~Verstraete: Phys. Rev. B
  {\bfseries 94} (2016) 155123.

\bibitem{AutoDiff2019}
H.-J. Liao, J.-G. Liu, L.~Wang, and T.~Xiang: Phys. Rev. X {\bfseries 9} (2019)
  031041.

\bibitem{Vanderstraeten2019}
L.~Vanderstraeten, J.~Haegeman, and F.~Verstraete: Phys. Rev. B {\bfseries 99}
  (2019) 165121.

\bibitem{Ponsioen2020}
B.~Ponsioen and P.~Corboz: Phys. Rev. B {\bfseries 101} (2020) 195109.

\bibitem{Lyu2021}
X.~Lyu, R.~G. Xu, and N.~Kawashima: arXiv:2102.08136  (2021).

\bibitem{HOSVD1}
L.~De~Lathauwer, B.~De~Moor, and J.~Vandewalle: SIAM J. Matrix Anal. Appl.
  {\bfseries 21} (2000) 1253.

\bibitem{HOSVD2}
L.~De~Lathauwer, B.~De~Moor, and J.~Vandewalle: SIAM J. Matrix Anal. Appl.
  {\bfseries 21} (2000) 1324.

\bibitem{Carmen_Banuls_LG2020}
M.~C. Ba{\~{n}}uls and K.~Cichy: Rep. Prog. Phys. {\bfseries 83} (2020) 024401.

\bibitem{SRG2009}
Z.~Y. Xie, H.~C. Jiang, Q.~N. Chen, Z.~Y. Weng, and T.~Xiang: Phys. Rev. Lett.
  {\bfseries 103} (2009) 160601.

\bibitem{Zhao2010}
H.~H. Zhao, Z.~Y. Xie, Q.~N. Chen, Z.~C. Wei, J.~W. Cai, and T.~Xiang: Phys.
  Rev. B {\bfseries 81} (2010) 174411.

\bibitem{Liu_gauge2013}
Y.~Liu, Y.~Meurice, M.~P. Qin, J.~Unmuth-Yockey, T.~Xiang, Z.~Y. Xie, J.~F. Yu,
  and H.~Zou: Phys. Rev. D {\bfseries 88} (2013) 056005.

\bibitem{ATRG}
D.~Adachi, T.~Okubo, and S.~Todo: Phys. Rev. B {\bfseries 102} (2020) 054432.

\bibitem{Evenbly2017}
G.~Evenbly: Phys. Rev. B {\bfseries 95} (2017) 045117.

\bibitem{Matsueda2013}
H.~Matsueda, M.~Ishihara, and Y.~Hashizume: Phys. Rev. D {\bfseries 87} (2013)
  066002.

\bibitem{Evenbly2013quantum}
G.~Evenbly and G.~Vidal, Quantum criticality with the multi-scale entanglement
  renormalization ansatz, Strongly correlated systems, pp. 99--130. Springer,
  2013.

\bibitem{Giovannetti2008}
V.~Giovannetti, S.~Montangero, and R.~Fazio: Phys. Rev. Lett. {\bfseries 101}
  (2008) 180503.

\bibitem{Milsted2018}
A.~Milsted and G.~Vidal: arXiv:1805.12524  (2018).

\bibitem{FFT1}
G.~Danielson and C.~Lanczos: J. Franklin Inst. {\bfseries 233} (1942) 365.

\bibitem{FFT2}
J.~W. Cooley and J.~W. Tukey: Math. Comp. {\bfseries 19} (1965) 297.

\bibitem{NumRecipes}
W.~H. Press, W.~T. Vetterling, S.~A. Teukolsky, and B.~P. Flannery: {\em
  Numerical recipes} (Cambridge university press Cambridge, 1986), Vol. 818.

\bibitem{Ferris2014}
A.~J. Ferris: Phys. Rev. Lett. {\bfseries 113} (2014) 010401.

\bibitem{Fishman2015}
M.~T. Fishman and S.~R. White: Phys. Rev. B {\bfseries 92} (2015) 075132.

\bibitem{Evenbly_White}
G.~Evenbly and S.~R. White: Phys. Rev. Lett. {\bfseries 116} (2016) 140403.

\bibitem{SDRG1}
S.-k. Ma, C.~Dasgupta, and C.-k. Hu: Phys. Rev. Lett. {\bfseries 43} (1979)
  1434.

\bibitem{DSFisher1994}
D.~S. Fisher: Phys. Rev. B {\bfseries 50} (1994) 3799.

\bibitem{Hikihara1999}
T.~Hikihara, A.~Furusaki, and M.~Sigrist: Phys. Rev. B {\bfseries 60} (1999)
  12116.

\bibitem{Goldsborough2014}
A.~M. Goldsborough and R.~A. R\"omer: Phys. Rev. B {\bfseries 89} (2014)
  214203.

\bibitem{Seki2020}
K.~Seki, T.~Hikihara, and K.~Okunishi: Phys. Rev. B {\bfseries 102} (2020)
  144439.

\bibitem{Daul2000}
S.~Daul, I.~Ciofini, C.~Daul, and S.~R. White: Int. J. Quantum Chem. {\bfseries
  79} (2000) 331.

\bibitem{Chan2002}
G.~K.-L. Chan and M.~Head-Gordon: J. Chem. Phys. {\bfseries 116} (2002) 4462.

\bibitem{LegezaChem2003}
O.~Legeza, J.~R\"oder, and B.~A. Hess: Phys. Rev. B {\bfseries 67} (2003)
  125114.

\bibitem{Chan2011}
G.~K.-L. Chan and S.~Sharma: Annu. Rev. Phys. Chem. {\bfseries 62} (2011) 465.

\bibitem{Barcza2011}
G.~Barcza, O.~Legeza, K.~H. Marti, and M.~Reiher: Phys. Rev. A {\bfseries 83}
  (2011) 012508.

\bibitem{Nakatani2013}
N.~Nakatani and G.~K.-L. Chan: J. Chem. Phys. {\bfseries 138} (2013) 134113.

\bibitem{ChemRev2015}
S.~Szalay, M.~Pfeffer, V.~Murg, G.~Barcza, F.~Verstraete, R.~Schneider, and
  O.~Legeza: Int. J. Quantum Chem. {\bfseries 115} (2015) 1342.

\bibitem{Chan2016}
G.~K.-L. Chan, A.~Keselman, N.~Nakatani, Z.~Li, and S.~R. White: J. Chem. Phys.
  {\bfseries 145} (2016) 014102.

\bibitem{Stoudenmire2016}
E.~Stoudenmire and D.~J. Schwab: In D.~Lee, M.~Sugiyama, U.~Luxburg, I.~Guyon,
  and R.~Garnett (eds), {\em Advances in Neural Information Processing
  Systems}, Vol.~29, 2016, pp. 4799--4807.

\bibitem{Gallego2019}
A.~J. Gallego and R.~Orus: arXiv:1708.01525  (2017).

\bibitem{Chen2018}
J.~Chen, S.~Cheng, H.~Xie, L.~Wang, and T.~Xiang: Phys. Rev. B {\bfseries 97}
  (2018) 085104.

\bibitem{Li2018}
S.-H. Li and L.~Wang: Phys. Rev. Lett. {\bfseries 121} (2018) 260601.

\bibitem{Google2019}
F.~Arute, K.~Arya, R.~Babbush, D.~Bacon, J.~Bardin, R.~Barends, R.~Biswas,
  S.~Boixo, F.~Brandao, D.~Buell, B.~Burkett, Y.~Chen, J.~Chen, B.~Chiaro,
  R.~Collins, W.~Courtney, A.~Dunsworth, E.~Farhi, B.~Foxen, A.~Fowler, C.~M.
  Gidney, M.~Giustina, R.~Graff, K.~Guerin, S.~Habegger, M.~Harrigan,
  M.~Hartmann, A.~Ho, M.~R. Hoffmann, T.~Huang, T.~Humble, S.~Isakov,
  E.~Jeffrey, Z.~Jiang, D.~Kafri, K.~Kechedzhi, J.~Kelly, P.~Klimov, S.~Knysh,
  A.~Korotkov, F.~Kostritsa, D.~Landhuis, M.~Lindmark, E.~Lucero, D.~Lyakh,
  S.~Mandrà, J.~R. McClean, M.~McEwen, A.~Megrant, X.~Mi, K.~Michielsen,
  M.~Mohseni, J.~Mutus, O.~Naaman, M.~Neeley, C.~Neill, M.~Y. Niu, E.~Ostby,
  A.~Petukhov, J.~Platt, C.~Quintana, E.~G. Rieffel, P.~Roushan, N.~Rubin,
  D.~Sank, K.~J. Satzinger, V.~Smelyanskiy, K.~J. Sung, M.~Trevithick,
  A.~Vainsencher, B.~Villalonga, T.~White, Z.~J. Yao, P.~Yeh, A.~Zalcman,
  H.~Neven, and J.~Martinis: Nature {\bfseries 574} (2019) 505–510.

\bibitem{Kawaguchi2004}
A.~Kawaguchi, K.~Shimizu, Y.~Tokura, and N.~Imoto: quant-ph/0411205  (2004).

\bibitem{Dang2019}
A.~Dang, C.~D. Hill, and L.~C.~L. Hollenberg: {Quantum} {\bfseries 3} (2019)
  116.

\bibitem{Dumitrescu2017}
E.~Dumitrescu: Phys. Rev. A {\bfseries 96} (2017) 062322.

\bibitem{Zhou2020}
Y.~Zhou, E.~M. Stoudenmire, and X.~Waintal: Phys. Rev. X {\bfseries 10} (2020)
  041038.

\bibitem{Huang2020}
C.~Huang, F.~Zhang, M.~Newman, J.~Cai, X.~Gao, Z.~Tian, J.~Wu, H.~Xu, H.~Yu,
  B.~Yuan, M.~Szegedy, Y.~Shi, and J.~Chen: arXiv:2005.06787  (2020).

\bibitem{Liu2019}
J.-G. Liu, Y.-H. Zhang, Y.~Wan, and L.~Wang: Phys. Rev. Research {\bfseries 1}
  (2019) 023025.

\bibitem{RanMPS2020}
S.-J. Ran: Phys. Rev. A {\bfseries 101} (2020) 032310.

\bibitem{Peruzzo2014}
A.~Peruzzo, J.~McClean, P.~Shadbolt, M.-H. Yung, X.-Q. Zhou, P.~J. Love,
  A.~Aspuru-Guzik, and J.~L. O'Brien: Nat. Commun. {\bfseries 5} (2014) 4213.

\bibitem{VQE2021review}
J.~Tilly, H.~Chen, S.~Cao, D.~Picozzi, K.~Setia, Y.~Li, E.~Grant, L.~Wossnig,
  I.~Rungger, G.~H. Booth, and J.~Tennyson: ArXiv:2111.05176  (2021).

\bibitem{LiChenFisher2019}
Y.~Li, X.~Chen, and M.~P.~A. Fisher: Phys. Rev. B {\bfseries 100} (2019)
  134306.

\bibitem{Goto2020}
S.~Goto and I.~Danshita: Phys. Rev. A {\bfseries 102} (2020) 033316.

\bibitem{XianLiFisher2021}
X.~Chen, Y.~Li, M.~P.~A. Fisher, and A.~Lucas: Phys. Rev. Research {\bfseries
  2} (2020) 033017.

\bibitem{METTS2009}
S.~R. White: Phys. Rev. Lett. {\bfseries 102} (2009) 190601.

\bibitem{Stoudenmire2010}
E.~M. Stoudenmire and S.~R. White: New J. Phys. {\bfseries 12} (2010) 055026.

\bibitem{Iitaka2020}
T.~Iitaka: arXiv:006.14459  (2020).

\bibitem{Goto2021}
S.~Goto, R.~Kaneko, and I.~Danshita: Phys. Rev. B {\bfseries 104} (2021)
  045133.

\bibitem{Iwaki2021}
A.~Iwaki, A.~Shimizu, and C.~Hotta: Phys. Rev. Research {\bfseries 3} (2021)
  L022015.

\bibitem{Caputa2017}
P.~Caputa, N.~Kundu, M.~Miyaji, T.~Takayanagi, and K.~Watanabe: J. High Energy
  Phys. {\bfseries 11} (2017) 097.

\bibitem{Jefferson2017}
R.~Jefferson and R.~C. Myers: J. High Energy Phys. {\bfseries 10} (2017) 107.

\bibitem{Molina-Vilaplana2018}
J.~Molina-Vilaplana and A.~Del~Campo: J. High Energy Phys. {\bfseries 08}
  (2018) 012.

\bibitem{HaPPY2015}
F.~Pastawski, B.~Yoshida, D.~Harlow, and J.~Preskill: J. High Energy Phys.
  {\bfseries 2015} (2015) 1.

\bibitem{Hayden2016}
P.~Hayden, S.~Nezami, X.-L. Qi, N.~Thomas, M.~Walter, and Z.~Yang: J. High
  Energy Phys. {\bfseries 2016} (2016) 1.

\bibitem{Vassur2019}
R.~Vasseur, A.~C. Potter, Y.-Z. You, and A.~W.~W. Ludwig: Phys. Rev. B
  {\bfseries 100} (2019) 134203.

\bibitem{Hu2020}
H.-Y. Hu, S.-H. Li, L.~Wang, and Y.-Z. You: Phys. Rev. Research {\bfseries 2}
  (2020) 023369.

\end{thebibliography}

\appendix
\section{TN Packages}

As discussed so far, the TN algorithms basically consist of contraction and operation of various tensors.
In developments of their source cords from scratch, one often suffers from bug fixing of complicated tensor operation.
Recently, it becomes standard to use TN packages.
For readers' convenience, we list up some of TN packages widely used as quantum simulators below, which also include lower-level functions of tensor contraction and operation.
\begin{itemize}

\item ``Itensor library"  by  M. Fishman, E. M. Stoudenmire, and S. R. White.\\
{\tt http://itensor.org/}
\item "Uni10: an opensource library for tensor network algorithms” by  Y.-J. Kao, Y.-D. Hsieh, and P. Chen,\\
{\tt https://uni10.gitlab.io/} 
\item ``Tensoroperations" by J. Haegeman.\\
{\tt https://github.com/Jutho/TensorOperations.jl} 
\item ``Tensor Network Python (TeNPy)" by J. Hauschild and F. Pollmann,\\
{\tt https://tenpy.readthedocs.io/en/latest/}
\item ``Tensortrace: an application to contract tensor networks" by G. Evenbly.\\
{\tt https://www.tensortrace.com/}
\item ``Tenes: Massively parallel tensor network solver" by T. Okubo, S. Morita, Y. Motoyama, K. Yoshimi, T. Kato, and N.Kawashima, \\
 {\tt https://www.pasums.issp.u-tokyo.ac.jp/tenes/}
\end{itemize}

\end{document}